%% file: main.tex
\documentclass[12pt, draftclsnofoot, onecolumn]{IEEEtran}
\usepackage{cite}
\usepackage{amsmath,amssymb,amsfonts}
\usepackage{algorithm}
\usepackage{algpseudocode}%
\usepackage{graphicx}
\usepackage{hyperref}
\usepackage{textcomp}
\usepackage{xcolor}
\usepackage{subcaption}
\usepackage{multirow}
\usepackage{siunitx}
\usepackage{lipsum}
\usepackage{mathtools}
\usepackage{cuted}
\usepackage{stfloats}
\usepackage{bbm} % for indicator function

\usepackage{enumitem}

\usepackage{amsthm}
\usepackage{tabularx}
\makeatletter
\newcommand{\multiline}[1]{%
  \begin{tabularx}{\dimexpr\linewidth-\ALG@thistlm}[t]{@{}X@{}}
    #1
  \end{tabularx}
}
\makeatother

\usepackage[font=small]{caption}

\allowdisplaybreaks
\begin{document}

% Adaptive Intelligent Reflecting Surface Control with Limited Feedback 
% under Practical Reflection Behavior and Time-Varying Channels

\title{ 
% Codebook, Beamformer, Limited Feedback, Adaptive,....
%Intelligent Reflecting Surfaces-Assisted Communication in Varying Channels: A Deep Reinforcement Learning Approach
% Adaptive IRS Control with Limited Feedback 
% in Uplink IRS-assisted  Communications 
% under Practical Considerations
Learning-Based Adaptive IRS Control with Limited Feedback Codebooks
% Adaptive codebook-based limited feedback
%Exploiting Multiple Intelligent Reflecting Surfaces in Multi-Cell Uplink MIMO Communications
% Multiple Intelligent Reflecting Surface Aided 
% Multi-cell Multi-user Uplink MIMO Communication: Reinforcement Learning Approach
%\\
%{\footnotesize \textsuperscript{*}Note: Sub-titles are not captured in Xplore and
%should not be used}
%\thanks{Identify applicable funding agency here. If none, delete this.}
}
% \author{\IEEEauthorblockN{Junghoon Kim\IEEEauthorrefmark{1}, Seyyedali Hosseinalipour\IEEEauthorrefmark{1}, Taejoon Kim\IEEEauthorrefmark{2}, David J. Love\IEEEauthorrefmark{1} and Christopher G. Brinton\IEEEauthorrefmark{1}}}

% \author{Author 1, Author 2, Author 3, \\
% Author 4, and Author 5}

% \author{Junghoon~Kim,~\IEEEmembership{Student Member,~IEEE,}
%         Seyyedali Hosseinalipour,~\IEEEmembership{Member,~IEEE,}
%         Taejoon~Kim,~\IEEEmembership{Senior Member,~IEEE,}
%         David J. Love,~\IEEEmembership{Fellow,~IEEE} 
%         and
%         Christopher G. Brinton,~\IEEEmembership{Senior Member,~IEEE,}
%         % <-this
%         }

\author{Junghoon Kim,
        Seyyedali Hosseinalipour,
        Andrew C. Marcum,
        Taejoon Kim,
        David J. Love,
        and
        Christopher G. Brinton
        % <-this
        \thanks{
% This work was supported in part by the National Science Foundation (NSF) under grants CNS1642982, CCF1816013, and CNS 1955561, and National Spectrum Consortium (NSC) under grant W15QKN-15-9-1004.
% This work was presented in part at the 2020 IEEE International Conference on Computer Communications (INFOCOM) \cite{kim2020joint}.
An abridged version of this work is under review in 2022 IEEE International Conference on Communications (ICC).}
\thanks{J. Kim, S. Hosseinalipour, D. J. Love, and C. G. Brinton are with the Department
of Electrical and Computer Engineering, Purdue University, West Lafayette, IN, 47907 USA
(e-mail: kim3220@purdue.edu; hosseina@purdue.edu; djlove@purdue.edu; cgb@purdue.edu).}
\thanks{A. C. Marcum is with Raytheon BBN Technologies, Cambridge, MA, 02138 USA (email: andrew.marcum@raytheon.com).}
\thanks{T. Kim is with the Department of Electrical Engineering and Computer Science, University of Kansas, KS, 66045 USA (email: taejoonkim@ku.edu).}
        }

% \IEEEauthorblockA{\IEEEauthorrefmark{1}Electrical and Computer Engineering, Purdue University, West Lafayette, IN, USA}
% \IEEEauthorblockA{\IEEEauthorrefmark{2}Electrical Engineering and Computer Science, University of Kansas, Lawrence, KS, USA}
% \IEEEauthorblockA{\IEEEauthorrefmark{1}\{kim3220, hosseina, djlove, cgb\}@purdue.edu, \IEEEauthorrefmark{2}taejoonkim@ku.edu}}
\maketitle

\vspace{-6mm}

\begin{abstract}

\input{abstract}

\end{abstract}

\vspace{-2mm} 
\begin{IEEEkeywords}
Intelligent reflecting surface, limited feedback, adaptive codebook, deep reinforcement learning
% DNN policy learning
\end{IEEEkeywords}

\vspace{-2mm} 
\input{intro}

\input{signal}
\input{optimization}
\input{adjacent}

\input{DRL}
\input{eff}

\input{sim}

\input{conc}
\vspace{-1mm}
\bibliographystyle{IEEEtran}
\bibliography{ref}

\end{document}

%% file: abstract.tex
%Applications of intelligent reflecting surface (IRS) in wireless networks have attracted significant attention recently. 
% The intelligent reflecting surface (IRS) is a promising technology for 6G-and-beyond systems due to its low deployment cost and energy consumption.
% An IRS 
Intelligent reflecting surfaces (IRS) 
consist of configurable meta-atoms, which can change the wireless propagation environment through design of their reflection coefficients.
%
% Different from existing works, w
We 
% conduct adaptive IRS control in
consider a practical setting where
% while explicitly considering that
(i) the IRS reflection coefficients are achieved by adjusting \textit{tunable elements} embedded in the meta-atoms,
% i.e., their controllable capacitance;  %Due to the dependency of the amplitude and phase shift, some of reflection coefficients may not be achievable in practice.
(ii) the IRS reflection coefficients are affected by the  \textit{incident angles} of the incoming signals, 
(iii) the IRS is deployed in multi-path, time-varying channels, and
(iv) the feedback link from the base station to the IRS has a low data rate.
Conventional optimization-based IRS control protocols, which rely on channel estimation and conveying the optimized variables to the IRS, are not applicable in this setting due to the difficulty of channel estimation and the low feedback rate.
%
% make a sentence on basic IRS assumption. Make another sentence on how past work has not efficiently dealt with the control messaging required for IRS adaptation.  Then insert a sentence on how we use a finite-rate connection and propose codebooks.
% Motivated by the low-rate feedback requirement,
% To address these challenges,
Therefore,
we develop a novel adaptive codebook-based limited feedback protocol 
% for IRS control 
where only a codeword index is transferred to the IRS.
% Furthermore, instead of individual channel estimation, we only 
% propose two solutions for adaptive codebook design approaches 
% To address these challenges, 
% we develop a novel adaptive codebook-based limited feedback protocol for IRS control over time.
% By adapting to time-varying channels, we propose two solutions for updating the codebook,
We propose two solutions for adaptive codebook design: \textit{random adjacency (RA)} and \textit{deep neural network policy-based IRS control (DPIC)}, both of which only require the end-to-end compound channels.
% We propose two solutions for adaptive codebook design,
% which can be readily obtained at the BS. 
% These solutions are
% \textit{random adjacency (RA)}, which is a perturbation-based approach,
% % utilizes correlations across the channel instances, 
% and  \textit{deep neural network policy-based IRS control (DPIC)}, which is based on a deep reinforcement learning.
% Both approaches only require  
% the end-to-end compound channels, which can be readily obtained at the BS.
% We further leverage multiple agents for DPIC and combine the RA and DPIC.
We further develop several augmented schemes based on the RA and DPIC.
% by leveraging multiple agents for DPIC and combining the RA and DPIC.
% multiple agents DPIC and hybrid approaches
% For numerical experiments, we consider two scenarios -- indoor UE and outdoor UE -- 
Numerical evaluations show that the data rate and average data rate over one coherence time are improved substantially by our schemes.
% Through numerical simulations, we show that RA and DPIC outperform the baseline in terms of data rate and average data rate over one coherence time and reveal the superior performance of our augmented strategies.

% in both of the simulation scenarios.

%% file: intro.tex
%\vspace{-1.15mm}
\section{Introduction}
\label{sec:intro}

%% Point-to-point communication

The intelligent reflecting surface (IRS), also called a reconfigurable intelligent surface, is one of the innovative technologies for 6G-and-beyond \cite{zhang2020prospective,hosseinalipour2020federated}. 
An IRS  is a software-controlled meta-surface, which is composed of configurable meta-atoms with flexible reflection coefficients.
% By fine-tuning the meta-atoms, an IRS can reflect the incident signals to a desired direction to enhance the communication system performance in terms of power savings, throughput, etc.
By fine-tuning the meta-atoms, an IRS can reflect the incident signals to a desired direction to enhance the communication system performance in terms of power savings, throughput, etc.
% system, by changing the effective end-to-end channel, to yield performance enhancements in terms of...
%
Compared to a traditional antenna array with one or more radio frequency (RF) chains for
active relaying/beamforming,
% techniques, which rely on antennas equipped with RF chain, 
the IRS is built of low cost meta-surfaces which require low energy consumption for adaptive tuning~\cite{liaskos2018new}.
% can reflect the signals with lower hardware cost and energy consumption due to its low-cost passive structure. 
These benefits have motivated active research on utilizing an IRS
in communications/signal processing literature~\cite{wu2019towards,liaskos2018new}. 
%pan2021reconfigurable

% Most of the works in communications area 
Most communication and signal processing IRS works
focus on designing the {\it reflection coefficients} of the IRS meta-atoms considering different performance metrics of interest~\cite{wu2019intelligent,huang2019reconfigurable,cui2019secure}, e.g., sum-rate, power saving, secrecy rate, etc. However, these works either neglect the practical reflection behavior of the IRS meta-atoms or the nature of the channels under which the IRS is deployed.
In this paper, we aim to control the IRS adaptively considering the practical reflection behavior and realistic channel environment.
We take the first step towards this direction via focusing on a \textit{point-to-point communication model}, commonly used in the IRS literature, e.g., \cite{wu2019intelligent,huang2019reconfigurable,cui2019secure},
which we follow to construct a fundamental system model 
that encompasses the practical considerations.

\vspace{-4mm}
\subsection{Related Work and Shortcomings of Current Methods}
\subsubsection{Practical Reflection Behavior of IRS}
Much of the prior works on IRS reflection coefficient design for communications
% in communications/signal processing
% , e.g., \cite{wu2019intelligent,huang2019reconfigurable,cui2019secure,kim2021multi,he2020adaptive,pei2021ris,psomas2021low}, 
have assumed to control either (i) only the phase shift with full/lossless signal reflection (i.e., assuming no signal attenuation upon reflecting from the IRS), or (ii) both the phase shift and attenuation of reflection, which are independently controlled from one other.
However, it is practically difficult to implement either of these approaches.
% due to two reasons.
% it is practically difficult to realize either of them.
First, the full/lossless signal reflection cannot be realized in practice due to the inevitable
% hardware limitations leading to 
energy loss caused by the dielectric loss, metallic loss, and ohmic loss~\cite{rajagopalan2008loss}.
Second, the reflection phase shift and attenuation cannot be controlled independently 
because the reflection behavior is determined by adjusting the \textit{tunable elements} inside the meta-atoms.
% because the reflection behavior of an IRS is inherently determined by the \textit{impedance} of its meta-atoms, which is tuned via  \textit{configuration} of the tunable elements inside the meta-atoms. 
This fact 
implies that the IRS reflection phase shift and attenuation are \textit{interdependent} as revealed in the physics literature~\cite{zhu2013active,shao2021electrically}.
% some reflection coefficients cannot be realized.
% which do not admit arbitrary realization of reflection coefficients.
% In other words, the phase shift and amplitude are dependent to each other depending on the tunable element.
%
This interdependency has only been considered in a few works in the communication area~\cite{abeywickrama2020intelligent}.

Another aspect overlooked in prior works is the dependency between the IRS reflection behavior and the \textit{incident angles} of the incoming electromagnetic (EM) waves.
% of the impedance of the IRS surface on the angle of arrival of the signals. 
This fact was first revealed in recent work~\cite{tang2020wireless}, which demonstrates that
the IRS reflection coefficient is sensitive to the  incident angles of the incoming EM waves.
Motivated by this observation,
the authors of~\cite{chen2020angle} propose an angle-dependent reflection coefficient model for each IRS meta-atom using an equivalent circuit model.
% Motivated by this result, the authors of~\cite{chen2020angle}
% further study the angle-dependent characteristics of the IRS reflection coefficient and introduce an angle-dependent reflection model for the IRS with varactor-based meta-atoms using an equivalent circuit model for each meta-atom.
In parallel, the authors in \cite{pei2021ris,costa2021electromagnetic} also
demonstrate the reflection response varies with the incident angle of the EM wave.
% and explain this phenomenon from the fact that the reflectance behavior of the IRS elements depends on the incident angle \cite{zhang2020controlling}.
%
% Note that, when the incident angle is within a given range such that the reflection coefficient is constant, it can be reduced to the amplitude-dependent phase shift model in \cite{abeywickrama2020intelligent}.
% \rev{Although recent work \cite{lee2021single} in physics literature presents
% a single-layer phase gradient metasurface lens
% to achieve angle independency up to $30^o$ for mmWave focusing within the frequency range between $35$ and $40$ GHz, the current art for the IRS implementation suffers from the angle dependency in various frequency ranges and  incident angles.}
To the best of our knowledge, the angle-dependent property of the IRS reflection coefficient has not been 
incorporated into uplink/downlink signal transmission models for wireless communication systems.

% considered
% in the canonical system model used in the communications and signal processing literature.

% \rev{We consider a \textit{feedback link} from the BS to the control board of the IRS to support real-time IRS configuration to adapt to the varying channels~\cite{pei2021ris}.
% Typically, the feedback link has low data rate because the channel state information (CSI) of the feedback link is unknown at the BS. Therefore, the BS is encouraged to feed back only small amount of necessary information to the IRS, to have low overhead for the feedback in each coherence time.
% Most of existing works assume that the optimized variables are fed back to the IRS without any communication overhead. Although recent work \cite{zappone2020overhead} discussed the feedback overhead caused by the transmission of the designed phase shifts from the BS to the IRS, the practical IRS reflection behavior and multi-path time-varying channels are not discussed jointly with the time overhead.} 

\subsubsection{Communication Overhead for IRS Control under Realistic Channel Environment}
To develop  solutions for IRS reflection design, existing works either assume perfect knowledge of the channel state information (CSI)~\cite{wu2019intelligent,huang2019reconfigurable,cui2019secure,abeywickrama2020intelligent} or estimate the CSI before IRS reflection design~\cite{yang2020intelligent}. In both cases, 
for adaptive IRS control under time-varying channels, a successive channel estimation at the base station (BS) and feedback of information from the BS to the IRS should be conducted. This successive procedure incurs communication time overhead.
The work \cite{zappone2020overhead} takes into account the communication time overhead required for channel estimation and feedback for the IRS phase shift design, and shows that the \textit{average} data rate over a channel coherence time is decreased by the overhead. Nevertheless, in \cite{zappone2020overhead}, the practical IRS reflection behavior and successive IRS control under time-varying channels have not been considered.

% \q{The traditional techniques for IRS control typically follows the procedure~\cite{zappone2020overhead}: (i) channel estimation (at base station (BS) in uplink), (ii) optimization at the BS for IRS reflection coefficients, (iii) feedback of the optimized reflection coefficients to the IRS, and (iv) data transmission.}
% % enables real-time IRS configuration that adapts to time-varying channels~\cite{pei2021ris}.
% \q{
% Since channels are time-varying in realistic channel environment, the above procedure should be conducted periodically, e.g., per channel coherence time, to adapt to the dynamic channels.
% As the time overhead grows, which is caused by (a) the pilot symbol transmission for channel estimation and (b) the feedback of information to the IRS, the available time for data transmission will be shorter 
% within a coherence time. Therefore, the time overhead should be considered as an important performance metric under time-varying channel environment.
% The work \cite{zappone2020overhead} takes into account the communication time overhead required for channel estimation and feedback, incorporate the overhead into the data rate expression, and solve for the IRS phase shift design in a single coherence time.
% Further, the work shows that the \textit{average} data rate over a coherence time is affected by the overhead. 
% Nevertheless, a successive IRS control in time-varying channels and
% the practical IRS reflection behavior have not been taken into account.}

To reduce the overhead for IRS control,
some recent works consider a low overhead \textit{feedback} link from the BS to the IRS
% achieve a low overhead feedback 
by either employing codebook structures~\cite{he2020adaptive,pei2021ris,kim2021multi} or one-bit feedback~\cite{psomas2021low}.
% In IRS-assisted communications, a \textit{feedback} link refers to a data link from the BS to a control board of the IRS, where the control board receives the feedback signal from the BS and changes the IRS configuration properly~\cite{pei2021ris}.
% The BS is generally considered as a processing entity since the BS has high computing power while the IRS does not, and therefore the BS feeds back some processed information to the IRS over the feedback link.
The feedback link typically has a low data rate because the CSI of the feedback link is unknown at the BS~\cite{wu2019towards}.
% , and thus may cause a bottleneck for adaptive IRS control. 
%
In general, codebooks are known to provide high performance gains in limited feedback systems~\cite{love2008overview}, and thus are widely used in 
wireless communications, such as Long-Term Evolution (LTE), LTE-Adv, LTE-Adv Pro, and 5G New Radio (NR)~\cite{dahlman20205g}.
In IRS-assisted communications,
a codebook refers to a set of IRS reflection coefficients, which are shared at both the BS and the IRS \cite{he2020adaptive,pei2021ris,kim2021multi}. 
% In this setting, the BS 
% only feeds back a specific codeword index to the IRS, 
% which the IRS uses to recover
%  the desired reflection coefficient from the codebook. 
%  with the codeword index from the codebook.
% the BS feeds back only an index information to the IRS, which enables a low overhead feedback. 
The work \cite{he2020adaptive}
considers the codebook construction for
uniform linear arrays (ULA). In \cite{pei2021ris} and \cite{kim2021multi}, the codebook is constructed via discrete Fourier transform (DFT) quantization~\cite{love2003equal} and random vector quantization (RVQ)~\cite{au2007performance,santipach2009capacity}, respectively.
In these works, the BS feeds back a specific codeword index to the IRS, which the IRS uses to recover the desired reflection coefficients from the codebook.
% The codebook construction with discrete Fourier transform (DFT) quantization \cite{pei2021ris} and random vector quantization (RVQ) \cite{kim2021multi} are considered, respectively.
The work \cite{psomas2021low} adapts the random perturbation-based method with one-bit feedback for IRS control, previously proposed in traditional wireless communications \cite{mudumbai2007feasibility}. All of these works, however,
% While all of these works achieve a low overhead feedback, they =
directly design the IRS {\it reflection coefficients} without considering
% and thus do not consider 
the practical IRS reflection behavior.

% Why conventional protocol does not work
% -Channel estimation not working - rely on known set of reflection coefficients
% -Reflection coefficient design is not enough : Practical ref. beh. not considered , incident angle

% -Capacitance control
% -Codebook based approach : Low rate feedback
% -Adaptive codebook : Address dynamic channels
% -New protocol (new adaptive codebook-based protocol): Not rely on individual channel estimation - Use channel sounding and Observe its response (Do not rely on incident angles, multi-path)

\vspace{-4mm}
% \subsection{Main Motivations and Summary of Contributions}
\subsection{Our Methodology and Summary of Contributions}
\label{ssec:intro_method}

In this paper, we consider adaptive IRS control in the practical setting where 
(i) the IRS reflection coefficients are achieved by adjusting tunable elements embedded in the meta-atoms, i.e., their controllable  capacitance,
(ii) the  IRS  reflection  coefficients are affected by the incident angles of the incoming EM wave,
(iii) the IRS is deployed in an environment with multi-path, time-varying channels, and 
(iv) the feedback link from the BS to the IRS has a low data rate.

The joint consideration of the practical IRS reflection behavior and realistic channel environment makes the contemporary optimization-based methods used for IRS control~\cite{wu2019intelligent,huang2019reconfigurable,cui2019secure,abeywickrama2020intelligent,yang2020intelligent,zappone2020overhead}, which rely on channel estimation, inefficient. This is
because channel estimation  in turn requires known IRS reflection coefficients, which cannot be obtained in a real-world system
since (i)
incoming signals in a multi-path channel  have different angles of arrival to the IRS and thus experience different reflection responses caused by
angle-dependent reflection behavior of the meta-atoms and (ii)
it is difficult to measure the incident angles of multiple incoming signals at the IRS since the IRS is typically a passive device without active sensors.

For effective  IRS  control  under the aforementioned  practical  setting,
we propose a novel \textit{adaptive codebook-based limited feedback protocol.}
% The proposed protocol consists of four steps: (i) IRS channel sounding/reconfiguration, (ii) data rate calculation and codeword selection at BS, (iii) feedback from BS to IRS, and (iv) data transmission.
There are several novelties in our proposed protocol.
First, we directly design the meta-atom {\it capacitance values}  for IRS configuration, different from the current methods that design IRS reflection coefficients, some of which  may be not feasible for implementation. 
Second,  we adopt a codebook structure, where the codebook is a set of  capacitance values for IRS configuration and employed at the IRS. 
Third, we develop two \textit{adaptive} codebook design methods, where the codebook is updated to account for time-varying channels. 
% The key of the proposed protocol is how to update the codebook.
These methods are
(i) \textit{random adjacency (RA)}, which utilizes the correlation across the channel instances, and (ii) \textit{deep neural network (DNN) policy-based IRS control (DPIC)}, which is a deep reinforcement learning-based method.
% The adaptive codebook enables an successive IRS control adapting to the dynamic channels.
Both of these approaches only require the end-to-end compound channels from the user equipment (UE) to the BS, which can be readily obtained at the BS.
% containing specific IRS configuration. 
Thus our IRS control methodology
% The proposed protocol 
does not require any complicated estimation/tracking process for the channels, UE location, and incident angles.

The contributions of this paper are summarized as follows:

\begin{itemize}[leftmargin=4.5mm]
    \item We introduce a novel signal model that considers the practical IRS reflection behavior for the IRS-assisted uplink communications system, where the reflection coefficient of each IRS meta-atom is a function of the incident angle of the EM wave and its controllable capacitance.
    % (Sec. \ref{sec:signal}).
    \item We formulate the data rate maximization problem and discuss the challenges associated with solving the problem in the practical setting.
    % due to the IRS reflection behavior and multi-path, time-varying wireless channels.
    Motivated by the existence of a low-rate feedback channel between the IRS and BS and the requirement of successive IRS control under time-varying channels,
    % time varying channels 
    % IRS-UE and IRS-BS effective capacitance configuration, 
    we propose a novel adaptive codebook-based limited feedback protocol.
    % (Sec. \ref{sec:optimization}).
    % , where the codebook is updated periodically to adapt to the time-varying channels
%    each codeword in the codebook is a possible realization of the capacitance configuration. 
%    Furthermore, we propose an {\it adaptive} codebook structure where the IRS updates the codebook adaptively.
%    (Sec. IV).
    \item 
%    We propose an {\it adaptive} codebook structure where the IRS updates the codebook adaptively.
    For adaptive codebook design, we propose RA and DPIC. 
    % adaptive codebook strategies.
    % , both of which require minimal measurements for the codebook updates.
    % \textit{random adjacency (RA)}, which utilizes the correlation across the channel instances, and (ii) \textit{deep neural network (DNN) policy-based IRS control (DPIC)}, which is a deep reinforcement learning-based method.
    % Both of these approaches only require minimal measurements for the codebook updates.
    % , and thus enable us to simplify the procedure of the IRS control and adaptation over time.
    % compatible with the limited feedback protocol: (i) random adjacency (RA), which is a channel-estimation free method, and (ii) deep neural network (DNN) policy-based IRS control (DPIC), which is a policy learning method through a deep reinforcement learning (DRL). 
    In DPIC, we tailor an actor-critic network for the DNN policy learning to make it compatible with the limited feedback protocol. 
    % Especially, w
    We incorporate the RVQ process into the behavior policy, 
    which allows low-overhead feedbacks. 
    % for adaptive codebook updates.
    Further, we develop several augmented strategies 
    based on DPIC, which incorporate multi-agent learning and a hybrid of RA and DPIC approaches.
    We analyze the computational complexity and the total time overheads of the proposed approaches.
    % by leveraging multiple agents for DPIC and combining the RA and DPIC.
    % (Sec. \ref{sec:codebook}). 
    % \item 
    % We analyze the time complexity for computations and the total time overheads of the proposed approaches. We define a new metric, called effective data rate, which incorporates the time overheads with the data rate, 
    % to measure the system performance under time-varying channels.
    % BREAK THIS DOWN INTO TWO AND THEN ADD THAT BEHAVIOR POLICY WITH RVQ AND QUANTIZATION!
%     Then, we define the effective data rate as an average data rate over one coherence time for the RA and DPIC, where the effective data rate incorporates the overheads into the data rate performance 
    %  (Sec. \ref{sec:effective_rate}).
    \item 
%    We recover the practical reflection coefficient function of the IRS unit element  from the experiments of the work \cite{chen2020angle}. 
    For simulations, we consider two practical scenarios in multi-path fading channels: (i) indoor UE with no line-of-sight (LoS) link to the IRS, and (ii) outdoor UE with a LoS link.
    % between the UE and the IRS.
    For both scenarios, we evaluate the performances of the data rate and 
    average data rate over one coherence time,
    % effective data rate, 
    and demonstrate that
    RA and DPIC outperform the baseline.
    Also, 
    simulations show
    % we reveal 
    the superior performance of our augmented strategies as compared to their counterparts.
    % in both of the simulation scenarios. 
    % (Sec. \ref{sec:sim}).
\end{itemize}

Numerical evaluations show that the data rate and average data rate over one coherence time are improved substantially by the proposed schemes.

%%
% For numerical experiments, we consider two practical scenarios -- indoor UE and outdoor UE -- and show that RA and DPIC out-perform the baseline in terms of data rate and average data rate over one coherence time. Also, we reveal the superior performance of our augmented strategies as compared to their counterparts in both of the simulation scenarios.
%%

%\rev{Emphasis 1: The DNN policy learning is an unsupervised learning, and yields better performances than RA. If a supervised learning is used to train DNN, its maximum performance will be just that of RA. There are no ways to have an optimal label or action.}

%\rev{Emphasis 2: The proposed algorithms are the channel estimation free methods. It is hard to estimate the channels due to the IRS properties (reflection coefficients is dependent on the incident angles) and time-varying nature of the channels.}

%ASsa

%% file: signal.tex
\vspace{-2mm}
\section{System Model for IRS-assisted Uplink Communications}
\label{sec:signal}
% In this section,  we first discuss the reflection behavior of the IRS meta-atoms in
We begin by formalizing IRS meta-atom reflection behavior in
Sec.~\ref{ssec:IRS}.
Then, we describe the signal model of IRS-assisted uplink communications in Sec. \ref{ssec:formulation}.

\begin{figure}[t]
  \includegraphics[width=.95\linewidth]{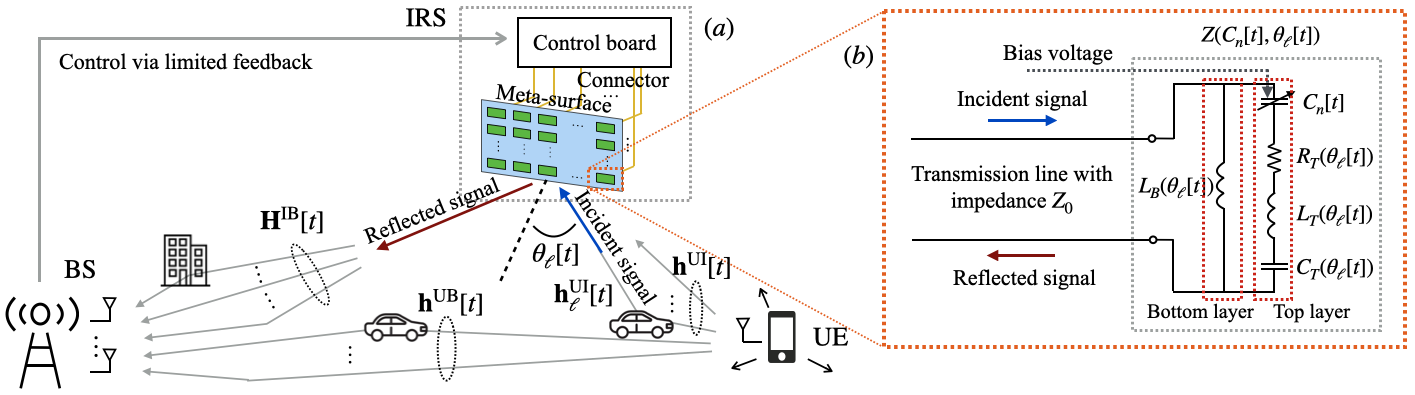}
  \centering
  \vspace{-2mm}
  \caption{The system model consisting of a UE, IRS, and BS in uplink point-to-point communication, where
  the IRS is controlled by the BS via a limited feedback link.
  (a) Depiction of the IRS as two interconnected systems: meta-surface and control board.
  (b) Equivalent circuit model of the signal response at each IRS meta-atom.}
  \label{fig:system_merged}
  \vspace{-7mm}
\end{figure}

\vspace{-3mm}
\subsection{Reflection Behavior of IRS Meta-atoms}
\label{ssec:IRS}

% IRS represents a {\it software-controlled meta-surface}. It is also called reconfigurable intelligent surface (RIS), intelligent reflective surfaces, or programmable meta-surface. 

An IRS consists of two interconnected systems shown in Fig. \ref{fig:system_merged}(a): a meta-surface 
% as a physical core 
and control board. 
% as an algorithm core. 
%
%\subsection{Meta-surface as a Physical Core}
%\label{subsec:physics}
%\textbf{Meta-surface and meta-atoms.} 
A meta-surface is 
an ultra-thin sheet composed of 
periodic sub-wavelength metal/dielectric structures, i.e., meta-atoms.
%the meta-atoms as elementary units. 
The size of each meta-atom is typically from $\lambda/10$ to $\lambda/5$~\cite{liaskos2018new}, where $\lambda$ denotes the wavelength of the EM wave. Each meta-atom generally contains a semiconductor device as a tunable element, e.g., positive-intrinsic-negative (PIN) diode, variable capacitor (varactor), metal-oxide-semiconductor field-effect transistor (MOSFET) \cite{shao2021electrically}.
% zhang2020dynamic : Graphene
% Each meta-atom contains a tunable element, e.g., positive-intrinsic-negative (PIN) diode, variable capacitor (varactor), 
% metal-oxide-semiconductor field-effect transistor (MOSFET),
% and graphene \cite{shao2021electrically,cui2014coding}.
% By adjusting the tunable elements, 
By adjusting the bias voltages applied to these tunable elements,
we can change the impedances over the meta-surface 
% can be changed 
to have a desired functionality,
%, and therefore change the electric current on the surface. 
%Versatile functionalities, 
e.g., perfect absorption, anomalous reflection, and polarization of the incoming signal.
We focus on the \textit{reflection} behavior, where the reflected signals from the meta-surface become constructive at a desired angle/direction, i.e., beamforming.
%each {\it meta}-atom can change its behavior, such that it absorbs, reflects, refracts, and changes the polarization of the incoming signals. 
%By tuning the capacitances of varactors, we can adjust the impedances over the meta-surface, and therefore change the electric current on the surface. With a proper tuning of elements, the signals would become constructive at a desired angle, and thus a beamforming is achieved. 
%
%
% For this application, we adopt the varactors as tunable elements 
% since continuously variable capacitance provides a full $360^o$ reflection phase
%  at each meta-atom~\cite{shao2021electrically,zhu2013active}.

% The impedance of the meta-surface can be modeled via an equivalent circuit model~\cite{}. 

%Recent research, e.g., \cite{abeywickrama2020intelligent,zhu2013active, tang2020wireless,cai2020practical,chen2020angle}, focus on the varactor-implemented meta-atom, since the varactor can offer each meta-atom almost a full $360^o$ reflection phase \cite{zhu2013active}.
%Two main approaches have been proposed for the im- plementation of programmable metasurfaces, namely, (i) by interfacing the tunable elements through an external Field- Programmable Gate Array (FPGA) [17], [18], or (ii) by integrating sensors, control units, and actuators within the metasurface structure [19]–[22].

%\subsection{Control Circuit as an Algorithm Core}
%\label{subsec:physics}

A control board that is connected to the meta-surface enables flexible configuration of the tunable elements.
% by adjusting the bias voltages.
% by adjusting the bias voltages via the implementation of different logic functions.}
% adjust the bias voltages by providing reprogrammed to implement different logic functions, allowing flexible configuration of the meta-surface.
A field programmable gate array (FPGA)-based  control board is generally considered for IRS control due to the flexible implementation of different logic functions~\cite{abadal2020programmable}.
%yang2016programmable
A control board adjusts the bias voltage applied to the semiconductor in each meta-atom and changes the capacitance of the semiconductor, i.e., the tunable element.
To adapt to dynamic channels, the control board can flexibly tune the capacitance over time.
% To adapt to varying channels,
% a control board can flexibly adjust the bias voltage over time, which results in changing the charge accumulation of the semiconductor (tunable element), i.e., capacitance, in each meta-atom.
%
%, which is connected to the meta-surface (physical core) to control the varactors
%Specifically, the meta-surface (physical core) is connected to the FPGA (algorithm core) via connectors so that the varactors are controlled by the FPGA-based circuit signals (See Fig.6 (a) in \cite{abadal2020programmable}).
%In each meta-atom (unit cell), 
%The capacitance $C_n[t]$ is controlled by a bias voltage applied to the corresponding unit \rev{ cell}.
%, normally with the relationship \cite{} as
%$C_n[t] \propto 1/\sqrt{V_n[t]}$.
% For each meta-atom, g
Given a range of potential bias voltage values,
% lies in interval $[0,V_{\max}]$, 
the capacitance $C_n[t]$ at meta-atom $n$ at time $t$ satisfies
\vspace{-1.5mm}
\begin{equation}
    C_{\min} \le C_n[t] \le C_{\max},
    \label{eq:Crange}
    \vspace{-1.5mm}
\end{equation}
where $C_{\min}$ and $C_{\max}$ may vary for different types of semiconductor devices.
% Although a resistance can be adjusted via a bias voltage control, we focus on the capacitance tuning in this paper since controlling variable capacitance provides a full $360^o$ reflection phase at each meta-atom~\cite{shao2021electrically,zhu2013active}.
% While the PIN diode or varactor diode are the semiconductor considered in this paper, our system model and methodologies can be straightforwardly applied to other types of semiconductor.

Through tuning the capacitance of the meta-atoms, 
their impedance can be adjusted. 
However, the impedance is also dependent on the incident angle of the incoming EM wave~\cite{costa2021electromagnetic}.
% As a result, the corresponding reflection behavior of the meta-surface with embedded semiconductors depends on both the capacitance at the meta-atoms and incident angles of the EM waves.
Both of these factors should be considered in IRS reflection behavior design.
To explicitly describe the reflection behavior of the meta-surface, we will next investigate the impedance and reflection coefficient at meta-atom level.\footnote{Since the physical size of the meta-atom is usually smaller than the wavelength of the incident signal, the signal response of the meta-atom can be described by an equivalent circuit model
\cite{koziel2013surrogate}.}
As an example, 
% we provide 
% the impedance  and reflection model of the equivalent circuit at each meta-atom proposed in \cite{chen2020angle}, where the meta-atom structure is composed of the symmetrical metallic patches \rev{equipped with} a varactor diode.
we provide the impedance and
reflection coefficient of a meta-atom equipped with a varactor using its equivalent circuit model~\cite{chen2020angle} depicted
in Fig.~\ref{fig:system_merged}(b).
Denote $\theta_\ell[t]$ as the incident angle of the $\ell$-th channel path to the IRS.\footnote{
In this section, we discuss the angle-dependent impedance (in \eqref{eq:Z}) and reflection coefficient model (in \eqref{eq:reflectioncoeff}) provided in \cite{chen2020angle} where only azimuth coordinates of the incident angle are considered. 
% Thus, we consider only the azimuth coordinates in \eqref{eq:Z} and \eqref{eq:reflectioncoeff}.
Our proposed signal model and methodologies can be readily applied to the case including the elevation angle.}
% , which will be discussed in detail with the signal model in Sec. \ref{ssec:formulation}.
Under a far-field assumption where $\theta_\ell[t]$ is the same across all the meta-atoms,
% We consider a far-field assumption, where the incident angle $\theta_\ell[t]$ is the same across all the meta-atoms.
the impedance of meta-atom $n$ can be described as~\cite{chen2020angle} 
%$Z(C_n[t], \theta[t])$ of such equivalent circuit is expressed as follows:
% \vspace{-1mm}
\begin{equation}
    Z(C_n[t], \theta_\ell[t]) =
    \frac{j2 \pi f L_B(\theta_\ell[t]) \big( R_T(\theta_\ell[t]) + j2 \pi f L_T(\theta_\ell[t]) + \frac{1}{j 2 \pi f C_T(\theta_\ell[t])}   + \frac{1}{j 2 \pi f C_n[t]} \big) }
    {
    j2 \pi f L_B(\theta_\ell[t]) + \big( R_T(\theta_\ell[t]) + j2 \pi f L_T(\theta_\ell[t]) + \frac{1}{j 2 \pi f C_T(\theta_\ell[t])}   + \frac{1}{j 2 \pi f C_n[t]} \big)    
    },
    \label{eq:Z}
    % \vspace{-1mm}
\end{equation}
where $L_T(\theta_\ell[t])$, $C_T(\theta_\ell[t])$, and $R_T (\theta_\ell[t])$ are the inductance, capacitance, and loss resistance of the top layer, respectively,
 $L_B(\theta_\ell[t])$ is the bottom layer inductance,
 $C_n[t]$ is the variable capacitance, and $f$ is the operating frequency of the incident EM waves.
Except $C_n[t]$, 
% which is the capacitance of a varactor, 
all of these parameters
% i.e., $L_T(\theta_\ell[t])$, $C_T(\theta[t])$, $R_T(\theta[t])$, and $L_B(\theta[t])$, 
 are dependent on the incident angle $\theta_\ell[t]$, which makes the reflection behavior of meta-atom angle-dependent. This phenomenon is also observed in~\cite{tang2020wireless,pei2021ris,costa2021electromagnetic}.
%
% The work \cite{chen2020angle} demonstrates that,
% when EM waves illuminate the IRS from different directions, the circuit parameters change due to the spatial dispersion in the meta-surface. The authors in \cite{pei2021ris} explain this phenomenon from the fact that the reflectance of the IRS meta-atom depends on the incident angle \cite{zhang2020controlling}. This  results in variable equivalent circuit parameters depending on the incident angles.

% The reflection behavior of the equivalent circuit model can be represented by the reflection coefficient. 
Considering the impedance discontinuity between the free space impedance $Z_0  \approx  \SI{376.73}{\ohm}$ and the meta-atom impedance $Z(C_n[t], \theta_\ell[t])$, the reflection coefficient\footnote{In fact, the impedance in \eqref{eq:Z} and the reflection coefficient in \eqref{eq:reflectioncoeff} are dependent on the frequency $f$.
%i.e., $Z(C_n[t], \theta[t], \omega)$. 
However, since we consider 
a fixed frequency with narrowband of a few tens of MHz bandwidth,
we can approximate
the IRS reflection coefficients as constant across $f$~\cite{chen2020angle, pei2021ris}, and thus do not consider the dependency of $f$.}
% if the signal bandwidth $W$ is limited to a few tens of MHz, as shown in the plots of the reflection coefficients along $w$ in the works~\cite{chen2020angle, pei2021ris}.
% In this paper, we consider a fixed operating frequency and narrowband with a few tens of MHz, and therefore we do not consider the dependency of $\omega$.
%
% The same holds for the incident angle $\theta[t]$.
% can be dependent on the operating frequency, because the channel propagation varies in frequencies. Given the narrowband assumption, it is reasonable to consider that  the channel propagation properties (including $\theta[t]$) are the same across the utilized frequency band.}
 of meta-atom $n$ is~\cite{pozar2011microwave}
% \vspace{-1.5mm}
\begin{equation}
    \Gamma(C_n[t], \theta_\ell[t]) = \frac{Z(C_n[t], \theta_\ell[t]) - Z_0}
    {Z(C_n[t], \theta_\ell[t]) + Z_0}.
    \label{eq:reflectioncoeff}
    % \vspace{-1.5mm}
\end{equation}
The expressions in \eqref{eq:Z}\&\eqref{eq:reflectioncoeff} reveal two practical considerations for tuning the meta-atoms.

\vspace{-1mm}
\begin{itemize}[leftmargin=4.5mm]
\item {\it \textbf{Consideration 1.} Dependency between amplitude/attenuation and phase shift.}
The amplitude/attenuation $|\Gamma(C_n[t], \theta_\ell[t])|$ and phase shift $\angle \Gamma(C_n[t], \theta_\ell[t])$ of the reflection are jointly controlled by the semiconductor with capacitance $C_n[t]$. In other words, the amplitude and the phase shift at a meta-atom cannot be controlled independently, which is also reported in~\cite{abeywickrama2020intelligent}.
Thus, it is beneficial to design the variable capacitance instead of the reflection coefficient since some combinations of attenuation and phase shifts may not be feasible.

% \rev{This dependency is  taken into account for the reflection coefficient design in .}

\item {\it \textbf{Consideration 2.} Dependency between reflection coefficient and incident angle.} 
The reflection coefficient is a function of the incident angle $\theta_\ell[t]$.
This will pose new challenges for applications of IRS in practical wireless systems with multi-path and time-varying channels, which will be discussed in detail in Sec. \ref{ssec:problem}. 
% The incident angles thus must be incorporated in the signal model and also considered for IRS configuration.
This dependency is observed and explained in~\cite{costa2021electromagnetic,tang2020wireless,chen2020angle,pei2021ris}, but not yet incorporated in the canonical signal model for IRS-assisted communications.
% In multi-path channels, the incoming signals with different incident angles experience different reflection responses from the same IRS meta-atom.
% Also, due to dynamic channels, the reflection response will be unexpected unless the incident angles are carefully estimated in real-time, which is impractical.
% Especially, in uplink channel from the UE to IRS, the incident angles are  highly probable to vary because of the UE mobility.
\end{itemize}
We incorporate the above practical considerations into our signal model and methodology.

\vspace{-5mm}
\subsection{Signal Model for IRS-assisted Uplink Communications}
\label{ssec:formulation}

We consider the IRS-assisted uplink communications with a UE, a BS, and an IRS, depicted in Fig. \ref{fig:system_merged}. 
The UE is equipped with a single antenna, while the BS possesses $N_{\rm BS}$ antennas.
We assume a block fading channel model with time index  $t=0,1,...$, where channels are constant during each block.
Let $N_{\rm IRS}$ denote the number of the IRS meta-atoms.
We define the {\it capacitance vector} across the IRS meta-atoms at time $t$
as
\vspace{-1.5mm}
\begin{equation}
    {\bf c}[t] = \big[ C_{1}[t], ...,  C_{N_{\rm IRS}}[t] \big] \in \mathbb{R}^{N_{\rm IRS} },
    \label{eq:capvec}
    \vspace{-1.5mm}
\end{equation}
where $C_n[t]$ is the capacitance of the semiconductor in meta-atom $n$.
We also formulate the {\it reflection coefficient matrix} across the IRS meta-atoms as
\vspace{-1.5mm}
\begin{equation}
    {\boldsymbol \Phi}( {\bf c}[t] , \theta_\ell[t] ) = {\rm diag} \big( 
    \Gamma(C_1[t], \theta_\ell[t]), 
    \Gamma(C_2[t], \theta_\ell[t]), 
    ...,
     \Gamma(C_{N_{\rm IRS}}[t], \theta_\ell[t]) 
    %  , ..., \Gamma(C_{N_{\rm G}}[t], \theta[t])
      \big) \in \mathbb{C}^{N_{\rm IRS} \times N_{\rm IRS} },
     \label{eq:reflection_matrix}
     \vspace{-1.5mm}
\end{equation}
where the $n$-th diagonal entry $\Gamma(C_n[t], \theta_\ell[t])$ is the reflection coefficient at meta-atom $n \in \{1,..., N_{\rm IRS}\}$
% , $n = 1,...,N_{\rm IRS}$, 
given the incident angle $\theta_\ell[t]$.
The reflection coefficient matrix 
% incorporates the practical IRS reflection behavior in its mathematical form, and 
enables us to incorporate the practical IRS reflection behavior into the signal model for IRS-assisted communications.

We consider multi-path channels and adopt a geometric channel model representation~\cite{tse2005fundamentals}.
We represent the channel from the UE to the IRS (i.e., UE-IRS channel) as 
$ {\bf h}^{\rm UI}[t] = \sum_{\ell=1}^{L[t]} {\bf h}^{\rm UI}_\ell(\theta_\ell[t], t) \in \mathbb{C}^{N_{\rm IRS} \times 1} $, in which
${\bf h}^{\rm UI}_\ell(\theta_\ell[t], t)$ is the $\ell$-th path channel with the incident angle $\theta_\ell[t]$ and $L[t]$ is the number of paths. 
We assume a narrowband system, 
where $\theta_\ell[t]$, $\forall \ell$, is the same across the utilized frequency band and consider a single tap channel model.
Subsequently, the received signal at the BS at time $t$ is given by
\vspace{-1mm}
\begin{equation}
    {\bf y}[t] =
    \bigg( 
    {\bf h}^{\rm UB}[t] 
    +
     {\bf H}^{\rm IB}[t] \bigg( 
    \sum_{\ell =1}^{L[t]} {\boldsymbol \Phi}( {\bf c}[t], \theta_\ell[t] ) {\bf h}^{\rm UI}_\ell(\theta_\ell[t], t) \bigg) \bigg)
     \sqrt{P} x[t] + {\bf n}[t] \in \mathbb{C}^{N_{\rm BS} \times 1},
    \label{eq:signal_model}
    \vspace{-1mm}
\end{equation}
where $P \in \mathbb{R}^+$ denotes the transmit power and $x[t] \in \mathbb{C}$ denotes the transmit symbol of the UE, where $\mathbb{E}[|x[t]|^2]=1$.
The noise vector ${\bf n}[t] \in \mathbb{C}^{N_{\rm BS} \times 1}$ follows the complex Gaussian distribution $\mathcal{CN} ( {\bf 0}, \sigma^2 {\bf I} )$,
% i.e., ${\bf n}[t] \sim \mathcal{CN} ( {\bf 0}, \sigma^2 {\bf I} )$, 
where ${\bf I}$ denotes the identity matrix and $\sigma^2$ is the noise variance.
%
%the first term denotes the direct signal from UE to BS and the second term denotes the signal from UE to BS via IRS.
${\bf h}^{\rm {UB}}[t] \in \mathbb{C}^{N_{\rm BS} \times 1}$ is the direct channel from the UE to the BS (i.e., UE-BS channel) and ${\bf H}^{\rm IB}[t] \in \mathbb{C}^{N_{\rm BS} \times N_{\rm IRS} }$ is the channel from the IRS to the BS (i.e., IRS-BS channel).
We define the end-to-end compound channel in \eqref{eq:signal_model}  as the \textit{effective channel} given by
\vspace{-1mm}
\begin{equation}
    {\bf h}_{\rm eff}({\bf c}[t],t) \triangleq {\bf h}^{\rm UB}[t]
    +
     {\bf H}^{\rm IB}[t] \bigg( 
    \sum_{\ell =1}^{L[t]} {\boldsymbol \Phi}( {\bf c}[t] , \theta_\ell[t] ) {\bf h}^{\rm UI}_\ell(\theta_\ell[t], t) \bigg) \in \mathbb{C}^{N_{\rm BS} \times 1},
    \label{eq:effchan}
    \vspace{-1mm}
\end{equation}
which 
% ${\bf h}_{\rm eff}({\bf c}[t],t ) \in \mathbb{C}^{N_{\rm BS} \times 1} $ 
encapsulates all the channels (i.e.,  ${\bf h}^{\rm {UB}}[t]$, ${\bf H}^{\rm IB}[t]$, and ${\bf h}^{\rm UI}[t]$) and 
the specific IRS configuration (i.e., ${\bf c}[t]$). 

%% file: optimization.tex
\vspace{-1mm}
\section{Problem Formulation, Challenges, and Limited Feedback Protocols}
\label{sec:optimization}
\vspace{-1mm}

We first formulate the data rate maximization problem for IRS control and discuss the challenges associated with solving it in Sec.~\ref{ssec:problem}. To address the challenges, we propose a novel adaptive codebook-based limited feedback protocol for IRS control in Sec.~\ref{ssec:protocol}. Finally, we 
discuss how the IRS codebook differs from traditional  precoding codebooks in Sec.~\ref{ssec:naive}.
% discuss
% the impracticality of using the current art in codebook design to our setting in Sec.~\ref{ssec:naive}.

\vspace{-4mm}
\subsection{Problem Formulation and Challenges}
\label{ssec:problem}

% The BS is generally considered as a processing unit to solve  an optimization for IRS configuration~\cite{pan2021reconfigurable}, and only transmits the optimized variables to the IRS via a feedback link.
% This is because 
% (i) the IRS does not have any active sensors to measure some sort of signal information, while the BS can measure them with active sensors;
% (ii) the BS has high computational capabilities while the IRS does not; 
% (iii) transmitting only optimized variables requires less feedback overhead than transmitting massive data information (e.g., full CSI).
% In this paper, we also consider the BS as the main processing unit while the IRS just performs simple computations.

% z = h^H/|h|
% |z^h|
% We assume that the BS adopts maximum ratio combining (MRC) to recover the transmit symbol. 
% Thus, the instantaneous data rate maximization problem at time $t$
% can be formulated  as
We aim to maximize the capacity of the channel as a performance metric.
Therefore, we formulate the achievable data rate maximization problem at time $t$ as
% \vspace{-1mm}
% the following optimization problem:
\begin{align}
    & \underset{ {\bf c}[t] }{\text{maximize}} & & 
 R({\bf c}[t],t) =
\log_2 
\bigg( 
1+  \frac{ P \| {\bf h}_{\rm eff}( {\bf c}[t],t) \|_2^2} {\sigma^2}
\bigg) 
\label{eq:obj:rate}
\\
& \text{subject to}
& &  
C_{\min} \le {C}_n[t] \le C_{\max}, \; n=1,..., N_{\rm IRS}.
\label{eq:con:rate}
\end{align}
\vspace{-9mm}
% where 
% ${\bf h}_{\rm eff}({\bf c}[t],t)$ is given in \eqref{eq:signal_model}, and ${\bf c}[t] = \big[ C_1[t], ...,  C_{N_{\rm IRS,ctr}}[t] \big]$. 

\noindent 
% where $W$ (Hz) is the bandwidth.
The constraint \eqref{eq:con:rate} states that each capacitance $C_n[t]$, $n=1,...,N_{\rm IRS}$, should reside in the allowed region discussed in  \eqref{eq:Crange}. 
The objective is to adapt ${\bf c}[t]$ based on the time-varying channels.

Operationally, we aim for the optimization \eqref{eq:obj:rate}-\eqref{eq:con:rate} to be solved at the BS since (i) the BS can obtain measurements and exploit them in deriving the solution for IRS control while the IRS has no sensing capability, and (ii) the BS usually has abundant computing resources while the IRS is often not equipped with powerful processing units.
The BS would then generate feedback information for the IRS, used to reconfigure the capacitance at the meta-atoms via the IRS control board. 
% However, solving \eqref{eq:obj:rate}-\eqref{eq:con:rate} presents three key challenges.
However, solving \eqref{eq:obj:rate}-\eqref{eq:con:rate} and tuning the IRS meta-atoms via the feedback link are faced with the following challenges.

\begin{enumerate} [label={({C}}{{\arabic*})}]
\item {\it Impracticality of optimization-based methods due to non-triviality of channel estimation.}
% Conventional optimization-based
% methods, relying on channel estimations, cannot be applied: the channel estimation requires the IRS reflection coefficients, which depend on the incident angles of incoming signals that cannot readily be measured (the IRS has no active sensors)
Conventional optimization-based methods to solve \eqref{eq:obj:rate}-\eqref{eq:con:rate} require the BS to estimate all the channels, 
${\bf h}^{\rm {UB}}[t]$, ${\bf H}^{\rm IB}[t]$, and ${\bf h}^{\rm UI}[t]$, and incident angles $\{\theta_\ell[t]\}_\ell$ in real-time, which are encapsulated in ${\bf h}_{\rm eff}({\bf c}[t],t)$.
However, channel estimation techniques require known IRS reflection coefficients, which cannot be obtained in a real-world system due to multi-path nature of the channels and angle-dependent behavior of meta-atoms (refer to Sec.~\ref{ssec:intro_method}). \label{C1}
% Therefore, conventional optimization methods cannot be readily exploited.
    % \item {\it Difficulty of estimation for incident angles and channels.}
    % Since the IRS typically has no active sensors,
    % it is challenging to measure the incident angles. Then, the angle-dependent IRS reflection is also unknown. This results in the difficulty for the channel estimation.
    % This renders the channel estimation challenging because the techniques for channel estimation rely on a set of known IRS reflection coefficients. 
    % \item {\it Unknown incident angles.}
    % In the uplink, 
    % the number of angles and the angles themselves will vary across time.
    % Due to the existence of no active sensor at the IRS,
    % it is challenging to measure multiple time-varying incident angles at the IRS in real-time. 
    % \label{C1}
    % %
    % \item {\it Difficulty of channel estimation.}
    % Since the incident angles are unknown, the angle-dependent IRS reflection is also unknown. 
    % This renders the channel estimation challenging because the techniques for channel estimation rely on a set of known IRS reflection coefficients. 
    % \label{C2}
    %
    \item {\it Dynamic channels and overhead requirements.}
     Adaptive control of ${\bf c}[t]$ is necessary to have an efficient IRS operation in time-varying channels. Such control requires periodic information acquisition from the BS. The time overhead of information acquisition should be  a small fraction of channel coherence time to ensure reasonable data transmission time.~\label{C2}
    \item {\it Low data rate of feedback link.}
    % We consider a \textit{limited feedback} for IRS-assisted communications.
    A feedback link refers to the data link from the BS to a control board of the IRS~\cite{pei2021ris}. Typically, the feedback link has low data rate because the channel state information (CSI) of the feedback link is unknown at the BS~\cite{wu2019towards}.
    % The feedback link has low data rate, and t
    Therefore, the BS must feed back only small amount of necessary information to the IRS.~\label{C3}
    % , to have low overhead for the feedback in each coherence time.
    % \rev{\item {\it Low computational overhead requirement at IRS.}
    % The IRS is allowed to have low power consumption and storage to maximize its practicality. Individual control for  meta-atoms at the IRS would incur high computational overhead.
    % \label{C5}}
\end{enumerate}

% Also, we take into account the low data rate of the feedback link (from the BS to an antenna-equipped control module of the IRS) in the IRS-assisted communication \cite{wu2019towards}, which is caused by unknown channel state information of the feedback link at the BS.

These challenges render the existing IRS control protocols ineffective since they mostly 
rely on either full CSI or channel estimates,
% rely on full CSI knowledge, 
neglect
the overhead of information acquisition from the BS, and
% overlook the characteristics of the feedback channel.
overlook the behavior of meta-atoms and the characteristics of the feedback channel.
%
% \rev{and the IRS}.
% of the optimization variables  
% from the BS to the IRS.
% where the protocol consists of (i) the pilot-based channel estimation (ii) solving the optimization with the estimated channels, and (iii) the data transmission with the optimized variables.
The main contribution of our work is developing a methodology to jointly address these challenges.
% To the best of our knowledge, we are the first to consider all of the above challenges.

\vspace{-4mm}
\subsection{Adaptive Codebook-based Limited Feedback Protocols for IRS-assisted Communication}
\label{ssec:protocol}

% \rev{We consider a \textit{group control} scheme with $N_{\rm G}$ groups where the same capacitance value is set for the meta-atoms belonging to the same group.
% We define the {\it capacitance vector} for a group control across the IRS as
% \vspace{-1.5mm}
% \begin{equation}
%     {\bf c}[t] = \big[ C^{\rm g}_{1}[t], ...,  C^{\rm g}_{N_{\rm G}}[t] \big] \in \mathbb{R}^{N_{\rm G} },
%     \label{eq:capvec}
%     \vspace{-1.5mm}
% \end{equation}
% where $C^{\rm g}_i[t]$ is the common capacitance value 
% of the semiconductor in 
% every meta-atom belonging in group $i$.
% Then, we focus on controlling $N_{\rm G}$ capacitance, i.e., ${\bf c}[t]$, to configure 
% $N_{\rm IRS}$ capacitance $\{ C_n[t] \}_{n=1}^{N_{\rm IRS}}$ over the meta-surface, where $N_{\rm G} \leq N_{\rm IRS}$.
% A group control reduces the control overhead and complexity at the IRS when a large number of meta-atoms are considered~\cite{yang2020intelligent}.}

Motivated by the low overhead feedback requirement (see \ref{C3}), we propose to exploit a \textit{codebook} structure for IRS control, where the BS sends only a quantized codeword index to the IRS.
% only a quantized codeword index
% a few quantized scalars
%
% Then, the codebook the codebook refers to set of feasible IRS configuration, which stored at the IRS.
% The BS sends only the index information to the IRS, and the IRS reconfigures its meta-surface by recovering configuration with the index information from the codebook.
% We define a codebook as a set of $M$ feasible capacitance vectors, where $M$ is the codebook size. 
% Note that the IRS owns the codebook. The IRS selects the best capacitance vector in the codebook using the feedback information from the BS, and finally tunes the meta-atoms with the selected capacitance vector.}
Furthe, 
% we consider an {\it adaptive codebook}, 
% where we change the codebook to adapt to the channel variations.
we consider adaptive design of this codebook based on channel variations.
We denote the adaptive codebook as ${\mathcal C}[t] = \{ {\bf q}_m [t] \}_{m=1}^M $, where ${\bf q}_m [t] \in \mathbb{R}^{N_{\rm IRS}}$ is the $m$-th codeword (capacitance vector) in the codebook and $M$ is the codebook size.
% for the meta-atom groups.
The codebook is stored and its updates are conducted at the IRS through its control board \cite{pei2021ris} (See Fig.~\ref{fig:system_merged}(a)).
We propose a novel \textit{limited feedback protocol}  consisting of four steps conducted per each coherence time block $t$ depicted in   Fig.~\ref{fig:timeline}:

\begin{figure}[t]
  \includegraphics[width=.5\linewidth]{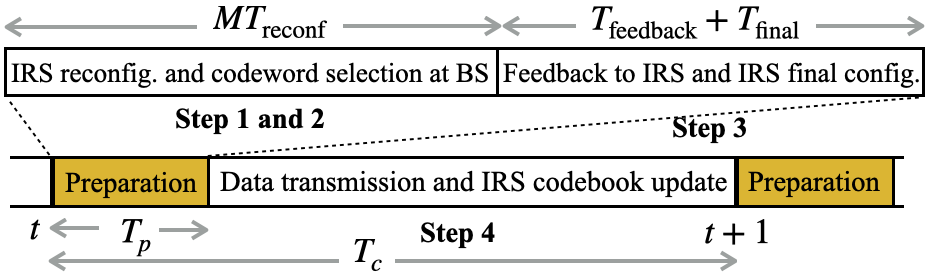}
  \centering
    \vspace{-2mm}
  \caption{Time frame structure of the proposed limited feedback protocol for IRS-assisted communication.
  }
  \label{fig:timeline}
  \vspace{-7mm}
\end{figure}      

%\rev{Attach a timeline figure}

\begin{enumerate}[label={{\bf Step} }{{\arabic*}},leftmargin=15mm]
% \begin{itemize}[leftmargin=4.5mm]
\item {\it \textbf{IRS channel sounding and reconfiguration}.}
While the UE transmits pilot symbols,
the IRS explores all of the $M$ capacitance vectors, i.e., ${\bf q}_m[t]$, in $\mathcal{C}[t]$, $m=1,...,M$.\label{step1}

\item {\it \textbf{Codeword selection at BS.}}
The BS measures the effective channel ${\bf h}_{\rm eff}({\bf q}_m[t],t)$ and calculates the data-rate $ R({\bf q}_m[t],t)$  in \eqref{eq:obj:rate}, as IRS applies  ${\bf q}_m[t]$, $m=1,...,M$. 
The BS obtains the codeword index 
$m^\star[t]= \underset{m \in \{1,...,M \}}{\arg\max} \; R({\bf q}_m[t],t)$.\label{step2}
% , such that the data rate is maximized.
%
\item {\it \textbf{Feedback to IRS and IRS final configuration.}}
The BS feeds back the index $m^\star[t] \in \{1,...,M \}$ to the IRS. Then, the IRS tunes the meta-atoms with
% sets its final configuration with
 $ { \bf q}_{\star}[t] = {\bf q}_{m^\star[t]}[t] \in \mathcal{C}[t]$.\label{step3}
\item {\it \textbf{Data transmission and IRS codebook update.}} The data transmission is conducted during the rest of the coherence time. During this time period, the IRS obtains the next codebook $\mathcal{C}[t+1]$ either locally or with assistance from the BS.\label{step4}
%  , which will be used for Step 1 of the next time $t+1$.
\end{enumerate}

The benefits of our protocol include in its (i)  simple procedure for IRS configuration in limited coherence times, (ii) low-overhead feedback, and (iii) adaptation to dynamic channels. 
In regards to (i), our protocol does not entail the complicated processes required for the estimation/tracking of the channels, incident angles, and UE location~(see \ref{C1} in Sec.~\ref{ssec:problem}).
% in Sec. \ref{ssec:problem}). 
In particular, it only requires partial CSI (i.e., the compound channel in \eqref{eq:effchan}) for the IRS control.
In regards to (ii), we consider digital feedback, rather than feeding back continuous vectors or matrices, which reduces the feedback time overhead~(see \ref{C3} in Sec.~\ref{ssec:problem}).
In regards to (iii), we consider adaptive codebook updates in accordance to the channels~(see~\ref{C2} in Sec.~\ref{ssec:problem}).
% in Sec. \ref{ssec:problem}).
% , where the codebook is updated over time by adapting to the time-varying channels.

% In the proposed protocol, 
Careful design of the codebook ${\mathcal C}[t]$ is critical to obtaining high data rates, since the codewords $\{{\bf q}_m[t]\}_{m=1}^M$ are the  solution candidates and the best one (${ \bf q}_{\star}[t]$ in \ref{step3}) among them is selected as a solution.\footnote{The data rate performance also depends on the codebook size $M$.
% as well as the composition of the codewords in the codebook.
However, $M$ should be limited because of the finite coherence time and non-negligible IRS reconfiguration time.
% Since $M$ is limited due to the time length of the channel coherence time and the non-negligible time required for IRS reconfiguration in practice,
We consider that $M$ is predetermined and fixed in the protocol.}
% , and focus on how to construct/update the codebook.}
% Therefore, to maximize the data rate, we aim to design (or update) the codebook ${\mathcal C}[t]$.
Shown in \ref{step4}, the IRS obtains the next codebook $\mathcal{C}[t+1]$ at time $t$, and therefore the codebook update can be regarded as a \textit{prediction and refinement} problem.
% The codebook update can be regarded as a {\it prediction} problem, where at time $t$, the IRS should obtain the next codebook $\mathcal{C}[t+1]$
% in \ref{step4}.
The BS exploits the effective channels estimated up to time $t$ as available information, denoted by $\mathcal{G}[t] =  \{ {\bf h}_{\rm eff}({\bf q}_m[t'],t') \}_{1\leq m\leq M,t' \le t}$, for the codebook update. 
We develop our methodology under the assumption that the BS obtains the effective channels without any noise. However, for simulations in Sec.~\ref{sec:sim}, we consider that the BS measures a noisy version of the effective channels (i.e., the received pilot signal that contains noise) to have more realistic results.
% We denote the available information at time $t$ as
% , which can be used as an available information for the codebook update. We denote the available information upto time $t$ as 
% $\mathcal{G}[t] =  \{ {\bf h}_{\rm eff}({\bf q}_m[t'],t') \}_{1\leq m\leq M,t' \le t}$.
We make use of $\mathcal{G}[t]$ to obtain $\mathcal{C}[t+1]$
% by considering that
by understanding that the channels are in practice correlated between consecutive coherence times. Thus, the past channel observations can be effectively incorporated to 
predict the codebook over time.
% in the instantaneous design of the codebook.
% In other words, the past observations and decisions can be effectively used to determine current variable decision.

% Since the data rate maximization problem in \eqref{eq:obj:rate}-\eqref{eq:con:rate} is an optimization conducted independently per time, w

% The data rate maximization problem in \eqref{eq:obj:rate}-\eqref{eq:con:rate} is difficult to solve due to the aforementioned challenges.

% is an optimization conducted independently per time,
% Therefore,

We finally reformulate the optimization \eqref{eq:obj:rate}-\eqref{eq:con:rate} to be compatible with the proposed protocol and change the design variable to the codebook.
Because the channels at time $t+1$ are unknown, we formulate the problem as a stochastic optimization:
% by incorporating the prediction process and the utilization of the previous history of information, given by
%
\begin{align}
    & \underset{ \mathcal{C}[t+1] = \{ {\bf q}_m[t+1] \}_{m=1}^M }{\text{maximize}} & & 
    \mathbb{E}_{{\mathcal{E}}[t+1]} \left[ \underset{{\bf q}_m[t+1]  \in \mathcal{C}[t+1] }{\max} R({\bf q}_m[t+1],t+1) \Big| \mathcal{G}[t] \right]
    \label{eq:obj:maxrate}
\\
& \hspace{0.7cm} \text{subject to}
& &  
C_{\min} \le {q}_{m,n}[t+1] \le C_{\max}, \; m=1,...,M, \; n=1,..., N_{\rm IRS},
\label{eq:con:maxrate}
\end{align} 
where 
${q}_{m,n}[t+1]$ denotes the $n$-th entry
% the capacitance for the meta-atom $n$ 
of the $m$-th codeword ${\bf q}_m[t+1] = [q_{m,1}[t+1], ..., q_{m,N_{\rm IRS}}[t+1]]$ at time $t+1$,
% ${q}_{m,i}[t+1]$ denotes the $i$-th capacitance (for the meta-atoms belonging in the group $i$) of the $m$-th codeword ${\bf q}_m[t+1] = [q_{m,1}[t+1], ..., q_{m,\q{N_{\rm IRS}}}[t+1]]$ at time $t+1$,
and $\mathcal{E}[t+1]$ denotes the channel distribution at time $t+1$, including the statistics of ${\bf H}^{\rm IB}[t+1]$, ${\bf h}^{\rm UI}[t+1]$, and ${\bf h}^{\rm UB}[t+1]$.
In the above problem,
we aim to obtain the next codebook $\mathcal{C}[t+1]$ with the available information $\mathcal{G}[t]$ such that the codebook contains good codewords to maximize the expectation 
of the data rate obtained from the \textit{best} codeword over
 the next channel statistics  
 $\mathcal{E}[t+1]$.
Note that the channel distributions
are not static over time due to temporal environment variations (e.g., temperature, precipitation, UE mobility, etc.). 
% because channels are in practice correlated between consecutive times. 
Also, the channel distribution $\mathcal{E}[t+1]$ is unknown to the BS, which adds another degree of difficulty to solve~\eqref{eq:obj:maxrate}-\eqref{eq:con:maxrate}. We will propose low-overhead adaptive codebook design to solve~\eqref{eq:obj:maxrate}-\eqref{eq:con:maxrate}.

\vspace{-2mm}
\subsection{Distinction of the IRS Codebook from the Current Art Precoding Codebook} 
% Impracticality of the Current Art in Codebook Construction to Our Setting}
\label{ssec:naive}

In traditional codebook-based wireless communications,
an encoding function and distortion measure are defined, and the codebook is designed such that the distortion measure is minimized
using points in the complex Grassmannian manifold~\cite{love2003grassmannian}.
When channel vectors  are  modeled as scaled versions of array response vectors,
codebooks are often constructed 
as a set of particular vector subspaces characterized by
an array manifold structure,
% \rev{The popular codebook schemes are constructed under the array manifold structure, 
such as DFT quantization-based codebooks~\cite{love2003equal}, LTE/5G NR codebooks~\cite{LTEstandard,ghosh20185g}, and beamforming quantization-based codebooks~\cite{lau2004design,narula1998efficient,hur2013millimeter}.
In \cite{kim2011mimo,mondal2006channel},  adaptive codebook design methods are proposed based on a specific manifold structure of the channels.
For such codebook construction, the channels or their statistics are assumed to be known and each codeword is designed on the complex vector space.
However, in our problem, we do not assume any knowledge of the channel statistics, and
each codeword in the IRS codebook resides in the $N_{\rm IRS}$-dimensional hypercube where each entry of the codeword ranges in $[C_{\min},C_{\max}]$.

% However, in our problem, the solution manifold cannot be characterized because each codeword is the capacitance vector, which exists in the hypercube in real domain.
% In this procedure, the channels or their statistics are assumed to be known, and each codeword is designed on the complex vector space. 

% Some recent works consider the codebook or low-overhead feedback structure for  IRS-assisted communications \cite{he2020adaptive,pei2021ris,kim2021multi,psomas2021low}.
% % with an aim to design the {\it IRS reflection coefficients}.
% These works  
% % \cite{he2020adaptive,pei2021ris,kim2021multi} 
% consider the codebook construction for
%  uniform linear arrays (ULA) \cite{he2020adaptive},  discrete Fourier transform (DFT) quantization \cite{pei2021ris}, and random vector quantization (RVQ) \cite{kim2021multi}.
% %  , are used to construct the codebook.
% % considers the FFT-based codebook, and the work \cite{he2020adaptive} the hierarchical codebook.
% The work \cite{psomas2021low} adapts the random perturbation-based method with one-bit feedback, previously proposed in traditional wireless communications \cite{mudumbai2007feasibility} for IRS control.
% % Although this method is not a codebook approach, it can have low time overhead required for the variable optimization and feedback. 
% All of these works design the IRS {\it reflection coefficients} and do not consider the practical IRS reflection behaviors. Also, time-varying channels are not taken into account except in \cite{kim2021multi}.

The RVQ codebook \cite{au2007performance,santipach2009capacity} can be exploited in our proposed protocol, and thus we use it as a baseline.
% build our baseline based on it.
% We take these approaches as baselines that can be operated in our protocol.
Specifically, in RVQ design each codeword is randomly generated such that each entry ranges in $[C_{\min},C_{\max}]$ at each time.
% in the allowed region,  i.e., each entry ranges in $[C_{\min},C_{\max}]$ at each time. 
% The codebook can be either fixed or varying over time: (i) a codebook is generated at the beginning and can be fixed over time, (ii) 
% The codebook is generated with a {\it seed} at each time. We take this as a baseline.
%
% Second, the random perturbation based-method with one-bit feedback can be used. Although this is not a codebook approach, this can be exploited to update the solution with limited feedback. The high-level idea is that (i) the IRS generates random perturbation (with a seed) and adds the perturbation to the previous solution, and then (ii) the BS probes the data rate. If the performance is better than before, the BS sends a bit with `1' to the IRS so that the solution is updated. Otherwise, the BS sends a bit with `0' to undo the update. This process can be conducted $M$ times in each coherence time, which require total $M$ bits.
% We take these two approaches, a RVQ codebook and the random perturbation-based method, as baselines to be operated in our protocol.
%
Although this baseline can be operated in the proposed limited feedback protocol, it would not adapt to the varying channels properly. Ideally, it is best to update the codebook by predicting how the optimal solution \textit{changes} according to the next-time channel statistics in \eqref{eq:obj:maxrate}-\eqref{eq:con:maxrate}.
% , given in the optimization \eqref{eq:obj:maxrate}-\eqref{eq:con:maxrate}.
Motivated by this, we next propose two {\it adaptive} codebook approaches, where the codebook is updated with the previous decisions and responses.

%% file: adjacent.tex
\section{Adaptive Codebook Design}
\label{sec:codebook}

For adaptive codebook design, we propose a low-overhead perturbation-based approach in Sec.~\ref{ssec:RA} and a deep neural network (DNN) policy-based approach in Sec.~\ref{ssec:DPIC}.
Then, we discuss the computational complexity of the approaches and present a group control strategy in Sec.~\ref{ssec:comp}. Finally, we quantify the time overhead and the average data rate over one channel coherence block in Sec.~\ref{ssec:effective_rate}.

% Then, we present a group control strategy, and quantify the time overhead and average data rate over one channel coherence block in

% We propose two adaptive codebook design approaches: random adjacency  in Sec.~\ref{ssec:RA}
% and deep neural network policy-based IRS control in Sec.~\ref{ssec:DPIC}.
% % , which is a deep reinforcement learning (DRL)-based method. 
% Then, we discuss the computational complexity of the approaches in Sec.~\ref{ssec:comp}. Finally, we study the time overhead and define a new metric to evaluate the average data rate over one channel coherence block in Sec.~\ref{ssec:effective_rate}.

% We propose a deep learning-driven adaptive IRS control, leveraging a deep reinforcement learning (DRL) technique. First, we discuss the motivation of how the DRL can be leveraged for the IRS-assisted communication in Sec. \ref{ssec:DRL}. 
% The entire procedures for leveraging DRL can be divided into (i) training period and (ii) utilization period. 
% We first discuss Markov decision process (MDP) formulation in Sec. \ref{ssec:MDP}, and how to train the policy in the training period in Sec. \ref{ssec:training}.
% Then, we describe how to leverage the trained policy in the utilization period in Sec. \ref{ssec:utilization}.

\vspace{-2mm}
%%%%%%%%%%%%%%%%%%%%%%%%%%%%%%%%
\subsection{Random Adjacency (RA) Approach}
\label{ssec:RA}

%\begin{figure}[t!]
%%
%\centering
%\begin{subfigure}{.45\linewidth}
%  \centering
%  \includegraphics[width=.5\linewidth]{figures/RA.png}
%  \caption{RA approach.}
%  \label{fig:adjacent}
%\end{subfigure}
%%
%\begin{subfigure}{.45\linewidth}
%  \centering
%  \includegraphics[width=.5\linewidth]{figures/DPIC.png}
%  \caption{DPIC approach.}
%  \label{fig:DDPG}
%\end{subfigure}
%%
%\caption{
%Depiction for the adaptive codebook update of the two proposed approaches. The RA approach updates the capacitances codeword around the previous solution, 
%%i.e., ${\bf c}_m[t+1] = {\bf c}[t] + {\bf d}_m[t] $, $m=1,...,M$, 
%while DPIC approach updates it codeword-by-codeword.}
%%i.e., ${\bf c}_m[t+1] = {\bf c}_m[t] + {\bf d}_m[t]$.
%\label{fig:codebook_update}
%\end{figure}

% The RA approach is simple to implement because it only utilizes the correlation across the channel instances. 

One of the natural ways to construct an adaptive codebook is to use random perturbation-based methods used in obtaining the solutions to beamforming design~\cite{mudumbai2007feasibility}, which determine the current solution by adding a random perturbation to the previous solution. 
We accordingly propose a {\it random adjacency} (RA) approach, which 
can be viewed as a random perturbation-based method for codebook design,
% can be viewed as a codebook version of the random perturbation-based method,
to solve the optimization~\eqref{eq:obj:maxrate}-\eqref{eq:con:maxrate}.
Since the optimization~\eqref{eq:obj:maxrate}-\eqref{eq:con:maxrate} is conducted successively over time in time-correlated channels, the optimal solutions in adjacent time blocks are expected to be close to one another. 
The RA approach exploits this intuition by generating multiple solution candidates (for the codebook at time $t+1$) around the previous solution.
% The high-level idea of the RA approach is to generate multiple solution candidates (for the codebook at time $t+1$) around the previous solution.
% % at $t$ is expected to be around its optimal solution evaluated at $t-1$.
% The high-level idea of the RA approach is to generate multiple solution candidates (for the next time) around the previous solution.
The codebook resides and is updated at the IRS, which requires no feedback overhead for the codebook update (the feedback is still used to transfer the index of the best codeword deployed for data transmission in each coherence time in \ref{step3} in Sec.~\ref{ssec:protocol}).

% while the BS does not need to know the codebook.

Formally, the IRS obtains the codebook $\mathcal{C}[t+1] = \{{\bf q}_m[t+1]\}_{m=1}^M $, where the $m$-th codeword is updated by adding a random perturbation ${\bf z}_m[t] \in \mathbb{R}^{N_{\rm IRS}}$ to the previous solution ${\bf q}_\star[t]$ (obtained in \ref{step4} in Sec.~\ref{ssec:protocol}) as
\vspace{-2.5mm}
\begin{equation}
    {\bf q}_m[t+1] = {\rm clip} ( {\bf q}_\star[t] + {\bf z}_m[t], [C_{\min}, C_{\max}] ), ~ m \in \{1,..., M\}, 
    \label{eq:RA:update}
    \vspace{-1.5mm}
\end{equation}
which we call the \textit{RA update} for the $m$-th codeword.
Here, ${\rm clip}(\cdot,[C_{\min}, C_{\max}])$ is an element-wise clip function ensuring constraint
% the solution to reside in the allowed region to satisfy  
\eqref{eq:con:maxrate}.
Each entry of ${\bf z}_m[t]$ is generated from the uniform distribution $\mathcal{U}(-\delta, \delta)$, where $\delta$ is the maximum step size for the entry update.
% The RA approach follows \ref{step1}–\ref{step4}, while in \ref{step4} the codewords are updated by \eqref{eq:RA:update}.
The RA approach is summarized in Algorithm \ref{al:random}.

The codebook update by the RA approach  incurs a small computation and communication overhead, which  will be discussed in Sec.~\ref{ssec:comp} and \ref{ssec:effective_rate}.
Intuitively, it becomes more effective as the number of codewords $M$ grows larger
since more random points increase the chance of obtaining better codewords.
However, $M$ is limited due to the non-negligible IRS reconfiguration time and finite coherence time.
This makes the performance of the RA approach restricted due to the nature of the randomness and 
motives us to develop the next codebook update algorithm.

%%%%%%%%% Algorithm
 \begin{algorithm}[t]
 \caption{Random adjacency (RA) codebook design in the limited feedback protocol}
 \label{al:random}
 \begin{algorithmic}[1]
 \footnotesize
%  \State \textbf{Initialize} 
\State \textbf{Input:} $N_{\rm timestep}$ (the duration of the algorithm), $C_{\min}$, and $C_{\max}$.
  \State The IRS randomly generates the initial codebook $\mathcal{C}[0] = \{ {\bf q}_m[0]\}_{m=1}^M$ 
  within the allowed region in \eqref{eq:con:maxrate}.
%   w ${\bf q}_m[0]=[{x}_{m,1}[0],...,{x}_{m,N_{\rm G}}[0]]$ and ${x}_{m,n}[0]  \in [C_{\min}, C_{\max}]$, $\forall m,n$.
  \For{$t =0, ..., N_{\rm timestep} -1$}
  %%%%%%%%%%%%%%%%%%%%%%%% 
    % \State \multiline{
    % \Statex \hspace{3mm}  $\triangleright${ \underline{Preparation via IRS reconfiguration, codeword selection, and feedback (Steps 1-3):}}
    \State \textit{\textbf{Step 1. IRS channel sounding and reconfiguration.}} 
    The IRS meta-atoms are tuned following $\{{\bf q}_m[t]\}_{m=1}^M$. \label{RA:line:step1}
    % reconfigures its capacitance $M$ times with ${\bf q}_m[t]$, $m=1,...,M$.
    % The IRS reconfigures its capacitances $M$ times, i.e., ${\bf q}_m[t] \in \mathcal{C}[t]$, $m=1,...,M$, while the UE transmits pilot symbols.
%    , where ${\bf c}_m[t]$ is the $m$-th codeword in the codebook $\mathcal{C}[t]$
    % }
    \State \textit{\textbf{Step 2. Codeword selection at BS.}}  The BS 
    % calculates and
    % the data-rate $R({\bf q}_m[t],t)$, $m=1,...,M$, and
    determines 
    % the index 
    $m^\star[t] = \underset{m \in \{1,...,M \}}{\arg\max} \;  R({\bf q}_m[t],t) $. \label{RA:line:step2}
    % \Statex \textit{**\underline{Preparation: Feedback to IRS:}}
    \State \multiline{ \textit{\textbf{Step 3. Feedback to IRS and IRS final configuration.}} The BS feeds back the index $m^\star[t] \in \{ 1,...,M \}$ to the IRS with total $\lceil \log_2 M \rceil$ feedback bits. The IRS tunes the meta-atoms with ${\bf q}_\star[t] = {\bf q}_{m^\star[t]}[t]$ for data transmission period. 
    }
%   \Statex \hspace{3mm}  $\triangleright$ {\underline{Data transmission and IRS codebook update:}}
    \State \multiline{ \textit{\textbf{Step 4. Data transmission and IRS codebook update.}} 
    The IRS obtains $\mathcal{C}[t+1]$ according to \eqref{eq:RA:update}.
    \label{RA:line:step4}
    % where ${\bf c}_m[t+1] = {\rm clip} ({\bf c}[t] + {\bf z}_m[t], [C_{\min}, C_{\max}])$ and each entry of ${\bf z}_m[t]$ is generated from $\mathcal{U}(-\delta, \delta)$, $m=1,..,M$.
%    the codewords around the selected codeword ${\bf c}[t]$, i.e., ${\bf c}_m[t+1] = {\bf c}[t] + {\bf z}_m[t]$, $m=1,..,M$, where ${\bf z}_m[t]$ is the random perturbation vector of which each entry is generated from $\mathcal{U}(-\delta, \delta)$. Then, IRS constructs the codebook $\mathcal{C}[t+1] = \{ {\bf c}_m[t+1] \}_{m=1}^M$.   
    }
     \EndFor
%  \RETURN $ \{ {a_{k,k'}}\} ,{\rm{ \{ }}{b_{k,i}}\} ,{\rm{ \{ }}{{\bf{f}}_k}\} , {\rm{ \{ }}{{\bf{z}}_{k',i}}\} ,{\rm{ \{ }}{F_{k,k'}}\} $ in ${{\cal G}^\star}$
 \end{algorithmic}
 \end{algorithm}

%\clearpage

%% file: DRL.tex
\vspace{-2mm}
\subsection{DNN Policy-based IRS Control (DPIC) Approach}
\label{ssec:DPIC}

% \rev{A deep neural network (DNN) has been exploited to capture implicit features in the observed data. Motivated by this,}
% we propose a DNN-based IRS control (DPIC) approach, which aims to learn a \textit{policy} to design adaptive codebook using the history of information.
% %
% In DPIC, we consider a {\it codeword-wise} codebook update, where
% the $m$-th codeword is updated as
%  \vspace{-1.5mm}
% \begin{equation}
%     {\bf q}_m[t+1] = {\rm clip}\left({\bf q}_m[t] + {\boldsymbol u}_m[t], [C_{\min}, C_{\max}]\right), ~1\leq m\leq M,
%     \label{eq:DPIC:update}
%     \vspace{-1.5mm}
% \end{equation}
% where ${\boldsymbol u}_m[t]$ is the \textit{direction vector} for the $m$-th codeword update.
% Our goal is to determine ${\boldsymbol u}_m[t]$ so that it is the right update direction adapting to the unknown/varying channels. 

DNNs have been exploited to capture implicit features in the observed data. Motivated by this,
we propose a DNN policy-based IRS control (DPIC) approach, aiming to learn \textit{policies} for updating the codebook using the history of observations.
In DPIC, the codebook resides at the IRS, as in RA. However, the IRS now updates the codebook via information reception from the BS through the feedback link.
% Similar to the RA, the codebook $\mathcal{C}[t]=\{ {\bf q}_m[t] \}_{m=1}^M$ resides and gets updated at the IRS.
% However, in DPIC, the codebook is updated with assistance from the BS via information transfer to the IRS over the feedback link.
% \rev{The ultimate goal is to update all the codewords $\{{\bf q}_m[t]\}_{m=1}^M$ for every time $t$. 
% -codeword - responses
% the BS measures the response of each codeword
% the BS has the codeword
% the BS trains the learning architecture (input: previous responses, output: next codeword)
% Why each codeword is updated by the response from the ame codeword?
% We consider that the learning architecture is trained with each codeword and responses ffrom the same codeword to capture 
% (i) the relationship between the codeword and response
% (ii) the successive codeword evolution over time (DRL) (related to future reward)
%
We consider that each codeword is updated \textit{independently} based on
% a portion of the history of observations (in \ref{step2} in Sec.~\ref{ssec:protocol}) obtained from 
its prior deployments.
Henceforth, without loss of generality, we focus on the updates of $m$-th codeword.

% we consider a scenario where the update conducted on each codeword is obtained based on a portion of history of observeations concerned with the enviroenemtn measurements when that particular codeword is used (Step XX)

% We consider a scenario in which the evolution of each codeword is conducted via the responses obtained from the environment (i.e., data rate measurement in Step XXX) upon using the same codeword.
% % \rev{To capture implicit features on each codeword evolution under time-correlated channels, we consider that each codeword is updated only with the responses obtained from the same codeword.}
% \rev{Henceforth, we focus on the $m$-th codeword update since each codeword is updated independently.}

\subsubsection{Low Overhead IRS Control via Direction Codebook}
To conduct the low overhead codeword update,
we introduce a fixed \textit{direction codebook} $\mathcal{D} = \{  {\bf d}_k \}_{k=1}^K$ where ${\bf d}_k \in \mathbb{R}^{N_{\rm IRS}}$, $k=1,...,K$.
The BS only transmits the index of a codeword in $\mathcal{D}$ to the IRS, which enables low feedback overhead for the codeword update.
We assume that $\mathcal{D}$ is generated once at the beginning of the policy learning and shared at both the BS and IRS.
The codebook $\mathcal{D}$ is constructed via RVQ for simulations in Sec.~\ref{sec:sim}.
The BS, as a processing entity, employs a \textit{learning architecture} consisting of a DNN policy and a subsequent quantization process.
In particular, the BS obtains a continuous direction vector ${\boldsymbol u}_m[t] \in \mathbb{R}^{{N_{\rm IRS}}}$ from the DNN policy, from which it finds the index $k_m[t] \in \{ 1,...,K \}$ through the quantization process,
such that $k_m[t]$-th codeword in $\mathcal{D}$, i.e., ${\bf d}_{k_m[t]}$, has the  highest similarity to ${\boldsymbol u}_m[t]$.
% In particular, the BS, as a processing entity, constructs a \textit{learning architecture} \rev{(discussed in Sec.~\ref{ssec:decisionmaking})} to first obtain a continuous direction vector ${\boldsymbol u}_m[t] \in \mathbb{R}^{{N_{\rm IRS}}}$, which is then used to obtain an index $k_m[t] \in \{ 1,...,K \}$ 
% such that $k_m[t]$-th codeword in $\mathcal{D}$, i.e., ${\bf d}_{k_m[t]}$, has 
% the  highest similarity to ${\boldsymbol u}_m[t]$.
The BS then feeds back the index $k_m[t]$ to the IRS,  which the IRS uses to
recover ${\bf d}_{k_m[t]}$
from $\mathcal{D}$ and then updates the $m$-th codeword as
\vspace{-2mm}
\begin{equation}
    {\bf q}_m[t+1] = {\rm clip}({\bf q}_m[t] + {\bf d}_{k_m[t]}, [C_{\min}, C_{\max}]), \; m \in \{1,...,M\},
    \label{eq:DPIC:update}
    \vspace{-2mm}
\end{equation}
which we call the \textit{DPIC update} for the $m$-th codeword.

% Operation of the learning architecture
% The operation of the learning architecture at the BS follows the successive decision making procedure.
% The BS determines $k_m[t]$ from the learning architecture using the used codeword ${\bf q}_m[t]$ and the effective channel ${\bf h}_{ {\rm eff} }({\bf q}_m[t], t )$, and then obtains the next codeword ${\bf q}_m[t+1]$ with $k_m[t]$ by \eqref{eq:DPIC:update}.
% In parallel, the BS trains the learning architecture using the previous observations (i.e., previously used codewords, effective channels and data rates).
\subsubsection{Successive Decision Making for Codeword Update}
% Learning-based IRS Control via Successive Decision Making}
\label{ssec:decisionmaking}
% The learning architecture at the BS consists of a DNN policy that determines ${\boldsymbol u}_m[t]$ and a quantization process that determines $k_m[t]$.
Our learning architecture consists of two phases: \textit{training phase} and \textit{utilization phase}.
In the training phase, the BS aims to train the DNN policy to have an improved ${\boldsymbol u}_m[t]$ over time, while in the utilization phase the BS exploits the trained DNN policy without additional training.
In both phases, the BS first determines ${\boldsymbol u}_m[t]$ with the DNN policy based on the current information (i.e., the codeword ${\bf q}_m[t]$ in use and the effective channel ${\bf h}_{ {\rm eff} }({\bf q}_m[t], t )$). Subsequently, the BS obtains $k_m[t]$ via a quantization process applied to ${\boldsymbol u}_m[t]$ (described in Sec.~\ref{sssec:learning}\&\ref{sssec:utilization}). The BS then feeds back $k_m[t]$ to the IRS, from which the IRS obtains the next codeword ${\bf q}_m[t+1]$ through \eqref{eq:DPIC:update}. The next codeword affects the subsequent information at the BS (i.e., ${\bf q}_m[t+1]$ and ${\bf h}_{ {\rm eff} }({\bf q}_m[t+1], t+1 )$). The codeword update can thus be formulated as a successive decision making process (Sec.~\ref{sssec:MDP}).
We then develop our learning architecture for training (Sec.~\ref{sssec:learning}) and utilization phases (Sec.~\ref{sssec:utilization}).

\subsubsection{Markov Decision Process (MDP) for Codeword Update}
\label{sssec:MDP}

We construct a Markov decision process (MDP) for the codeword update with the following state, action, and reward.

\textbf{State.} 
The state consists of information pertinent to the environment evolution, which we define as
% The state of the MDP consists of available information useful to capture the environment evolution and update the codeword, which is defined as
\vspace{-1.5mm}
\begin{equation}
    {\bf s}_m[t] = \{
    {  {\bf h}_{ {\rm eff} }({\bf q}_m[t], t ) },
    {\bf q}_m[t]
    \}  \in \mathcal{S} =  \mathbb{R}^{2 N_{\rm BS} + {N_{\rm IRS}}}, \; m \in \{1,...,M\},
    \label{eq:state}
    \vspace{-1.5mm}
\end{equation} 
where the real and imaginary parts of $ {\bf h}_{ {\rm eff} }({\bf q}_m[t], t )$ are stored as separate state dimensions.
% Since  is the complex-valued vector, 
% the real and imaginary parts of $ {\bf h}_{ {\rm eff} }({\bf q}_m[t], t )$ are stored individually into the states. The dimension of the state space is $2 N_{\rm BS} + {N_{\rm IRS}} $.
%The state ${\bf s}_m[t]$ will be used to determine the action at time $t$.
%update the $m$-th codeword for being used at time $t+1$.
%
% \hspace{-4mm}\textbf{Actions.}

\textbf{Action.} The action is the continuous direction vector ${\boldsymbol u}_m[t]$  described as:
\vspace{-1.5mm}
\begin{equation}
    {\bf a}_m[t]
    = {\boldsymbol u}_m[t] 
    \in \mathcal{A}  =  [-\delta, \delta]^{N_{\rm IRS}}, \; m \in \{1,...,M\},
    \label{eq:action}
    \vspace{-1.5mm}
\end{equation} 
% where the dimension of the action space is $ {N_{\rm IRS}}$. 
where entry of the action is bounded to the maximum step size, i.e., $[-\delta, \delta] \subset {\mathbb{R}}$. 
The action ${\bf a}_m[t]$ is used to determine the index $k_m[t]$ based on different processes in training (Sec.~\ref{sssec:learning}) and utilization  (Sec.~\ref{sssec:utilization}) phases.
The next codeword ${\bf q}_m[t+1]$ is then obtained from $k_m[t]$~by~\eqref{eq:DPIC:update}.
% and is discussed in Sec.~\ref{sssec:learning} and \ref{sssec:utilization}.
% \rev{The next codeword ${\bf q}_m[t+1]$ is obtained by \eqref{eq:DPIC:update} with the direction vector ${\bf d}_{k_m[t]}$ determined by the action ${\boldsymbol u}_m[t]$.}

\textbf{Reward.}
The reward provides an efficacy for desirable policy learning by evaluating an action at a given state.
% which provides a correct direction for the policy learning.
% so that the policy can determine the best action at a particular state.
We subsequently define the MDP reward as
\vspace{-1.5mm}
\begin{equation}
    r_m[t] =  R({\bf q}_m[t+1], t+1 )   
- N_{{\rm clip},m}[t]
\in \mathbb{R},\; m \in \{1,...,M\},
\label{eq:DPIC:reward}
\vspace{-1.5mm}
\end{equation}
%where $R({\bf c}_m[t+1], t+1 ) = \log_2 
%\bigg( 
%1+ \frac{p \| {\bf h}_{\rm eff}({\bf c}_m[t+1], t+1 ) \|_2^2 }{ \sigma^2}  
%\bigg)$.
where
$R({\bf q}_m[t+1], t+1 )$ denotes the data rate measured at the time $t+1$ using codeword ${\bf q}_m[t+1]$
%in \eqref{eq:DPIC:update}, 
and $N_{{\rm clip},m}[t]$ denotes the number of the clipped elements/dimensions in vector ${\bf q}_m[t+1] \in {\mathbb{R}}^{{N_{\rm IRS}}}$ 
that hit the clipping threshold in
% when applied to the clip function in
\eqref{eq:DPIC:update}.
% ${\bf q}_m[t+1] = {\rm clip} ({\bf q}_m[t] + {\bf a}_m[t], [C_{\min}, C_{\max}])$. 
$N_{{\rm clip},m}[t]$ is added as a  penalty  to
avoid actions that result in the capacitance vectors outside of the allowed region.
Note that the reward $r_m[t]$ is obtained at the next time $t+1$ since the data rate $R({\bf q}_m[t+1], t+1 )$ is calculated at time $t+1$.
% and (ii)  
% encourage more exploration within the allowed region. 
%This soft penalty prevents from badly converge
%the action from having the boundary values, $C_{\min}$ or $C_{\max}$, all the time.
%
%
%  A constant value could be multiplied to  $N_{{\rm clip},m}[t]$ to match with the magnitude of the data rate, depending on the system parameters and environments.
 %
 %
% Note that the reward value $r_m[t]$ is measured at the next time $t+1$ since the data rate $R({\bf q}_m[t+1], t+1 )$ is calculated at time $t+1$.

% that updates the capacitances to ${\bf c}_m[t+1] = {\rm clip} ({\bf c}_m[t] + {\bf a}_m[t], [C_{\min}, C_{\max}])$.

Based on the state, action, and reward, the MDP is defined as a tuple $ ( \mathcal{S},\mathcal{A},\mathcal{R}_{\bf s}^{\bf a},P_{{\bf s},{\bf s}'}^{\bf a},\gamma )$, where 
$P_{{\bf s},{\bf s}'}^{\bf a} = Pr \big[ {\bf s}_m[t+1]= {\bf s}' \big| {\bf s}_m[t]={\bf s}, {\bf a}_m[t]= {\bf a} \big]$ is the state transition probability for moving from state ${\bf s}$ to ${\bf s}'$  via action ${\bf a}$, 
$\mathcal{R}_{\bf s}^{\bf a} = \mathbb{E}\big[ r_m[t] \big| {\bf s}_m[t] \negmedspace = {\bf s}, {\bf a}_m[t]= {\bf a} \big]$ is
the reward function, and $\gamma$ is the discount factor used to take into account the rewards for the distant future.

\subsubsection{Training Phase for DNN Policy Learning}
\label{sssec:learning}
% In DPIC, the BS XXXX for a portion of the codewords.

We tailor a deep reinforcement learning (DRL) methodology to train the DNN policy with the formulated MDP.
% successive decision making process for codeword update 
% with the formulated MDP.}
We assume that the BS trains $M_{\rm A} \leq M$ different learning architectures, which are referred to as \textit{agents}. 
We consider that each agent is trained with a single codeword, where the codewords across the agents are non-overlapping. Thus $M_{\rm A}$ codewords are used during the training phase of DPIC.
% We \textit{partition} $M_A$ codewords across these agents, where each agent is trained with a single codeword, which is different than the other agents.
% We consider that the agents are trained with different codewords from each other.
Let $\mathcal{M}_{\rm A} \subset \{1,...,M\}$ denote the indices of the codewords associated with learning agents with  $|\mathcal{M}_{\rm A}|=M_{\rm A}$. 
% For $m \in \mathcal{M}_{\rm A}$, with the MDP of the  codeword $m$, the BS trains the agent $m$. 
We will use $m$ to denote a codeword and its associated agent interchangeably. We consider that agent $m\in \mathcal{M}_{\rm A}$ has the DNN policy $\pi( {\bf s}_m[t]; {\bf w}_{\pi,m})$, which outputs the continuous direction vector ${\boldsymbol u}_m[t] \in \mathbb{R}^{{N_{\rm IRS}}}$ given state ${\bf s}_m[t]$, where ${\bf w}_{\pi,m}$ is the respective DNN weight parameters.
% , $m \in \mathcal{M}_{\rm A}$.}

% We assume that the BS trains $M_{\rm A}  \leq  M$ agents. 
% Let $\mathcal{M}=\{1,2,...,M\}$ denote all the codeword indices, and $\mathcal{M}_{\rm A} \subset \mathcal{M}$ denote the indices of the codewords associated with learning agents with  $|\mathcal{M}_{\rm A}|=M_{\rm A}$.
% Each agent $m \negmedspace \in \negmedspace \mathcal{M}_{\rm A}$ is associated with updating the $m$-th codeword. We will use $m$ to denote a codeword and an agent interchangeably.  
% With the MDP of the $m$-th codeword,
% the BS trains the $m$-th DNN policy $\pi( {\bf s}_m[t]; {\bf w}_{\pi,m})$, where ${\bf w}_{\pi,m}$ is the respective DNN weight parameters, $m \in \mathcal{M}_{\rm A}$.
% We refer to the learning entity  dedicated to training the $m$-th DNN policy 
% as {\it agent $m$}.
% Agents are trained independently from one another.

%  with random noise addition and RVQ-based quantization
\textbf{Behavior policy.} We refer to $\pi( {\bf s}_m[t]; {\bf w}_{\pi,m})$ as a {\it target} policy, which is different from  the {\it behavior} policy that determines the actual action applied to the environment.
The actual action of the agent $m$ (i.e., ${\bf d}_{k_m[t]}$ in \eqref{eq:DPIC:update}) is determined at the BS via the two following steps.
%
% To determine the actual action, we consider the two following steps. 
First, the BS adds the random noise vector ${\bf v}_m[t]$ to the output of the target policy $\pi( {\bf s}_m[t]; {\bf w}_{\pi,m})$ to have more diverse responses and avoid getting trapped in local optima during training~\cite{sutton2018reinforcement},
% $\epsilon$-greedy
where ${\bf v}_m[t] \sim \mathcal{N}({\bf 0}, \epsilon[t] {\bf I}) $ with $\epsilon[t]$ denoting the exploration noise variance. We then use the clip function to confine the output result to the feasible action space.
% During training, the actual action is generally determined via random noise addition to the output of the target policy, e.g.,  using $\epsilon$-greedy, to have more exploration in the environment and avoid getting trapped in local optima~\cite{sutton2018reinforcement}.
% For our learning architecture, we add the random noise vector ${\bf v}_m[t]$, where ${\bf v}_m[t] \sim \mathcal{CN}({\bf 0}, \epsilon[t] {\bf I}) $ and $\epsilon[t]$ denotes the exploration noise variance, to the output of the target policy $\pi( {\bf s}_m[t]; {\bf w}_{\pi,m})$ and use the clip function to 
% confine it to the feasible action space.
%
Second, the BS adds the {\it quantization process} with the direction codebook $\mathcal{D}$, through which
% generated by RVQ in the action space $[-\delta, \delta]^{\q{N_{\rm IRS}}}$.
the BS determines the codeword index $k_m[t] \in \{1,...,K\}$ with closest Euclidean distance. 
In other words, the behavior policy $\mu_{\mathcal{D},\pi}(  {\bf s}_m[t] )$, yielding $k_m[t]$ as an output, is represented as
% such that the $k_m[t]$-th codeword is closest to the clipped output result among the codewords in $\mathcal{D}$ in Euclidean distance.
% With the two steps,
% Considering the random noise addition and quantization, we mathematically define the behavior policy $\mu_{\mathcal{D},\pi}(  {\bf s}_m[t] )$ as
\vspace{-1.5mm}
\begin{equation}
    k_m[t] = \mu_{\mathcal{D},\pi}(  {\bf s}_m[t] ) 
    = \underset{ k \in \{1,...,K\} }{\arg\min} \;  
    \| {\rm clip}(\pi( {\bf s}_m[t]; {\bf w}_{\pi,m}) + {\bf v}_m[t], [-\delta,\delta]) - {\bf d}_k \|_2.
    \label{eq:DPIC:behavior_policy}
    \vspace{-1.5mm}
\end{equation}
% The BS feeds back $k_m[t]$ to the IRS, which then
% recovers the actual action ${\bf d}_{k_m[t]}$ from $\mathcal{D}$ with $k_m[t]$ and 
% updates the codeword $m$ by~\eqref{eq:DPIC:update}.

\textbf{DNN policy learning with actor-critic network.}
For DNN policy learning, we exploit the actor-critic network using DNNs as function approximators that can learn policies in  continuous state and action spaces \cite{lillicrap2015continuous}. 
The actor-critic network consists of an 
actor network and a critic network, where 
the former selects an action using a policy, and the later evaluates/criticizes the action to guide the actor network to take better actions over time.
First, for a given policy $\pi(\cdot)$, we define the action-value function, called Q-function, with the discount factor $\gamma$ as
\vspace{-1.5mm}
\begin{equation}
    Q_m^{\pi}( {\bf s}, {\bf a}) =  \mathbb{E}_{\xi} \bigg[ \sum\nolimits_{i=0}^\infty \gamma^{i} r_m[t+i]  \big| {\bf s}_m[t] = {\bf s}, {\bf a}_m[t] = {\bf a}, \pi \bigg],
    \label{eq:Q}
\end{equation}
where $\xi$ encapsulates the state transition probability $P_{{\bf s},{\bf s}'}^{\bf a}$ and reward function $\mathcal{R}_{\bf s}^{\bf a}$.
% $R_{\bf s}^{\bf a}$
%
%where the action ${\bf a}_m[t]$ is determined by the target policy $\pi({\bf s}_m[t]; {\bf w}_{\pi,m})$.
%Since the behavior policy is different from the target policy, 
% Note that the state trajectory is determined by the behavior policy, not the target policy. Then, we define the performance objective as
With $Q_m^{\pi}( {\bf s}, {\bf a})$, we define the performance objective \cite{silver2014deterministic} as
\vspace{-1.5mm}
\begin{equation}
    J_m^\mu(\pi) = \int_\mathcal{S} \rho_m^\mu({\bf s}) Q_m^\pi({\bf s},\pi( {\bf s}; {\bf w}_{\pi,m}))  d {\bf s} 
    = \mathbb{E}_{ {\bf s} \sim \rho^\mu_m} \big[  Q_m^\pi( {\bf s},\pi({\bf s}; {\bf w}_{\pi,m}))  \big],
    \vspace{-1mm}
\end{equation}
where $J^\mu_m(\pi)$ denotes the expected cumulative discounted reward over all the states when the state trajectory is provided by behavior policy $\mu$. Here,
$\rho_m^\mu({\bf s}) = \int_\mathcal{S}  \sum_{i=1}^\infty \gamma^{i-1} p_1(\tilde {\bf s}) p(\tilde {\bf s} \rightarrow {\bf s},i,\mu) d \tilde {\bf s} $ is the discounted state distribution where $p(\tilde{\bf s} \rightarrow {\bf s},i, \mu)$ denotes the probability density at state ${\bf s}$ after transitioning for $i$ time steps from state $\tilde {\bf s}$ under behavior policy $\mu$, and $p_1(\tilde {\bf s})$ is 
the probability density of the initial state distribution.
The objective is to design the target policy $\pi$ (i.e., the DNN parameters ${\bf w}_{\pi,m}$) such that $J^\mu_m(\pi)$ is maximized.
%
%\textbf{Deterministic policy learning.}
%Our objective is to learn the parameters ${\bf w}_\pi$ for the policy $\pi(s;{\bf w}_\pi)$ by maximizing the performance objective $J_\mu(\pi)$.
To achieve this, the learning is conducted using gradient descent iterations on $J^\mu_m(\pi)$ where the DNN parameters are updated~as
\vspace{-7mm}
\begin{equation}
    {\bf w}_{\pi,m} \leftarrow {\bf w}_{\pi,m} - \alpha_{\pi} \nabla_{{\bf w}_{\pi,m}}  J^\mu_m(\pi),
    \label{eq:policyupdate}
    \vspace{-1.5mm}
\end{equation}
where  
% \begin{equation}
$\nabla_{{\bf w}_{\pi,m}}  J_m^\mu(\pi) \approx \mathbb{E}_{ {\bf s} \sim \rho^\mu_m} \big[ \nabla_{{\bf w}_{\pi,m}} \pi({\bf s};{\bf w}_{\pi,m}) \nabla_{\bf a}  Q_m^{\pi}({\bf s}, {\bf a})|_{{\bf a}=\pi({\bf s};{\bf w}_{\pi,m})} \big]$
    % \label{eq:dpg}
% \end{equation}
denotes the deterministic policy gradient (DPG),
the derivation of which is detailed in \cite{silver2014deterministic}, and
$\alpha_\pi$ is the learning rate.

%For learning the policy $\pi_\theta(s)$, a deep neural network with parameters $\theta$ can be used. 
%\rev{Learning the deterministic policy $\pi_\theta(s)$ with a deep neural network is called a deep deterministic policy gradient (DDPG) algorithm proposed by \cite{lillicrap2015continuous}.}
%
%
%\textbf{Actor-critic architecture for deterministic policy learning.}

We define another DNN as a function approximator for the Q-function, i.e., $Q({\bf s}, {\bf a}; {\bf w}_{Q,m}) \approx Q_m^{\pi}({\bf s}, {\bf a})$, with the parameters ${\bf w}_{Q,m}$, which is used to calculate the gradient in \eqref{eq:policyupdate}.
%
% This learning framework is called \textit{an actor-critic architecture}; An actor is referred to the policy $\pi({\bf s}; {\bf w}_{\pi,m})$, and the critic to the Q-function $Q({\bf s}, {\bf a}; {\bf w}_{Q,m})$. The nomenclature comes from their relationship  that the actor network selects an action, and the critic evaluates (criticizes) the action.
%
%
We obtain $Q({\bf s}, {\bf a}; {\bf w}_{Q,m})$ using Q-learning
% , a commonly used off-policy algorithm 
\cite{sutton2018reinforcement,silver2014deterministic} with minimizing the following loss function
\vspace{-1.5mm}
% Then, we can define the loss function as 
\begin{equation}
    \mathcal{L}_m = \mathbb{E} \big[(y - Q({\bf s}, {\bf a}; {\bf w}_{Q,m}))^2 \big],
    \label{eq:Qloss}
    \vspace{-1.5mm}
\end{equation}
where the target value $y$ is given by
%\begin{equation}
    $y = r + \gamma Q({\bf s}', \pi({\bf s}' ; {\bf w}_{\pi,m}) ; {\bf w}_{Q,m})$. Here, $r$ is the reward for action ${\bf a}$ given  state ${\bf s}$.
%\end{equation}
%$y = r + \gamma Q_w(s', \pi_\theta(s') )$ 
The learning for ${\bf w}_{Q,m}$ is followed by the gradient-based update as
\vspace{-1.5mm}
\begin{equation}
    {\bf w}_{Q,m} \leftarrow {\bf w}_{Q,m} - \alpha_Q \nabla_{{\bf w}_{Q,m}} \mathcal{L}_m,
    \label{eq:Qupdate}
    \vspace{-1.5mm}
\end{equation}
where $\alpha_Q$ is the learning rate and 
% $\nabla_{{\bf w}_{Q,m}} \mathcal{L}$ is the gradient of the loss function given by
$
     \nabla_{{\bf w}_{Q,m}} \mathcal{L}_m = -\mathbb{E} \big[
 (y-Q({\bf s}, {\bf a}; {\bf w}_{Q,m})) \nabla_{{\bf w}_{Q,m}} Q({\bf s},{\bf a};{\bf w}_{Q,m})
\big]$.
% \label{eq:gradQ}
% \end{equation}

%%%%%%%%%%%%%%%%%%%%
\begin{figure*}[t]
%   \vspace{-2mm}
  \includegraphics[width=\linewidth]{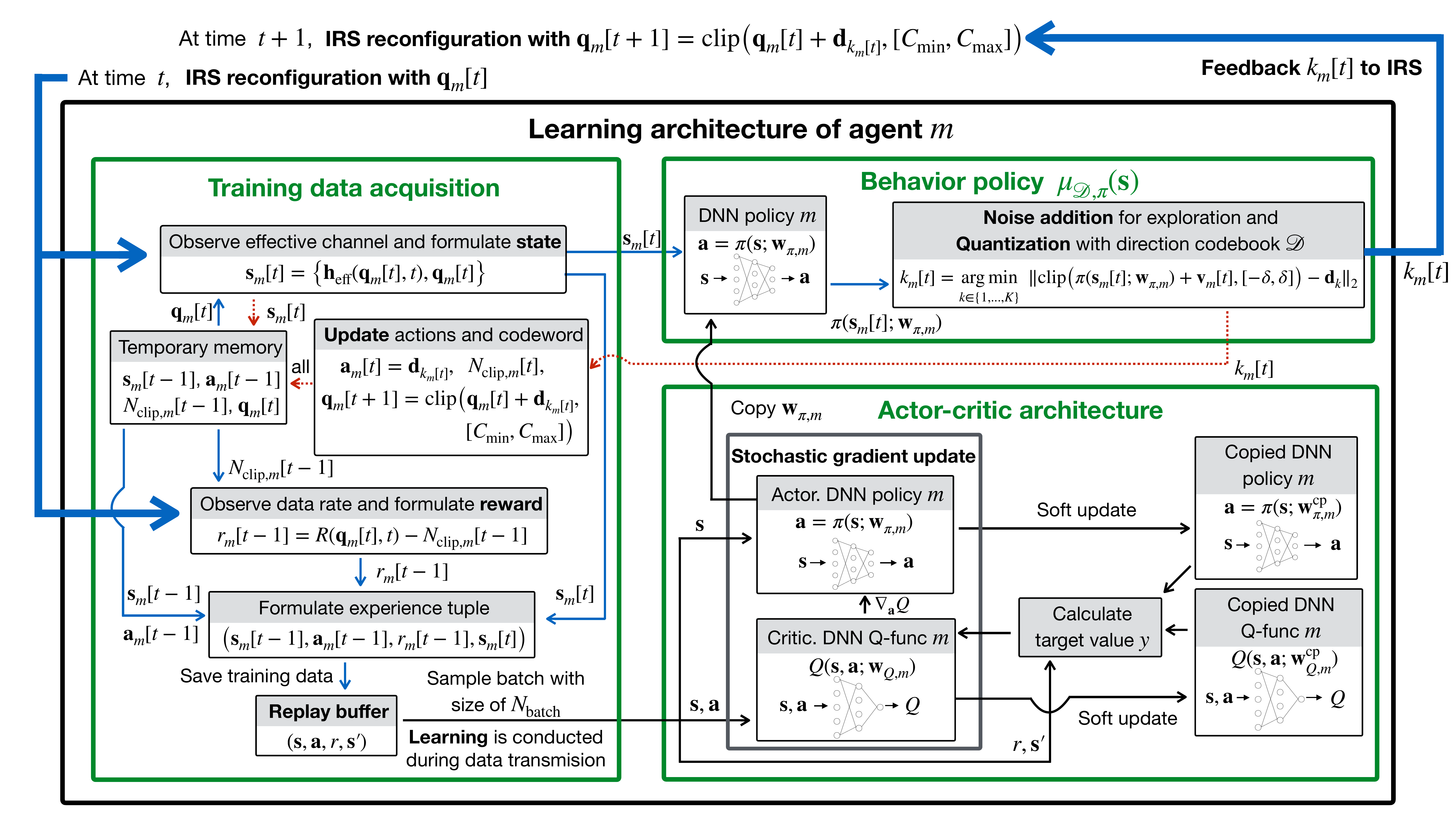}
  \centering
  \vspace{-11mm}
  \caption{The workflow for training each agent $m$ in our limited feedback protocol. 
  The agent collects the training data, updates the DNN policy (actor network) and DNN Q-function (critic network) in the actor-critic architecture via mini-batch learning, selects the action via the behavior policy, and feeds back the direction index to the IRS.
  }
  \label{fig:DPIC:actorcritic}
  \vspace{-7mm}
\end{figure*}    
%%%%%%%%%%%%%%%%%%%%

Using large and non-linear function approximators, such as DNNs, for reinforcement learning has been known to cause learning instability~\cite{sutton2018reinforcement}. 
%
% The authors in \cite{lillicrap2015continuous}  successfully resolve the instability issue by introducing several strategies.
% , where the alo called the DNN-implemented DPG as DDPG.
%
We make use of the strategies proposed in \cite{lillicrap2015continuous} to stabilize the learning.
% proposed in \cite{lillicrap2015continuous}. 
First, we use a \textit{replay buffer} to save the tuple $( {\bf s}_m[t], {\bf a}_m[t], r_m[t], {\bf s}_m[t+1] )$ over time, where the learning is conducted via random batch selection from the reply buffer. This makes the samples chosen for  learning uncorrelated which leads to stability of the learning.
Second, for updating 
% $\pi({\bf s}; {\bf w}_{\pi,m})$ and 
$Q({\bf s}, {\bf a}; {\bf w}_{Q,m})$, \textit{soft target} updates are used to improve the learning stability by making the target value $y$ in \eqref{eq:Qloss} slowly varying. 
This is enabled by constructing 
% To do that, we should construct 
two additional DNNs, called \textit{copied networks}: the copied policy $\pi({\bf s}; {\bf w}^{\rm cp}_{\pi,m})$ with  parameters ${\bf w}^{\rm cp}_{\pi,m}$, and the copied Q-function $Q({\bf s}, {\bf a}; {\bf w}^{\rm cp}_{Q,m})$ parameterized by ${\bf w}^{\rm cp}_{Q,m}$. 

% With the adopted strategies, 
% Then,

In particular, the training of  $\pi({\bf s}; {\bf w}_{\pi,m})$ and $Q({\bf s}, {\bf a}; {\bf w}_{Q,m})$ is conducted via mini-batch learning with size $N_{\rm batch}$ with sampled chosen from the replay buffer $\mathcal{B}_m$. To this end,
the agent $m$  samples a random batch $({\bf s}_i, {\bf a}_i, {r}_i, {\bf s}'_{i} )$ from $\mathcal{B}_m$, and sets $y_i = r_i + \gamma Q({\bf s}'_{i}, \pi({\bf s}'_{i} ; {\bf w}^{\rm cp}_{\pi,m} ); {\bf w}^{\rm cp}_{Q,m} ) $, $i=1,...,N_{\rm batch}$. 
Then, the gradients in  \eqref{eq:policyupdate} and \eqref{eq:Qupdate} are approximated as
\vspace{-1.5mm}
\begin{align}
    \nabla_{{\bf w}_{\pi,m}} J_m^\mu(\pi) & \approx \sum\nolimits_{i=1}^{N_{\rm batch}} \nabla_{{\bf w}_{\pi,m}} \pi({\bf s}_i; {\bf w}_{\pi,m} )\nabla_{\bf a}  Q({\bf s}_i,{\bf a}; {\bf w}_{Q,m}  )|_{ {\bf a}= \pi( {\bf s}_i; {\bf w}_{\pi,m} ) },
    \label{eq:dpg_batch}
    \\
    \nabla_{{\bf w}_{Q,m}} \mathcal{L}_m & \approx - \sum\nolimits_{i=1}^{N_{\rm batch}} (y_i - Q({\bf s}_i,{\bf a}_i); {\bf w}_{Q,m} )\nabla_{{\bf w}_{Q,m}} Q({\bf s}_i, {\bf a}_i; {\bf w}_{Q,m}).
    \label{eq:gradQ_batch}
    \vspace{-1.5mm}
\end{align}
The copied networks are then updated with the soft target update  parameter $\tau$ as
\vspace{-1.5mm}
\begin{equation}
    {\bf w}^{\rm cp}_{\pi,m}  \leftarrow \tau {\bf w}_{\pi,m} + (1-\tau) {\bf w}^{\rm cp}_{\pi,m},  \quad
    {\bf w}^{\rm cp}_{Q,m} \leftarrow \tau {\bf w}_{Q,m} + (1-\tau) {\bf w}^{\rm cp}_{Q,m}.
    \label{eq:soft}
    \vspace{-1.5mm}
\end{equation}

%{\small  sadas}

%\textbf{Algorithm for learning  $M_{\rm A}$ agents.}
%Incorporating the learning framework in the limited feedback protocol, we consider to train $M_{\rm A}$ agents among $M$ capacitances vectors in the training period, where $M_{\rm A} \le M$. Each agent $m$ is trained with the $m$-th capacitances codeword ${\bf c}_m[t]$, independently from other agents. 

%%%%%%%%% Algorithm
 \begin{algorithm}[t]
 \caption{Training $M_{\rm A}$ agents with actor-critic architecture in the proposed protocol}
 \label{al:policy}
 \begin{algorithmic}[1]
 \footnotesize
%  \State \textbf{Initialize} 
\State \textbf{Input.} $N_{\rm episode}$ (the number of learning episodes), $N_{\rm timestep}$ (the duration of each episode), 
$\epsilon_{0} = (C_{\max}-C_{\min})/5$ (the initial exploration variance), 
$\epsilon_{\min}=\epsilon_{0}/300$ (the minimum exploration variance)
$C_{\min}$, $C_{\max}$,
% the set of codeword indices associated with the training agents
$\mathcal{M}_{\rm A}$, and 
% the set of the rest codeword indices for RA update
${\mathcal{M}}_{\rm RA}$.
% ${\mathcal{M}}_{\rm RA}=\{1,...,M\}\setminus \mathcal{M}_{\rm A}$.
  \State 
  Initialize ${\bf w}_{Q,m}$, ${\bf w}_{\pi,m}$, ${\bf w}^{\rm cp}_{Q,m}$, and ${\bf w}^{\rm cp}_{\pi,m}$  for the DNN networks. 
%   the critic, actor, copied critic, and copied actor networks, respectively.
  Empty
  the replay buffer $\mathcal{B}_m$, $m \in \mathcal{M}_{\rm A}$. 
%   Set the exploration variance $\epsilon_{0} = (C_{\max}-C_{\min})/5$ and $\epsilon_{\min}=\epsilon_{0}/300$. 
  The direction codebook $\mathcal{D} = \{ {\bf d}_{k} \}_{k=1}^K$ is shared at both the BS and the IRS. 
   %%%%%%%%%%%%%%%%%%%%%%%% 
  \For{$e=0,...,N_{\rm episode}-1$}
   %%%%%%%%%%%%%%%%%%%%%%%% 
  \State \multiline{ Randomly generate the codebook $ \mathcal{C}[0] = \{ {\bf q}_m[0] \}_{m=1}^{M}$ satisfying \eqref{eq:con:maxrate}.
  Update $\epsilon_e = \max \{ \epsilon_{\min}, 0.99 \epsilon_{e-1} \} $, if $e \ge 1$. }
   %%%%%%%%%%%%%%%%%%%%%%%% 
  \For{$t=0,...,N_{\rm timestep}-1$}
  %%%%%%%%%%%%%%%%%%%%%%%% 
    %%% Step 1
    % \Statex \hspace{8mm}  $\triangleright$ \underline{Preparation via IRS reconfiguration, codeword selection, inference, and feedback (Steps 1-3):}
    \State \multiline{ \textit{\textbf{Step 1. IRS channel sounding and reconfiguration.}} 
    The IRS meta-atoms are tuned following $\{{\bf q}_m[t]\}_{m=1}^M$.
    }
    %%% Step 2
    \State \multiline{ \textit{\textbf{Step 2. Codeword selection and inference at BS.}} 
    The BS 
    % calculates data rate and 
    determines the index $m^\star[t] = \underset{m \in \{1,...,M \}}{\arg\max} \; R({\bf q}_m[t],t) $. 
%    }
%    \State \multiline{
    Each agent $m$, $m\in{\mathcal{M}}_{\rm A}$, at the BS 
    % measures ${\bf h}_{\rm eff}({\bf q}_m[t],t)$, 
    forms ${\bf s}_m[t] = \{  {\bf h}_{\rm eff}({\bf q}_m[t],t), {\bf q}_m[t] \}$ and
     determines ${k_m[t]} = \mu_{\mathcal{D},\pi}({\bf s}_m[t]) $ using \eqref{eq:DPIC:behavior_policy}, where ${\bf v}_m[t] \sim \mathcal{CN}({\bf 0}, \epsilon_e {\bf I}) $. \label{DPIC:tr:line:step2}
     }
    %%% Step 3
    \State \multiline{ \textit{\textbf{Step 3. Feedback to IRS and IRS final configuration.}} The BS feeds back 
    % the indices,
    $m^\star[t]$ and $\{k_m[t]\}_{m\in{\mathcal{M}}_{\rm A}}$ to the IRS with $\lceil \log_2 M \rceil  +  M_{\rm A} \lceil \log_2 K \rceil $ feedback bits.
    % where $ k_m[t] \in \{ 1,...,K \}$. 
    The IRS tunes the meta-atoms with ${\bf q}_\star[t]  =  {\bf q}_{m^\star[t]}[t]$ for data transmission. 
    }  
    % \Statex \hspace{8mm}  $\triangleright$ \underline{Data transmission, IRS codebook update, and BS training:}
    \State \multiline{ \textit{\textbf{Step 4. Data transmission, IRS codebook update, and BS training.}} 
    The IRS updates $\mathcal{C}[t+1] = \{ {\bf q}_m[t+1] \}_{m=1}^{M}$, where the DPIC update is conducted by \eqref{eq:DPIC:update} for $m\in \mathcal{M}_{\rm A}$, and the RA update by \eqref{eq:RA:update} for $m \in {\mathcal{M}}_{\rm RA} $.
    % $M_{\rm A}$ codewords are updated by the DPIC, i.e., ${\bf q}_m[t+1] = {\rm clip}({\bf q}_m[t] + {\bf d}_{k_m[t]}, [C_{\min}, C_{\max}]  )$, $m=1,...,M_{\rm A}$, and  $M-M_{\rm A}$ codewords are updated by the RA, i.e., ${\bf q}_m[t+1] = {\rm clip}({\bf c}[t] + {\boldsymbol u}_m[t], [C_{\min}, C_{\max}])$,  $m=M_{\rm A}+1,...,M$.
    %
    Each agent $m\in\mathcal{M}_{\rm A}$ at the BS 
    % saves ${\bf d}_{k_m[t]}$, ${\bf q}_m[t+1]$, and $N_{{\rm clip},m}[t]$, 
    computes $r_m[t-1]$ using \eqref{eq:DPIC:reward}, stores $( {\bf s}_m[t-1], {\bf a}_m[t-1], r_m[t-1], {\bf s}_m[t] )$ in $\mathcal{B}_m$, samples $({\bf s}_i, {\bf a}_i, {r}_i, {\bf s}'_{i} )$ from $\mathcal{B}_m$,
    and
    updates the DNN networks through \eqref{eq:policyupdate}, \eqref{eq:Qupdate}, \eqref{eq:dpg_batch}-\eqref{eq:soft}.
    \label{DPIC:tr:line:step4}
     }
     \EndFor
     \EndFor
 \end{algorithmic}
 \end{algorithm}

Fig. \ref{fig:DPIC:actorcritic} depicts the workflow for training each agent $m$.
% where the agent collects the training data, updates the DNN policy (actor) and DNN Q-function (critic) in the actor-critic architectures via batch learning, selects the action with the behavior policy, and feeds back the index $k_m[t]$ to the IRS.
We have so far focused on the training process of each agent $m$, $m\in \mathcal{M}_{\rm A}$. 
% Now, let us consider how the training process can be incorporated into the proposed protocol. 
Prior to training, the BS determines the number of agents  trained, i.e., $M_{\rm A}=|\mathcal{M}_{\rm A}|$. 
% The BS initially constructs $M_{\rm A}$ different  learning architectures of the agents, and trains the agent $m$  with the $m$-th codeword ${\bf q}_m[t]$ over time during the training period, where $m=1,...,M_{\rm A}$.
%
% Given in the proposed protocol, $M$ codewords are considered for different IRS configuration.
% Among $M$ codewords, the BS can choose the number of agents $M_{\rm A}$, i.e., how many agents are being trained during the training period.
%
% The workflow for training the learning architecture is  incorporated into the limited feedback protocol.
%
% The learning architecture is running independently for total $M_{\rm A}$ agents in the limited feedback protocol.
% Among $M$ codewords, the BS can choose the number of agents $M_{\rm A}$ being trained. 
The BS can train as many agents as possible, i.e., $M_{\rm A} = M$, or $M_{\rm A} < M$, while the rest of
$M-M_{\rm A}$ 
codewords are updated by the RA update, of which indices form the set
${\mathcal{M}}_{\rm RA}=\{1,...,M\}\setminus \mathcal{M}_{\rm A}$. 
Training different numbers of agents leads to different performance (see Sec.~\ref{sec:sim}) and incurs different computation/communication overhead (see Sec. \ref{ssec:comp}\&\ref{ssec:effective_rate}).
The overall algorithm for training $M_{\rm A}$ agents is given in Algorithm \ref{al:policy}. 

\subsubsection{Utilization Phase with Trained DNN Policies}
\label{sssec:utilization}

In the utilization phase, we utilize the trained agents to conduct the codebook update without additional  training of the agents.
% We develop several \textit{augmented strategies} for the codebook update.
%
Among $M$ codewords, we can select $M_{\rm DPIC}$ codewords to be updated by the DPIC update in \eqref{eq:DPIC:update} and $M_{\rm RA}$ codewords to be updated by the RA update in \eqref{eq:RA:update}, where $M = M_{\rm DPIC} + M_{\rm RA}$.
We develop four different strategies with different selections of $M_{\rm DPIC}$ and $M_{\rm RA}$:
% Four scenarios of the selection of $M_{\rm DPIC}$ and $M_{\rm A}$ are considered.
\begin{enumerate}[label=\alph*),leftmargin=5mm]
    \item $M_{\rm DPIC} \negmedspace = \negmedspace M$ and $M_{\rm A} \negmedspace = \negmedspace 1$: a single agent handles $M$ codeword updates. We call this case as {\it single-agent DPIC (SDPIC)}.
    \item $M_{\rm DPIC} \negmedspace = \negmedspace M$ and $M_{\rm A} \negmedspace > \negmedspace 1$: multiple agents handles $M$ codeword updates. We call this case as {\it multi-agents DPIC (MDPIC)}.
    \item  $M_{\rm DPIC} \negmedspace < \negmedspace M$ and $M_{\rm A} \negmedspace = \negmedspace 1$: a single agent handles $M_{\rm DPIC}$ codeword updates while $M_{\rm RA} = M - M_{\rm DPIC}$ codewords are updated by the RA update. We call this case as {\it RA+SDPIC}.
    \item $M_{\rm DPIC} \negmedspace < \negmedspace M$ and $M_{\rm A} \negmedspace > \negmedspace 1$: multi-agents handles $M_{\rm DPIC}$ codeword updatse while $M_{\rm RA} = M - M_{\rm DPIC}$ codewords are updated by the RA update. We call this case as {\it RA+MDPIC}.
\end{enumerate}
% The Algorithm \ref{al:DPIC} embraces all the possible strategies, i.e., SDPIC, MDPIC, RA+SDPIC, and RA+MDPIC.
%
When multiple codewords are updated with multiple agents in MDPIC and RA+MDPIC,
% i.e., $M_{\rm A}>1$ and $M_{\rm DPIC}>1$,
the BS allocates/partitions the codewords among the agents. 
Let $\mathcal{M}_{\rm DPIC}$ with $|\mathcal{M}_{\rm DPIC}| = M_{\rm DPIC}$ denote the set of indices of the codewords updated by the  DPIC update during the utilization phase. We let $j[m]
\in\mathcal{M}_{\rm A}$ denote the
 agent handling the codeword $m\in \mathcal{M}_{\rm DPIC}$.
%  and $j[m] \in \{ 1,...,M_{\rm A} \}$.
We take a simple round-robin strategy to allocate the codewords among the trained agents.
% as $j[m] = 1+ \mod(m-1,M_{\rm A})$.
If $M_{\rm A} \le M_{\rm DPIC}$, 
% some trained agents take charge of multiple codewords update where each agent uses its learned DNN policy to generate codeword updates for multiple codewords.
some of the agents may handle multiple codewords.
% take charge of multiple codewords update where
% each agent uses its learned DNN policy to generate codeword updates for multiple codewords.
If $M_{\rm A} > M_{\rm DPIC}$, some trained agents are not used while each of the rest takes charge of one codeword independently.
Utilizing more agents often improves
% data rate 
performance due to the ensemble learning principle~\cite{goodfellow2016deep}. 
% That is, multiple machine learning models that were trained independently with different initialization and training data are beneficial to the performance improvement.
% For the $m$-th codeword update as inference,
We represent the behavior policy during the utilization phase for the $m$-th codeword update, conducted by agent $j[m]$ with the learned parameters~${\bf w}_{\pi,j[m]}$, as
% In the utilization period, each agent $j[m]$  with the learned parameters ${\bf w}_{\pi,j[m]}$  updates the codeword $m$, with behavior policy given by
\vspace{-2mm}
\begin{equation}
     {k_m[t]} = \mu_{\mathcal{D},\pi}({\bf s}_m[t])  =
%     f_{\mathcal{D}}(\pi({\bf s}_m[t]; {\bf w}_{\pi,j[m]}) )
     \underset{ k \in \{1,...,K\} }{\arg\min} \;  
    \| \pi( {\bf s}_m[t]; {\bf w}_{\pi,j[m]}) - {\bf d}_k \|_2.
    \label{eq:DPIC:util_behavior_policy}
    \vspace{-1.5mm}
\end{equation}
% We can classify the strategies into two cases. 
%
% First, when $M_{\rm DPIC} = M$,
% the IRS updates all of $M$ codewords by the DPIC update with $M_{\rm A}$ trained agents. 
% % employed at the BS. 
% If $M_{\rm A} = 1$, a single agent takes charge of the $M$ codewords update. We call this case as {\it single-agent DPIC (SDPIC)}.
% If $M_{\rm A} > 1$, multiple agents take charge of the $M$ codewords update. We call this case as {\it multi-agents DPIC (MDPIC)}.
% Second, when $M_{\rm DPIC} <  M$,
% the IRS updates only $M_{\rm DPIC}$ codewords by the DPIC update, while $M_{\rm RA} = M - M_{\rm DPIC}$ codewords are updated by the RA update.
% % , where $M_{\rm RA} = M - M_{\rm DPIC}$.
% If $M_{\rm A} = 1$, a single agent takes charge of the $M_{\rm DPIC}$ codewords update. We call this case as {\it RA+SDPIC}. If $M_{\rm A} > 1$, multi-agents take charge of the $M_{\rm DPIC}$ codewords update. We call this case as {\it RA+MDPIC}.

%%%%%%%%% Algorithm
 \begin{algorithm}[t]
 \caption{DNN policy-based IRS control (DPIC) approach in the utilization phase. }
 \label{al:DPIC}
 \begin{algorithmic}[1]
 \footnotesize
%  \State \textbf{Initialize} 
\State \textbf{Input.} 
$N_{\rm timestep}$ (the duration of the algorithm), $C_{\min}$, $C_{\max}$, $\mathcal{M}_{\rm DPIC}$, and
% the set of rest of codeword indices for the RA update
${\mathcal{M}}_{\rm RA}=\{1,...,M\}\setminus \mathcal{M}_{\rm DPIC}$.
  \State The IRS randomly generates the codebook $\mathcal{C}[0] = \{ {\bf q}_m[0]\}_{m=1}^M$ satisfying \eqref{eq:con:maxrate}. The BS and IRS share $\mathcal{D} = \{ {\bf d}_{k} \}_{k=1}^K$.
  \For{$t =0, ..., N_{\rm timestep}-1 $}
  %%%%%%%%%%%%%%%%%%%%%%%% 
    \State \multiline{ \textit{\textbf{Step 1. IRS channel sounding and reconfiguration.}} 
    The IRS meta-atoms are tuned following $\{{\bf q}_m[t]\}_{m=1}^M$.}
    \State \multiline{ \textit{\textbf{Step 2. Codeword selection and inference at BS.}}  The BS 
    determines
    $m^\star[t] = \underset{m \in \{1,...,M \}}{\arg\max} \; R({\bf q}_m[t],t) $.
    Each agent $m\in \mathcal{M}_{\rm DPIC}$
    % the BS measures 
    % the effective channel 
    % ${\bf h}_{\rm eff}({\bf q}_m[t],t)$ and 
    constructs 
    ${\bf s}_m[t] = \{ {\bf h}_{\rm eff}({\bf q}_m[t],t), {\bf q}_m[t] \}$
    and determines
    $ {k_m[t]} = \mu_{\mathcal{D},\pi}({\bf s}_m[t])$
    by \eqref{eq:DPIC:util_behavior_policy}.
   }
    \State \multiline{ \textit{\textbf{Step 3. Feedback to IRS and IRS final configuration.}} The BS feeds back $m^\star[t]$ and $\{ k_m[t] \}_{m\in\mathcal{M}_{\rm DPIC}}$ to the IRS with $\lceil \log_2 M \rceil + M_{\rm DPIC} \lceil \log_2 K \rceil $ feedback bits. The IRS tunes its meta-atoms with ${\bf q}_\star[t] = {\bf q}_{m^\star[t]}[t]$ for data transmission.
    % , where $k_m[t] \in \{ 1, ..., K \}$. 
    }
    % \Statex \hspace{3mm}  $\triangleright$ \underline{Data transmission and IRS codebook update:}
    \State \multiline{ \textit{\textbf{Step 4. Data transmission and IRS codebook update.}} 
    The IRS updates  $\mathcal{C}[t+1] = \{ {\bf q}_m[t+1] \}_{m=1}^{M}$, 
    where the DPIC update is conducted by \eqref{eq:DPIC:update} for $m \in \mathcal{M}_{\rm DPIC}$, and the RA update by \eqref{eq:RA:update} for $m \in { \mathcal{M}}_{\rm RA}$.
    % where $M_{\rm DPIC}$ codewords are updated by the DPIC, i.e., ${\bf q}_m[t+1] = {\rm clip}({\bf q}_m[t] + {\bf d}_{k_m[t]}, [C_{\min}, C_{\max}])$, $m=1,...,M_{\rm DPIC}$, and  $M_{\rm RA}$ codewords are updated by the RA, i.e., ${\bf q}_m[t+1] = {\rm clip}({\bf c}[t] + {\boldsymbol u}_m[t], [C_{\min}, C_{\max}])$,  $m=M_{\rm DPIC}+1,...,M$.
    The BS calculates and stores $ \{ {\bf q}_m[t+1] \}_{m \in \mathcal{M}_{\rm DPIC}}$.
    }
     \EndFor
 \end{algorithmic}
 \end{algorithm}

While we refer to the aforementioned four cases as \textit{DPIC approaches}, we only refer to MDPIC, RA+SDPIC, and RA+MDPIC as \textit{augmented DPIC approaches} (SDPIC is excluded).
% , and all the four cases as the DPIC approach in a broad category.
The pseudo-code of the DPIC approach is given in Algorithm \ref{al:DPIC}. 

\vspace{-3mm}
% \subsection{Analysis of Time Complexity for Computations of the RA and DPIC Algorithms}
\subsection{{Computational Complexity and Group Control}}
\label{ssec:comp}

We analyze the computational/time complexity 
% \rev{(measured in terms of \textit{time})} 
of our approaches at the BS and IRS in each channel coherence block.
We first consider the RA approach (see Algorithm \ref{al:random}).
In line \ref{RA:line:step2},
the BS calculates the data rate with the measured effective channel 
% % by \eqref{eq:obj:rate} 
over total $M$ codewords with $\mathcal{O}(MN_{\rm BS})$ complexity.
% incurs the computational complexity with $\mathcal{O}(MN_{\rm BS})$.}
% once the IRS tunes its meta-atoms with ${\bf q}_m[t]$, 
% the BS measures the $m$-th effective channel and calculates the data rate $ R({\bf q}_m[t],t)$ sequentially for $m=1,...,M$,
% which incur the computational complexity for the data rate calculation over total $M$ codewords
% \rev{For these steps, the computational complexity at the BS for the data rate calculation over total $M$ codewords  is $\mathcal{O}(MN_{\rm BS})$.}
In line \ref{RA:line:step4}, the IRS updates the codebook with $\mathcal{O}(MN_{\rm IRS})$ complexity.
% by \eqref{eq:RA:update}.
% , which can be reasonably conducted within the data transmission time. 

We next consider the DPIC approach. The BS employs $M_{\rm A}$ agents each having four DNNs. For each DNN, we consider a fully connected neural network with two hidden layers, which have $L_1$ and $L_2$ neurons, respectively.
For the DNN policy and copied DNN policy, the sizes of the input and output layer are $2N_{\rm BS} + N_{\rm IRS}$ and $N_{\rm IRS}$, respectively.
For the DNN Q-function and copied DNN Q-function, the sizes of the input and output layer are $2N_{\rm BS} + N_{\rm IRS}$ and 1, respectively. The actions are included at the second hidden layer.
We first consider the training phase in Algorithm \ref{al:policy}.
In line \ref{DPIC:tr:line:step2}, the BS infers $k_m[t]$ with the agent $m$,
% immediately after it measures the $m$-th effective channel 
 $m\in \mathcal{M}_{\rm A}$.
The computational complexity for total inference with $M_{\rm A}$ agents is
thus $\mathcal{O}(M_{\rm A}((2N_{\rm BS} + N_{\rm IRS}) L_1 + L_1L_2 + L_2N_{\rm IRS} + K N_{\rm IRS}))$, which includes the quantization process per each inference with $\mathcal{O}(K N_{\rm IRS})$ complexity.
% Since the BS is assumed to have high computing power, each inference can be conducted within $T_{\rm reconf}$ together with the data rate calculation with  $\mathcal{O}(N_{\rm BS})$.
In line \ref{DPIC:tr:line:step4}, 
the complexity to train the $M_{\rm A}$ agents each with mini-batch size $ N_{\rm batch}$ is $\mathcal{O}(M_{\rm A} N_{\rm batch}((2N_{\rm BS} + N_{\rm IRS})L_1 + L_1 L_2 + L_2N_{\rm IRS}))$, which includes the updates for the DNN policy and DNN Q-function conducted via back-propagation and for the copied DNN policy and copied DNN Q-function conducted via soft target update.
% Since the data transmission time is not short, e.g., a few $ms$, and the BS has high computing power, we consider that the training can be conducted in each coherence time.
% If the BS cannot execute the training in each coherence time for some reason, the training in line \ref{DPIC:tr:line:step4} can be conducted intermittently, which only increases the total learning time without any modification of the algorithm. 
At the IRS, in line \ref{DPIC:tr:line:step4}, the codebook update has the complexity of  $\mathcal{O}(MN_{\rm IRS})$.
% , which can be handled within the data transmission.
%
The computational complexity at the BS and IRS during the utilization phase is the same as that of the training phase with
excluding the process of training the DNNs.

\textbf{From individual meta-atom control to group control.} Since the number of meta-atoms $N_{\rm IRS}$ is typically large, individual control for meta-atoms would incur high computational overhead at the BS and the IRS. To further reduce the computational overhead, we can consider a \textit{group control}~\cite{yang2020intelligent}, where IRS meta-atoms are partitioned into multiple groups and the same capacitance  is applied for the meta-atoms belonging to the same group. 
We denote the number of groups as $N_{\rm G}$, where $N_{\rm G} < N_{\rm IRS}$. 
We then focus on controlling $N_{\rm G}$ capacitance values to configure $N_{\rm IRS}$ meta-atoms over the meta-surface.
This implies that we can reduce the dimension of the design variables, i.e., capacitance vector and codeword, from $N_{\rm IRS}$ to $N_{\rm G}$.
% Specifically, the RA update in \eqref{eq:RA:update} and the DPIC update in \eqref{eq:DPIC:update} are changed. For the DPIC, the size of the learning architecture is further reduced, since the state and action space in \eqref{eq:state} and \eqref{eq:action} are reduced and the codeword dimension of the direction codebook is reduced.
Then, the computational complexity is reduced by replacing $N_{\rm IRS}$ with $N_{\rm G}$ in the complexity formula that we provided above. 
Furthermore, due to the group control, the DNN policy learning can be stabilized since the DNN policy learning has been successful when the action space size is not prohibitively large~\cite{lillicrap2015continuous}.
Due to these benefits, we incorporate the group control for our simulations in Sec.~\ref{sec:sim}.

%% file: eff.tex
\vspace{-3mm}
\subsection{Time Overhead and Effective Data Rate}
\label{ssec:effective_rate}
\vspace{-1.5mm}

% \q{Total $B$ bits. Think about how to describe it when we introduce $B$ bits constraint}

The implementation of our methods incurs (i) computation time, (ii) communication time, and (iii) IRS reconfiguration time overheads. 
For the computations carried out during the RA and the DPIC, it is reasonable to assume that 
the BS calculates the data rate over  $M$ codewords within the time duration for $M$ IRS reconfiguration by using its high computing power, and that
the IRS updates the codebook within the data transmission time.
% , which is not short, e.g., a few $ms$.
% during each IRS reconfiguration time $T_{\rm reconf}$ 
Also, in the DPIC, we assume that the BS with high computational capabilities can conduct the total inference within  $M$ IRS reconfiguration time
and
 the training in each coherence time.
We accordingly neglect the computation time overhead  and only focus on the communication time and IRS reconfiguration~time.

We define the {\it time overhead} $T_p$ as  the total time consumption except for data transmission
% as the {\it preparation} time $T_p$ 
shown in Fig. \ref{fig:timeline}, which is given by
% all the procedure for final IRS configuration, 
% given by
\vspace{-1.5mm}
\begin{equation}
    T_p = MT_{\rm reconf} + T_{\rm feedback} + T_{\rm final}.
    \label{eq:time_overhead}
    \vspace{-1.5mm}
\end{equation}
% where $T_p$ implies the total time consumption except for data transmission. %
First, $MT_{\rm reconf} $ denotes the total time for $M$ IRS reconfiguration used in both RA and DPIC approaches, where $T_{\rm reconf}$ is the time for  each IRS reconfiguration.
Second, $T_{\rm feedback} = B/R_{\rm feedback}$ is the time required for the feedback from the BS to the IRS, where $B$ is the number of feedback bits during one coherence time $T_c$ and $R_{\rm feedback}$ (bits/s) is the data rate for the feedback link. Note that $B < R_{\rm feedback}T_c$ since $T_{\rm feedback}$ should not exceed $T_c$. 
% $B$ is different for the RA and DPIC approach. 
For the RA approach, $B = \lceil \log_2 M \rceil$ for the feedback of $m^\star[t] \in \{ 1, ..., M \}$. For the DPIC approach, during the utilization time,
$B = \lceil \log_2 M \rceil + M_{\rm DPIC} \lceil \log_2 K \rceil$ for the feedback of $m^\star[t]$ and $\{k_m[t]\}_{m \in \mathcal{M}_{\rm DPIC}}$.
% ,
% where $k_m[t] \in \{1,...,K \}$, $\forall m\in \mathcal{M}_{\rm DPIC}$.
% and $K$ is the direction codebook size.
During the training period of DPIC, $B  = \lceil \log_2 M \rceil + M_{\rm A} \lceil \log_2 K \rceil$.
% can be obtained with a similar expression with replacing $M_{\rm DPIC}$ with $M_A$. 
%
% Second, $T_{\rm feedback}$ is the time required for the feedback from the BS to the IRS, which  is different for the RA and DPIC approach.
% For the RA approach, the feedback time is
% $T_{\rm feedback} \negmedspace = \negmedspace \frac{ \lceil \log_2 M \rceil} {W R_{\rm feedback}}$,
% where $\lceil \log_2 M \rceil$ is the number of bits required to feedback the index $m^\star[t] \in \{ 1, ..., M \}$, and
%  $R_{\rm feedback}$ (bits/s/Hz) is the unit data rate for the feedback link. 
% For the DPIC approach, during the utilization time, the feedback time is
% $T_{\rm feedback} = \frac{\lceil \log_2 M \rceil + M_{\rm DPIC} \lceil \log_2 K \rceil} {W R_{\rm feedback}}$,
% where $\lceil \log_2 M \rceil + M_{\rm DPIC} \lceil \log_2 K \rceil$ is the number of bits required to feedback $m^\star[t]$ and $\{k_m[t]\}_{m \in \mathcal{M}_{\rm DPIC}}$,
% where $k_m[t] \in \{1,...,K \}$, $\forall m\in \mathcal{M}_{\rm DPIC}$, and $K$ is the direction codebook size.
% During the training period of DPIC, $T_{\rm feedback}$ can be obtained with a similar expression with replacing $M_{\rm DPIC}$ with $M_A$. 
% For inference, we can set any $M_{\rm DPIC} \le M$, depending on the strategies described in Sec. \ref{ssec:utilization}.
%
Lastly, $T_{\rm final}$ denotes the execution time of the final IRS reconfiguration in \ref{step3} in Sec.\ref{ssec:protocol}.
% \begin{equation}
%     T_{\rm reset} =
%     \begin{cases}
%       T_{\rm reconf}, & \text{with probability} \;  (M-1)/M \\
%       0, & \text{with probability} \; 1/M
%     \end{cases}
% \end{equation}
If the selected index $m^\star[t]$ coincides with 
% indicates the capacitance of 
the last configuration in \ref{step1}, the IRS 
does not need to change the configuration, i.e., $T_{\rm final}=0 $; otherwise $T_{\rm final} = T_{\rm reconf}$.
% On average, the probability of this case is $1/M$. 
% Thus, we have $T_{\rm final} = 0$ with probability $1/M$, and $T_{\rm final} = T_{\rm reconf}$ with  $(M-1)/M$.

% After the preparation stage, the codeword ${\bf x}_{\max}[t] \in \mathcal{C}[t]$ will be used for IRS final configuration during the data transmission. 
To measure the average data rate during one coherence time $T_c$ under time-varying channels,
% Let $T_c$ denote the coherence time, and $T_c - T_p$ the actual data transmission time.
% Incorporating the time overhead $T_p$ into the data rate, w
we introduce a performance metric,
called {\it effective data rate}, according to
\vspace{-1mm}
\begin{equation}
    R_{\rm eff}[t] = \frac{T_c-T_p}{T_c} \log_2 
    \bigg(1+ \frac{P  \| {\bf h}_{\rm eff}( {\bf q}_\star[t], t) \|_2^2 }
    { \sigma^2 } \bigg) ,
    \label{eq:effrate}
    \vspace{-0.5mm}
\end{equation}
where $T_c - T_p$ is the actual data transmission time and ${\bf q}_\star[t] \in \mathcal{C}[t]$ is the selected codeword for the final IRS configuration.
The effective data rate captures the tradeoff between the data rate and the time overhead $T_p$.
As $M$ grows large, the data rate 
% of the system
may increase due to having larger number of reconfigurations. However, as $M$ increases, $T_p$ also increases, and thus $R_{\rm eff}[t]$ may decrease. In Sec. \ref{sec:sim}, we evaluate the data rate and effective data rate under different $M$.

%% file: sim.tex
\vspace{-2mm}
\section{Numerical Evaluation and Discussion}
\label{sec:sim}
\vspace{-1mm}

In this section, we describe the simulation setup in Sec.~\ref{ssec:sim:setup} and the channel model in Sec.~\ref{ssec:sim:channel}. We 
% introduce several evaluated methods including our proposed strategies and baselines in Sec. \ref{ssec:sim:method}, and 
conduct simulations
for two scenarios:
(i) existence of no LoS link between the UE and IRS in Sec. \ref{ssec:sim:indoor} and (ii) existence of an LoS link between them in Sec. \ref{ssec:sim:outdoor}. 
The former replicates a scenario with an indoor UE, while the later corresponds to an outdoor UE.
% \rev{We discuss a comprehensive strategy for the two scenarios in Sec.~\ref{ssec:sim:outdoor}.}
% the main takeaways of the simulations in Sec. \ref{ssec:sim:discussion}.
%Then, we present and discuss the results (Sec. IV-C).

\vspace{-4mm}
\subsection{Simulation Setup}
\label{ssec:sim:setup}
\vspace{-2mm}

\subsubsection{System parameters}
\label{sssec:symtem_parameters}

To emulate practical IRS reflection behavior, 
% we use the experimental results provided in~\cite{chen2020angle}. In particular,
we recover the phase shift $\angle \Gamma(C, \theta)$ and  attenuation $|\Gamma(C, \theta)|$ by the interpolation and extrapolation of the data in Fig.~4 and Table~1 of \cite{chen2020angle}.
Then, we obtain $\Gamma(C, \theta) = |\Gamma(C, \theta)| \exp(j \angle \Gamma(C, \theta))$ with the ranges of $C$ and $\theta$ as $(C_{\min} , C_{\max}) = (0.4 , 2.7)$ pF and $(0^o, 90^o)$, respectively.
We follow the same simulation setup as in~\cite{chen2020angle} to utilize the reflection coefficients: we set $f = 5.195$ GHz and consider only azimuth coordinates.
% where the elevation angles for incident signals are $0^o$.}
% \rev{Since we use the experimental results provided in \cite{chen2020angle} where only azimuth coordinates are considered, and we consider only azimuth coordinates where the elevation angles for incident signals are $0^o$ for simulations. The proposed signal model and methodologies can be readily applied to the case including the elevation angle.}
We consider $T_c=5$ ms, $N_{\rm BS} = 5$ and $N_{\rm IRS} = 200$, where the number of meta-atoms over the width and height of the IRS are  $N_{\rm IRS,w} = 50$ and $N_{\rm IRS,h} = 4$, respectively. 
We consider a group control with $N_{\rm G} = 10$, where $(N_{\rm IRS,w}/N_{\rm G}) \times  N_{\rm IRS,h}$ (i.e., $5 \times 4$) meta-atoms are controlled by each common capacitance.
The distance between adjacent BS antennas is $d_{\rm BS}=\lambda/2$, and the distance between adjacent IRS meta-atoms is $d_{\rm IRS} = \lambda/10$, where $\lambda = c/f$ is the wavelength and $c$
% =3 \times 10^8$ m/s 
is the speed of light.
The BS and IRS are assumed to have the same height and located at $ {\bf x}_{\rm BS} =  (0,0)$ m and ${\bf x}_{\rm IRS} = (90,30)$ m, respectively. 
The initial UE position ${\bf x}_{\rm UE}[0]$ is randomly generated within the circle with radius $r = 5$ m at the center point $(100,0)$ m, in each episode.
% The position of the UE is 
% % generated differently for episodes, and 
% varying over timesteps. 
%
The UE is moving with the velocity $v_{\rm UE} = 3$ km/h and constant azimuth angle $\eta$ over timesteps, i.e., ${\bf x}_{\rm UE}[t] = {\bf x}_{\rm UE}[t-1] + v_{\rm UE} T_c [\cos \eta, \sin \eta]^T $, where $\eta \sim \mathcal{U}(0,2\pi)$ is  generated in each episode.
% from uniform distribution on the interval $[a,b]$.
%
%\ref{step2}.
% We set $P = 20$ dBm and  $\sigma^2 = -80$ dBm. We consider the bandwidth $W=10$ MHz and \rev{$R_{\rm feedback}=10^6$ bits/s}.
We set $P = 20$ dBm, $\sigma^2 = -80$ dBm, and $R_{\rm feedback}=10^6$ bits/s.
To have more realistic results, we consider a noisy version of the effective channels in \eqref{eq:signal_model}.
%  we consider that the BS measures a noisy version of the effective channels in \eqref{eq:signal_model} (i.e., the received pilot signal that contains noise).
Also, we set $T_{\rm reconf} = 100 \mu s$ \cite{abadal2020programmable}, unless otherwise stated (we conduct simulations with different $T_{\rm reconf}$ in Figs. \ref{fig:NLoS}(\subref{fig:NLoS:effrate_Tr})\&\ref{fig:LoS}(\subref{fig:LoS:effrate_Tr})).\footnote{The reconfiguration time of IRS is determined by the characteristics of the control board and the internal communication between the control board and the meta-surface. The reconfiguration speed is typically known to be a few kHz \cite{abadal2020programmable}.} 

\subsubsection{Parameters for the proposed algorithms}
For the RA algorithm, we set $\delta = (C_{\max}-C_{\min})/5$.
For the DPIC algorithm, we set $\gamma = 0.9$,  $N_{\rm batch} = 32$, and  $|\mathcal{B}_m| = 5\times10^{5}$, $m \in \mathcal{M}_{\rm A}$.
% The rectified linear unit (ReLU) activation function is employed for the two hidden layers of the DNNs.
We consider $L_1=400$ and $L_2=300$ for the DNNs with ReLU activation function. 
% Note that the structure of DNNs are demonstrated in Sec. \ref{ssec:comp}.
% The number of neurons in the two hidden layers is 400 and 300, respectively.
% the DNN policy (actor) $\pi({\bf s}; {\bf w}_{\pi,m})$ and the DNN Q-function (critic) $Q({\bf s},{\bf a}; {\bf w}_{Q,m})$, $m=1,...,M_{\rm A}$, are composed of an input layer, two fully-connected hidden layers, and an output layer.
%
% The weights are initialized with $\mathcal{U}(-0.01,0.01)$, and the bias as $0$ for the DNN policy and $0.1$ for the DNN Q-function.
%
% The small initial weights/bias are intended to ensure the initial outputs for the DNN policy and DNN Q-function near zero.
We employ Adam optimizer for training.
% the neural network parameters 
We consider $\alpha_\pi = 3\times 10^{-4}$, $\alpha_Q = 3\times 10^{-3}$, and $\tau =0.005$.
For the DNN policy, the input and output size is $2N_{\rm BS}+ N_{\rm G}  = 20$ and $N_{\rm G} = 10$, respectively. In the output layer, the tanh function is employed, and the output is scaled by $\delta = (C_{\max} - C_{\min})/4$ to bound the actions. 
We set $|\mathcal{D}|=K=2048$, where each codeword in $\mathcal{D}$ is constructed by RVQ ranging within $[-\delta, \delta]^{N_{\rm G}}$.
For the DNN Q-function, the input and output sizes are $20$ and $1$, respectively.
% $d_s = 2N_{\rm BS} + N_{\rm IRS,ctr} = 20$
% , and actions are
% included at the second hidden layer.
% For action exploration during the training period, w
% We adopt a decreasing exploration noise $\epsilon_e$ given in Algorithm \ref{al:policy} 
% with
% a decreasing noise variance over the episodes, 
% = \max \{ \epsilon_{\min}, 0.99 \epsilon(e-1) \} $, where $e$ is the episode number, 
% initialization
% $\epsilon_0 = (C_{\max} - C_{\min})/5 $ and $ \epsilon_{\min} = \epsilon_0/300$.
% is the minimum variance.
% We set $|\mathcal{D}|=K=2048$. 
%The state and action space size are $2N_{\rm BS} + N_{\rm IRS,control} = 20$ and $N_{\rm IRS,control} = 10$, respectively.
%
We normalize the values for the state and action to match with the scale of the values, such that $ {\bf h}_{\rm eff}(\cdot) \leftarrow \sqrt{{P}/({\sigma^2 N_{\rm BS} N_{\rm G} }) } \times {\bf h}_{\rm eff}(\cdot) $ and ${\bf q}_m \leftarrow 10^{12} \times {\bf q}_m $ in \eqref{eq:state}, and ${\bf a}_m \leftarrow 10^{13} \times {\bf a}_m $ in \eqref{eq:action}.
% We normalize the values for the state, action, and reward to match with the scale of the values, such that $ {\bf h}_{\rm eff}(\cdot) \leftarrow \sqrt{{P}/({\sigma^2 N_{\rm BS} N_{\rm G} }) } \times {\bf h}_{\rm eff}(\cdot) $ and ${\bf q}_m \leftarrow 10^{12} \times {\bf q}_m $ in \eqref{eq:state}, ${\bf a}_m \leftarrow 10^{13} \times {\bf a}_m $ in \eqref{eq:action}, and $R(\cdot) \leftarrow R(\cdot)/W$ in \eqref{eq:DPIC:reward}.

\vspace{-2mm}
\subsection{Models for Channels  and Their Variations}
\label{ssec:sim:channel}

We adopt a multi-path geometric channel model  \cite{tse2005fundamentals} for the IRS-BS, UE-BS, and UE-IRS channels. In this model, a \textit{vector channel} ${\bf h}[t]$
is constructed as the sum of the signals over multiple paths as ${\bf h}[t] = \sum_{\ell} {\bf h}_\ell[t]$ where the $\ell$-th path channel ${\bf h}_\ell[t]$ is constructed with the path gain $g_\ell[t]$ and the angle (angle of arrival (AoA) or angle of departure (AoD)) $\theta_\ell[t]$ as
${\bf h}_\ell[t] = g_\ell[t] {\rm ARV}(\theta_\ell[t])$ with the array response vector ${\rm ARV}(\theta_\ell[t])$. 
% $\theta_\ell[t]$ denotes 
% the angle of direction (AoD) in the case of multiple antennas at the transmitter and single antenna at the receiver, and 
% angle of arrival (AoA) where the transmitter has single antenna and the receiver has multiple antennas.
%
Subsequently, a \textit{matrix channel} is constructed similarly with path gains, AoAs and AoDs of multiple paths.
To model channel variations, we consider that $g_\ell[t]$ evolves over time according to a first-order Gauss-Markov process \cite{sklar2001digital}, and $\theta_\ell[t]$ varies via random perturbation addition given by
\vspace{-1.5mm}
\begin{equation}
    g_\ell[t] = \rho g_\ell[t-1] +  \sqrt{1- {\rho}^2} {\nu}_\ell[t], \quad 
    \theta_\ell[t] = \theta_\ell[t-1] + \Delta\theta_\ell[t].
    \label{eq:evolution}
    \vspace{-1.5mm}
\end{equation}
In~\eqref{eq:evolution}, the time correlation coefficient $\rho$ obeys the Jakes model \cite{sklar2001digital}, i.e., $\rho = J_0 ( 2 \pi f_d T_c)$, where $J_0(\cdot)$ is the zeroth order Bessel function of the first kind and $f_d = v f/c$ is the maximum Doppler frequency, with the  velocity $v$ of the UE or scatterer.
For simulations, we set $v =3$ km/h and accordingly obtain $\rho=0.95$. Also, $\Delta\theta_\ell[t] \sim \mathcal{U}(-0.1^o,0.1^o)$ and we set ${\nu}_\ell[t] \sim \mathcal{CN}(0,\beta[t])$ and ${g}_\ell[0] \sim \mathcal{CN}(0,\beta[0])$ with $\beta[t]$ denoting the instantaneous large-scale fading factor, which is defined with Euclidean distance $d[t]$ between two network elements as
\vspace{-1.5mm}
\begin{equation}\label{eq:beta_Large}
    \beta[t] = \beta_0 - 10 \alpha \log_{10}(d[t]/d_0) ~(\textrm{dB}),
    \vspace{-1.5mm}
\end{equation}
where $\beta_0 = -30$ dB is the path loss at distance $d_0=1$ m, and $\alpha$ is the path loss exponent.

\subsubsection{IRS-BS channel}
Since the IRS is deployed to have an LoS path to the BS~\cite{wu2019towards,liaskos2018new},
we model the IRS-BS channel with the Rician channel~\cite{tse2005fundamentals} as
${\bf H}^{\rm IB}[t] =
\Big[
\sqrt{\frac{K^{\rm IB}}{1+K^{\rm IB}}} {\bf H}^{\rm IB}_0$ 
$+
\sqrt{ \frac{1}{1+K^{\rm IB}} } 
\sum_{\ell=1}^{L^{\rm IB}}
{\bf H}^{\rm IB}_{\ell}[t] \Big] $
where $K^{\rm IB} = 5$.
The LoS channel ${\bf H}^{\rm IB}_0$ is characterized by the AoA and AoD, set to $0^o$ and $-60^o$, respectively.
The number of non-LoS (NLoS) paths is $L^{\rm IB} =10$.
% For each NLoS channel ${\bf H}^{\rm IB}_{\ell}[t]$, 
The variations of the path gain, AoA, and AoD for each NLoS channel ${\bf H}^{\rm IB}_{\ell}[t]$ are modeled by \eqref{eq:evolution}, where the initial AoA and AoD are generated from $\mathcal{U}(-90^o,90^o)$, and
the large-scale fading coefficient is modeled by \eqref{eq:beta_Large} using the fixed distance between the IRS and the BS, and~$\alpha= 2$.

\subsubsection{UE-BS channel}
Assuming that there exists a blockage between the UE and the BS, we model the UE-BS channel with only NLoS signals as
${\bf h}^{\rm UB}[t] =  \sum_{\ell=1}^{L^{\rm UB}} {\bf h}^{\rm UB}_{\ell}[t]$, where $L^{\rm UB} = 10$. 
The variations of the path gain and AoA for each channel ${\bf h}^{\rm UB}_{\ell}[t]$ are modeled by \eqref{eq:evolution},
where the initial AoA is generated from $\mathcal{U}(-90^o,90^o)$, and
the large-scale fading factor is modeled by the varying distance between the UE and the BS and $\alpha = 3.75$.

\subsubsection{UE-IRS channel}
We consider two different channel scenarios. 
In the first scenario, there exists no LoS link between the UE and the IRS, for which
we model the UE-IRS channel with only NLoS signals as
${\bf h}^{\rm UI}[t] =  \sum_{\ell=1}^{L} {\bf h}^{\rm UI}_{ \ell}[t]$ with $L = 10$.
The variations of path gain and AoA for each channel ${\bf h}^{\rm UI}_{\ell}[t]$ are modeled by \eqref{eq:evolution}, where the initial AoA is generated from $\mathcal{U}(0^o,90^o)$, and
the large-scale fading factor is modeled by the varying distance between the UE and the IRS and $\alpha = 2.2$.
In the second scenario, there exists an LoS between the UE and IRS, for which
we model the UE-IRS channel with the Rician channel  as
${\bf h}^{\rm UI}[t] = 
\Big[
\sqrt{\frac{K^{\rm UI}}{1+K^{\rm UI}}} {\bf h}^{\rm UI}_0[t]$ 
$+ 
\sqrt{ \frac{1}{1+K^{\rm UI}} } \sum_{\ell=1}^{L} {\bf h}^{\rm UI}_{\ell}[t]
\Big]$,
where $K^{\rm UI}=1$ and $L = 10$.
The LoS channel ${\bf h}^{\rm UI}_0[t]$ is varying 
according to the UE movement described in Sec.\ref{sssec:symtem_parameters} modeled 
via the changing AoA and distance between the UE and the IRS. 
% obtained 
% according to the UE movements described in Sec.\ref{sssec:symtem_parameters}. 
% because the AoA of the LoS signal is varying.
The NLoS channel ${\bf h}^{\rm UI}_{\ell}[t]$ is modeled as in the first scenario.

\vspace{-3mm}
%%%%%%%%%%%%%%%%%%%%%%%%%%%%%
\subsection{Scenario 1. No LoS Link between the UE and the IRS}
\label{ssec:sim:indoor}

Scenario 1
represents an {\it indoor UE}, which does not have a LoS link to the IRS. 
We first evaluate the performance of the proposed algorithms in the utilization phase with $2000$ episodes, where each episode contains $30$ timesteps (coherence blocks).
Each episode has different realizations of the UE-IRS channel, UE-BS channel, IRS-BS channel, UE initial location, and UE moving direction.
Our baseline method is the RVQ codebook
% , where there are $M$ codewords in the codebook. 
% As a baseline, we adopt the random RVQ codebook 
described in Sec. \ref{ssec:naive}, denoted by ``RVQ" in the figures. The specific configuration of the proposed schemes is
$M_{\rm A}=1$ for SDPIC,
$M_{\rm A}=8$ for MDPIC,
$M_{\rm A}=1$, $M_{\rm DPIC}=1$, and $M_{\rm RA} = M-1$ for RA+SDPIC, and
$M_{\rm A}=4$, $M_{\rm DPIC}= \min\{ M, M_{\rm A} \} $, and $M_{\rm RA} = M - M_{\rm DPIC}$ for RA+MDPIC.

%%%%%%%%%%%%%%%%%%%%
\begin{figure*}[t]
\vspace{-2mm}
\centering
\begin{subfigure}{.34\linewidth}
  \centering
  \includegraphics[width=\linewidth]{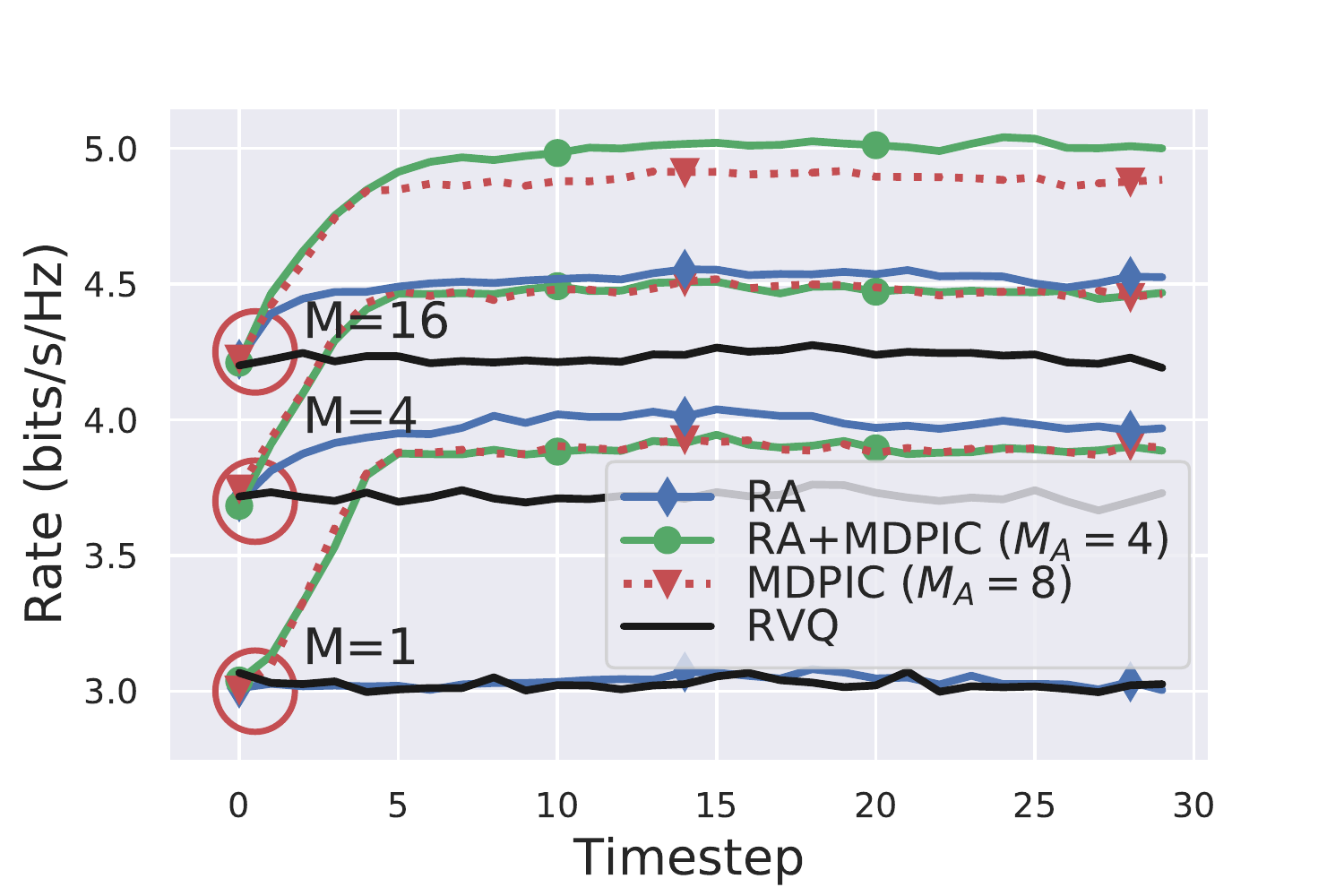}
  \vspace{-10mm}
  \caption{ Data rate along timesteps  
  }
  \label{fig:NLoS:rate_time}
\end{subfigure}
\begin{subfigure}{.34\linewidth}
  \centering
  \includegraphics[width=\linewidth]{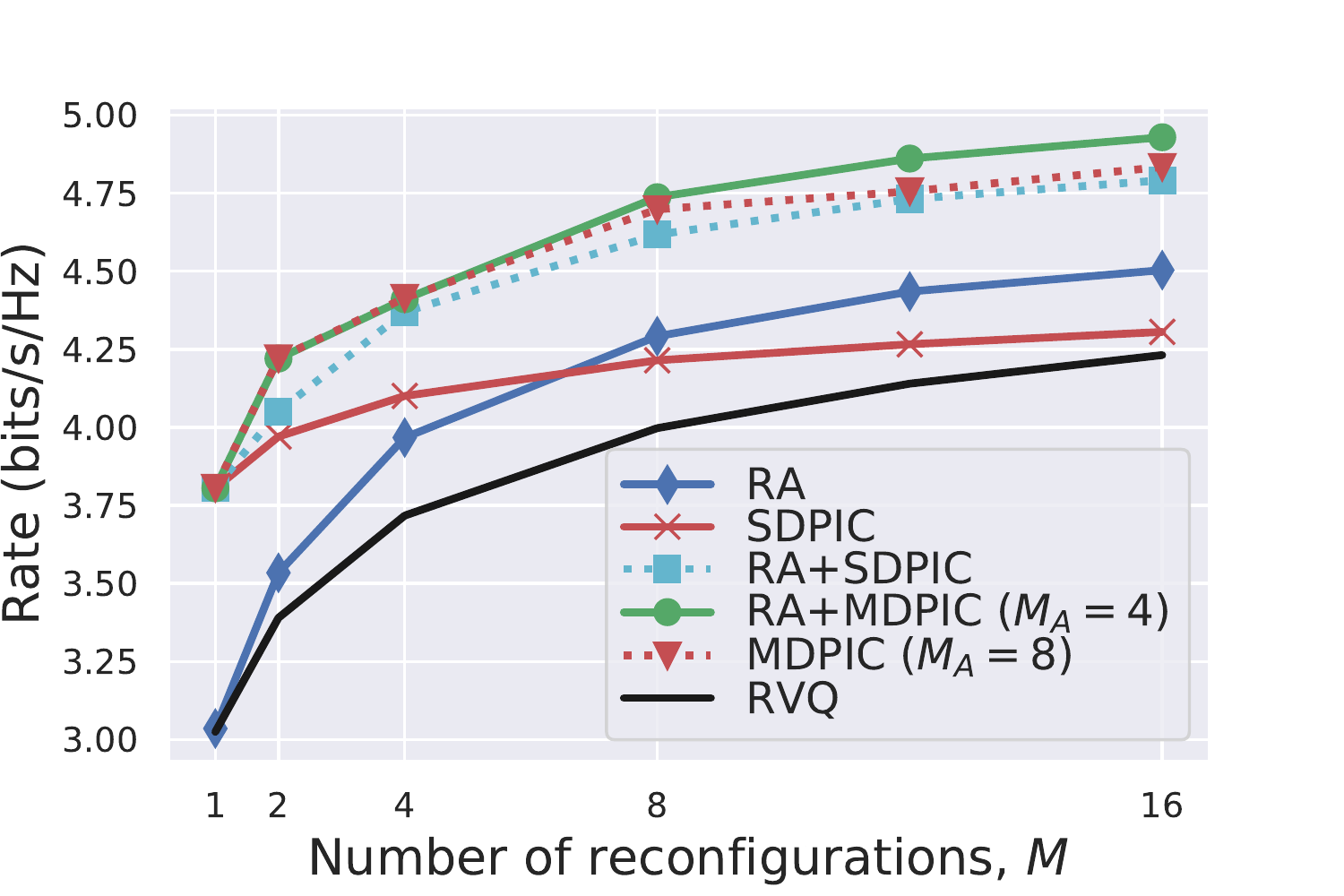}
  \vspace{-10mm}
  \caption{ Data rate along $M$  
  }
  \label{fig:NLoS:rate_M}
\end{subfigure}
\begin{subfigure}{.32\linewidth}
  \centering
  \includegraphics[width=\linewidth]{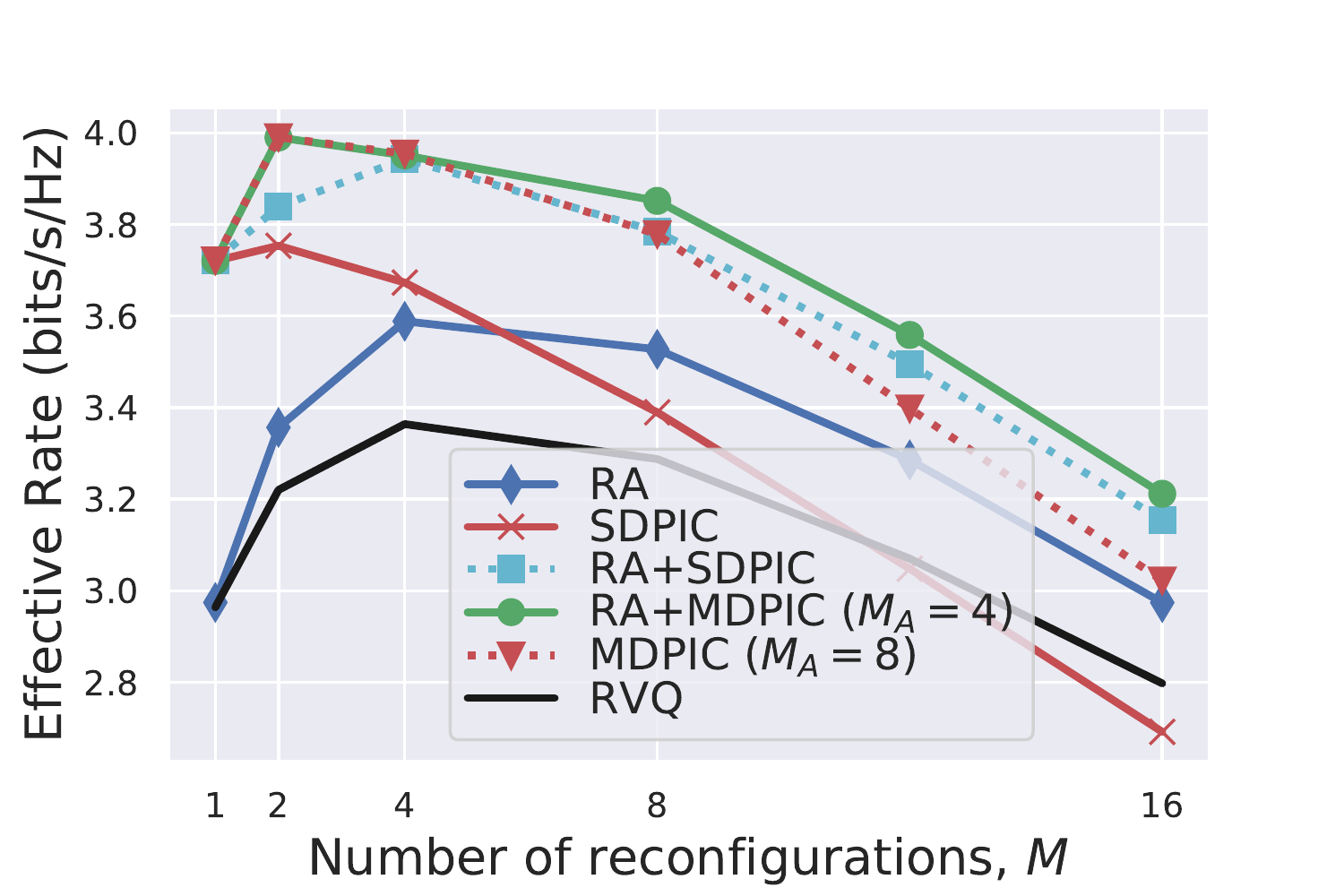}
  \vspace{-10mm}
  \caption{ Effective data rate along $M$ 
  }
  \label{fig:NLoS:effrate_M}
\end{subfigure}
\begin{subfigure}{.333\linewidth}
  \centering
  \includegraphics[width=\linewidth]{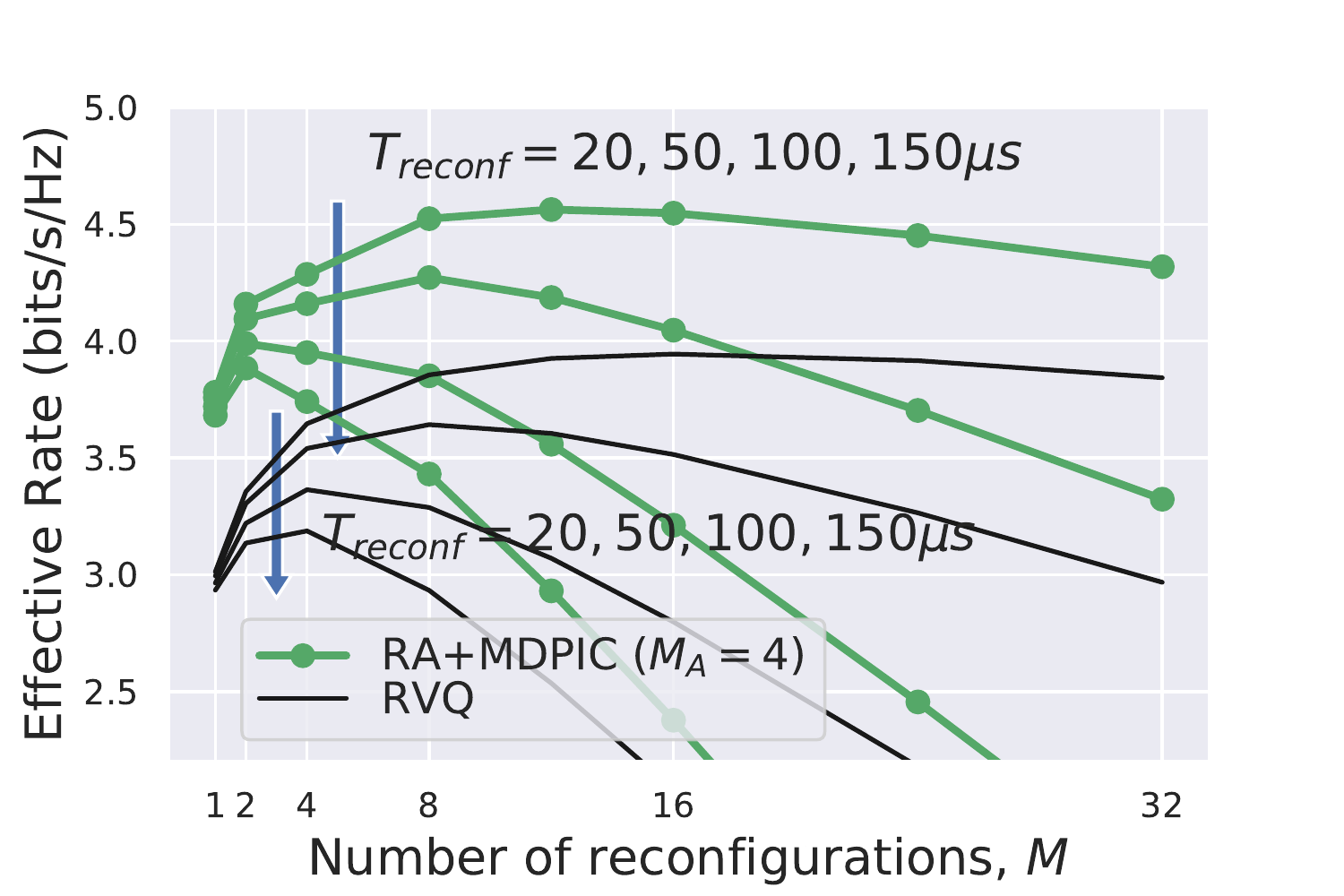}
  \vspace{-10mm}
  \caption{ Effective data rate with different $T_{\rm reconf}$ 
  }
  \label{fig:NLoS:effrate_Tr}
\end{subfigure}
\begin{subfigure}{.32\linewidth}
  \centering
  \includegraphics[width=\linewidth]{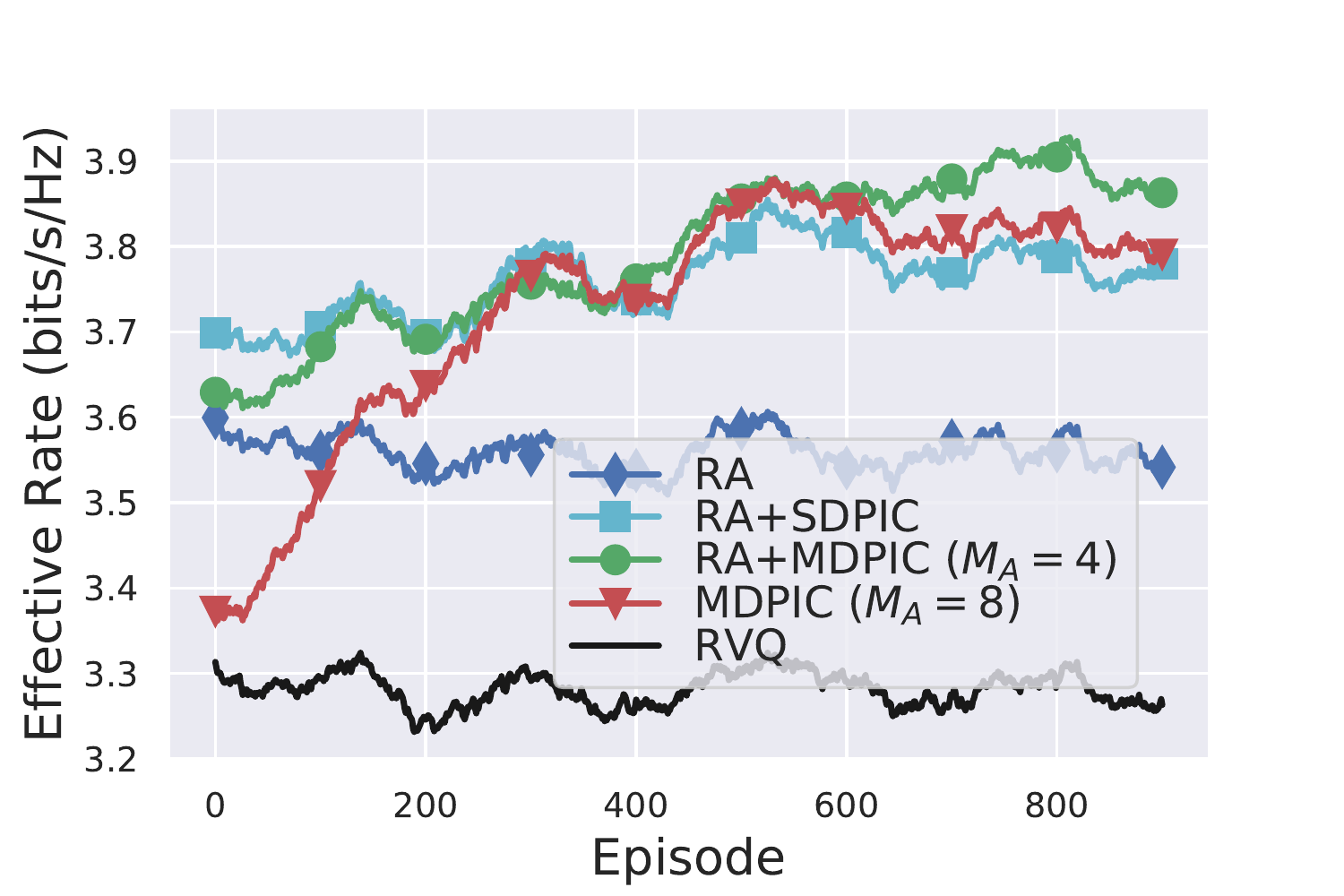}
  \vspace{-10mm}
  \caption{ Effective data rate along  episodes 
}
  \label{fig:NLoS:effrate_train}
\end{subfigure}
\vspace{-3mm}
\caption{
Performance evaluation of our methodology in Scenario 1. The plots in (\subref{fig:NLoS:rate_time})-(\subref{fig:NLoS:effrate_Tr}) correspond to the utilization phase, while the plot in (\subref{fig:NLoS:effrate_train}) describes the training phase.
% In (a), the proposed algorithms obtain high data rate within 4-5 timesteps. In (b), The data rate increases as $M$ increases. In (c), however, the effective data rate peaks around $M^\star=4$ because the time overhead is increasing over $M$. 
%From (b) and (c), the hybrid approaches (RA+SDPIC and RA+MDPIC) yield the better performances than others.
% In (d), the peak point varies for the IRS reconfiguration time $T_{\rm reconf}$.
%% because the increased time overhead outweighs the data rate improvement when $M>4$. 
% In (e), the hybrid approaches not only supplement the data rate at the beginning of the training, but also yield higher effective data rate than the others.
}
\label{fig:NLoS}
\vspace{-6mm}
\end{figure*}
%%%%%%%%%%%%%%%%%%%%

%
% We first consider the execution of our methods in the utilization period. 
Fig. \ref{fig:NLoS}(\subref{fig:NLoS:rate_time}) shows the average data rate along the timesteps over $2000$ episodes. 
The proposed schemes -- RA, RA+MDPIC, and MDPIC --   update the codebook adaptively at every timestep $t$ by using current observations (i.e., previously used codewords, effective channels, and data rates) to improve the data rate for the next timestep $t+1$.
The performances of the proposed schemes are improved over time and converge only within $4$-$5$ timesteps.
% in the utilization period.
%Only 4-5 timesteps are needed to obtain high data rate.
%The performance of MDPIC is bettern than Random and RA, because the MDPIC can adaptively control capacitances by predicting the direction vector for capacitances update from previous observations.
%
Overall, as the number of IRS reconfiguration $M$ increases, a higher data rate is achieved.
% , because the IRS is more probable to find better capacitance configuration.
Interestingly, RA+MDPIC yields better data rate compared to that of the MDPIC and RA
% This implies that 
because the multiple trained agents give good update directions, and the RA further improves the performance via random exploration for diverse update directions.
% around the good directions.
%
%
Fig. \ref{fig:NLoS}(\subref{fig:NLoS:rate_M}) shows the data rate along $M$, where each data point is averaged over 2000 episodes and 30 timesteps.
The MDPIC yields better data rate than the SDPIC, due to the advantage of using multiple agents.
Among the methods, the RA+MDPIC yields the best performance as the same in Fig.~\ref{fig:NLoS}(\subref{fig:NLoS:rate_time}).
% The MDPIC yields better data rate than the SDPIC, due to the advantage of using multiple agents.
% The augmented schemes, RA+SDPIC, RA+MDPIC, and MDPIC, reveal the superior performance as compared to the RA and SDPIC.
%

Fig. \ref{fig:NLoS}(\subref{fig:NLoS:effrate_M}) shows the  effective data rate
along $M$.
The effective data rate in \eqref{eq:effrate} captures the tradeoff between the data rate and the time overhead: as $M$ gets large, the data rate 
may increase due to having larger number of reconfigurations; however, at the same time, larger $M$ increases the total time overhead.
% \rev{As $M$ grows, the data rate is increasing due to the increased likelihood to select a good final codeword among $M$ codewords (see \ref{step2}). However, larger $M$ increases the total time overhead in \eqref{eq:time_overhead} due to more IRS reconfiguration required. Overall, the effective data rate in \eqref{eq:effrate} is determined by the data rate and the total time overhead.}
The RA+MDPIC shows the best performance in effective data rate for any $M$.
% than the MDPIC because the time overhead for the feedback in the MDPIC is higher.
We obtain the highest effective data rate when $M^\star=2$ or $M^\star=4$ depending on the method.
% As $M(>8)$ increases
As $M$ grows larger than $2$ or $4$, the increased time overhead outweighs the improvement of the data rate, leading to the decrease of the effective data rate. 
This finding agrees with recent results from \cite{zappone2020overhead}, where the performance of the overhead-aware metric is degraded as the overhead for the channel sounding and feedback increases.
%
%
%...
%
%
Fig. \ref{fig:NLoS}(\subref{fig:NLoS:effrate_Tr}) shows the effective data rate of the RA+MDPIC along $M$ with different $T_{\rm reconf}$. 
% Note that $T_{\rm reconf}$ is determined by the hardware characteristics of the control circuit and the internal communication. 
For $T_{\rm reconf}=20, 50, 100, 150 \mu s$, the best $M$ yielding the highest effective data rate is $M^{\star}=12,8,2,2$, respectively. 
%This shows that the time overhead highly depends on the IRS reconfiguration time from \eqref{eq:time_overhead}. 
% As can be seen, f
For larger
$T_{\rm reconf}$, smaller $M$ is preferred since a large $T_{\rm reconf}$ implies a large time overhead for each IRS reconfiguration.
% more IRS reconfiguration increases the overhead largely. 
For smaller $T_{\rm reconf}$, larger $M$ is preferred since more IRS reconfiguration increases the data rate.
Although finding optimal $M^\star$ in advance is challenging due to the difficulty of the analysis on agents' inferences, we could set a proper range of $M$ empirically from the value of $T_{\rm reconf}$. 
% The utilization of the DPIC algorithm is further discussed in Sec. \ref{ssec:sim:discussion}.

%\begin{figure}[t]
%  \includegraphics[width=.9\linewidth]{figures/sim/NLoS/Training_effrate_episode.pdf}
%  \centering
%  \caption{The effective data rate along the episodes in the training period of Scenario 1 where $M=8$. Each data point in the plots is a moving average over the previous 100 episodes.
%  }
%  \label{fig:NLoS:train}
%\end{figure}  

We next focus on the training phase with $1000$ episodes each containing $500$ timesteps and $M \negmedspace = \negmedspace 8$.
% during the training period. 
Fig. \ref{fig:NLoS}(\subref{fig:NLoS:effrate_train}) shows the effective data rate averaged over $500$ timesteps for each episode. 
Each data point is a moving average over the previous $100$ episodes.
For the training, the BS determines the number of agents being trained, $M_{\rm A}$. If $M_{\rm A} \negmedspace=\negmedspace M \negmedspace =\negmedspace 8$, all codewords are dedicated to training the agents, denoted by MDPIC ($M_{\rm A}\negmedspace=\negmedspace8$).
At the beginning of the training, the performance of the MDPIC is similar to the RVQ but is improved over time.
If $M_{\rm A} \negmedspace < \negmedspace M$, $M_{\rm A}$ codewords are dedicated to training $M_{\rm A}$ agents while $M-M_{\rm A}$ codewords are  updated by the RA update, which are the cases of RA+SDPIC
and RA+MDPIC ($M_{\rm A} \negmedspace = \negmedspace 4$). These \textit{hybrid} approaches use random exploration that supplements the low initial data rate of the MDPIC. 
% \rev{Since the channel models in Sec. \ref{ssec:sim:channel} are in general unknown to the BS, the training at the BS is
% conducted via \textit{actual} interactions with the IRS under the unknown channels over time.
% To this end, there should exist a UE or multiple UEs (each UE in a disjoint time period) for training, called \textit{training} UEs.}
% If the BS \rev{is concerned about the effective data rate performance of the training UEs}, it is encouraged to adopt one of the hybrid approaches to provide high effective data rate during training. 
The performance of the RA+MDPIC is better than that of the MDPIC despite of using less agents, even after the completion of the training, which agrees with the result in the utilization phase in Fig. \ref{fig:NLoS}(\subref{fig:NLoS:effrate_M}).

\vspace{-3mm}
%%%%%%%%%%%%%%%%%%%%%%%%%%%%%
\subsection{Scenario 2. LoS Link between the UE and the IRS}
\label{ssec:sim:outdoor}

%%%%%%%%%%%%%%%%%%%%
\begin{figure*}[t]
\vspace{-2mm}
\centering
\begin{subfigure}{.34\linewidth}
  \centering
  \includegraphics[width=\linewidth]{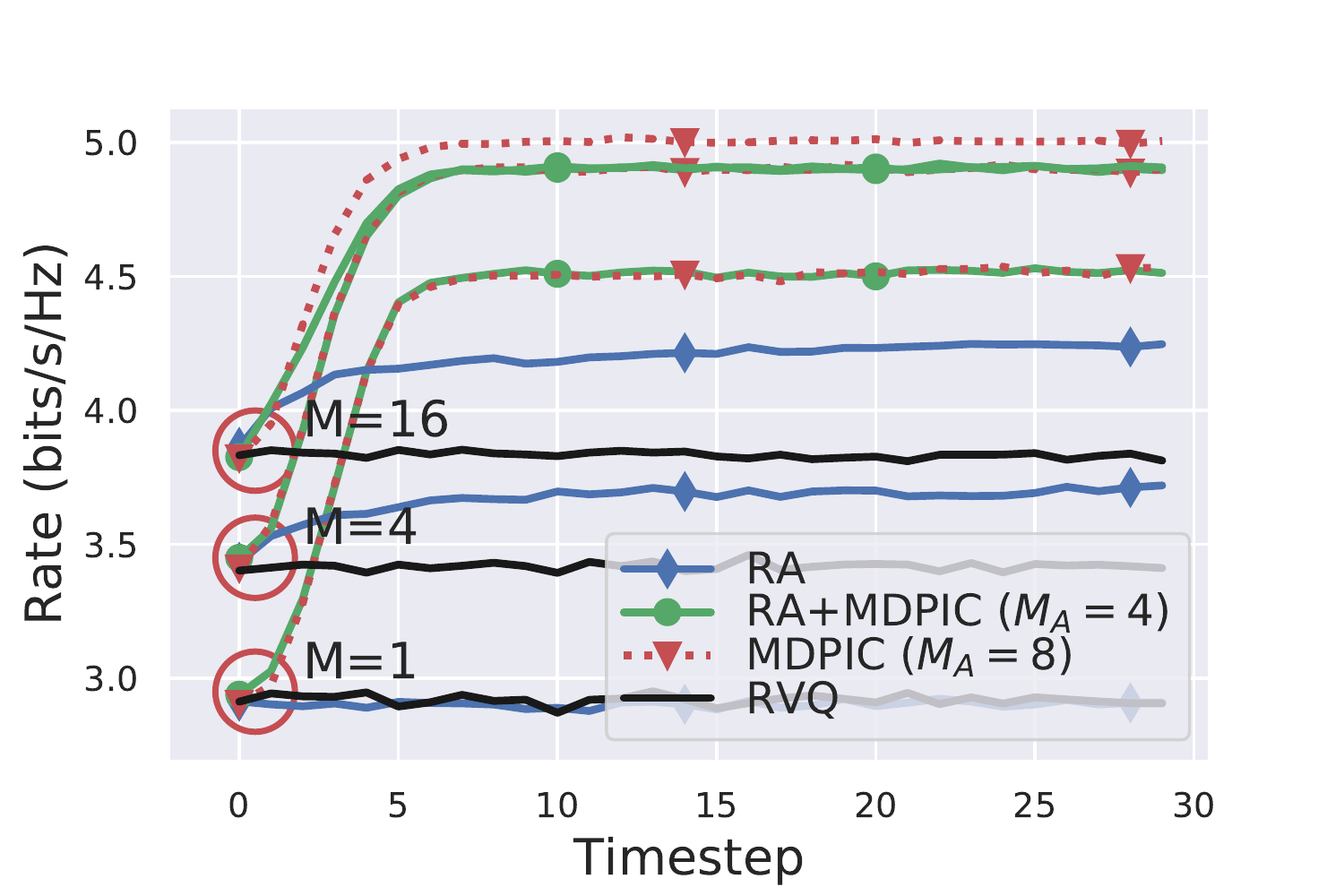}
  \vspace{-10mm}
  \caption{ Data rate along timesteps
  }
  \label{fig:LoS:rate_time}
%   \vspace{-2mm}
\end{subfigure}
\begin{subfigure}{.34\linewidth}
  \centering
  \includegraphics[width=\linewidth]{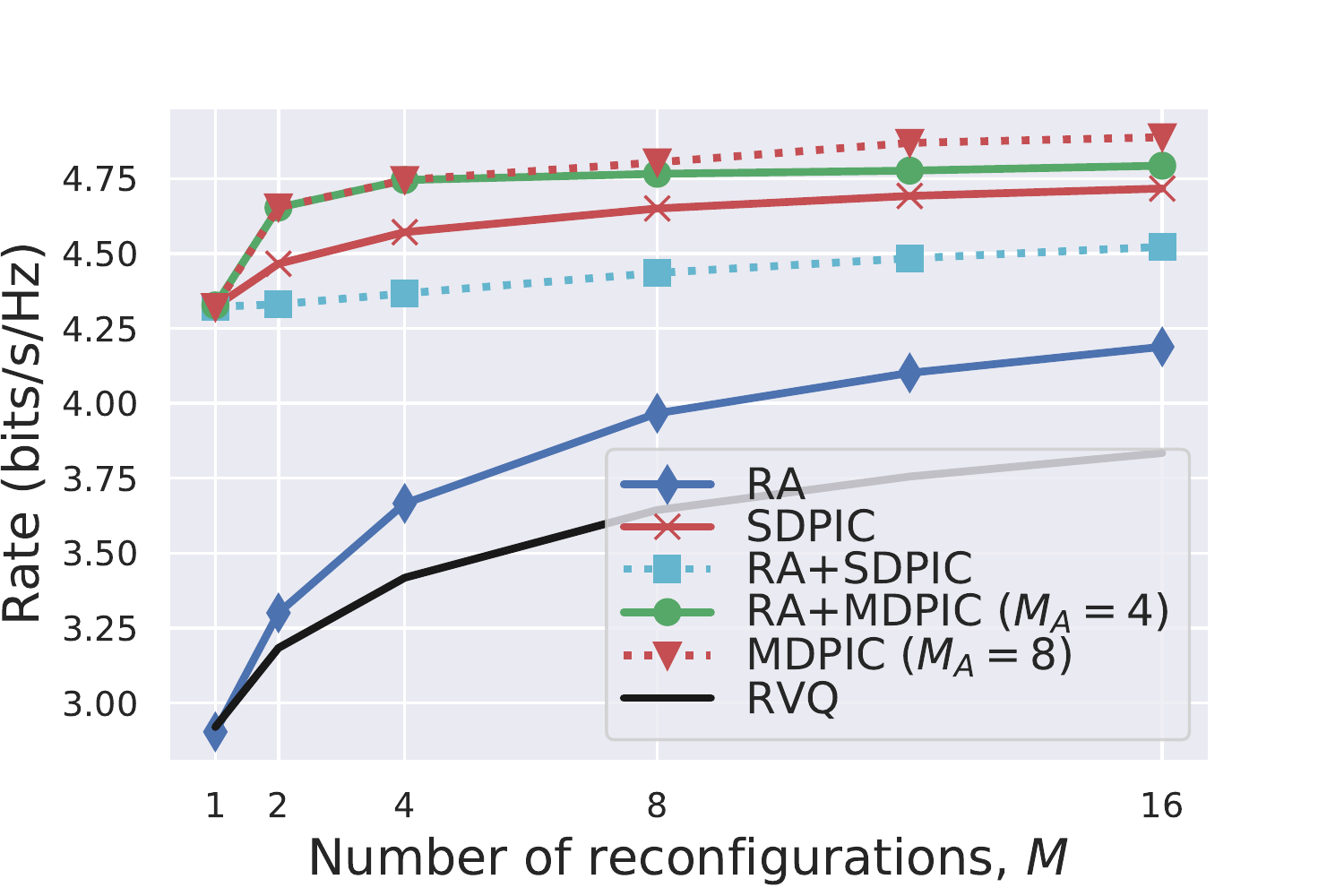}
  \vspace{-10mm}
  \caption{ Data rate along $M$
  }
  \label{fig:LoS:rate_M}
%   \vspace{-2mm}
\end{subfigure}
\begin{subfigure}{.32\linewidth}
  \centering
  \includegraphics[width=\linewidth]{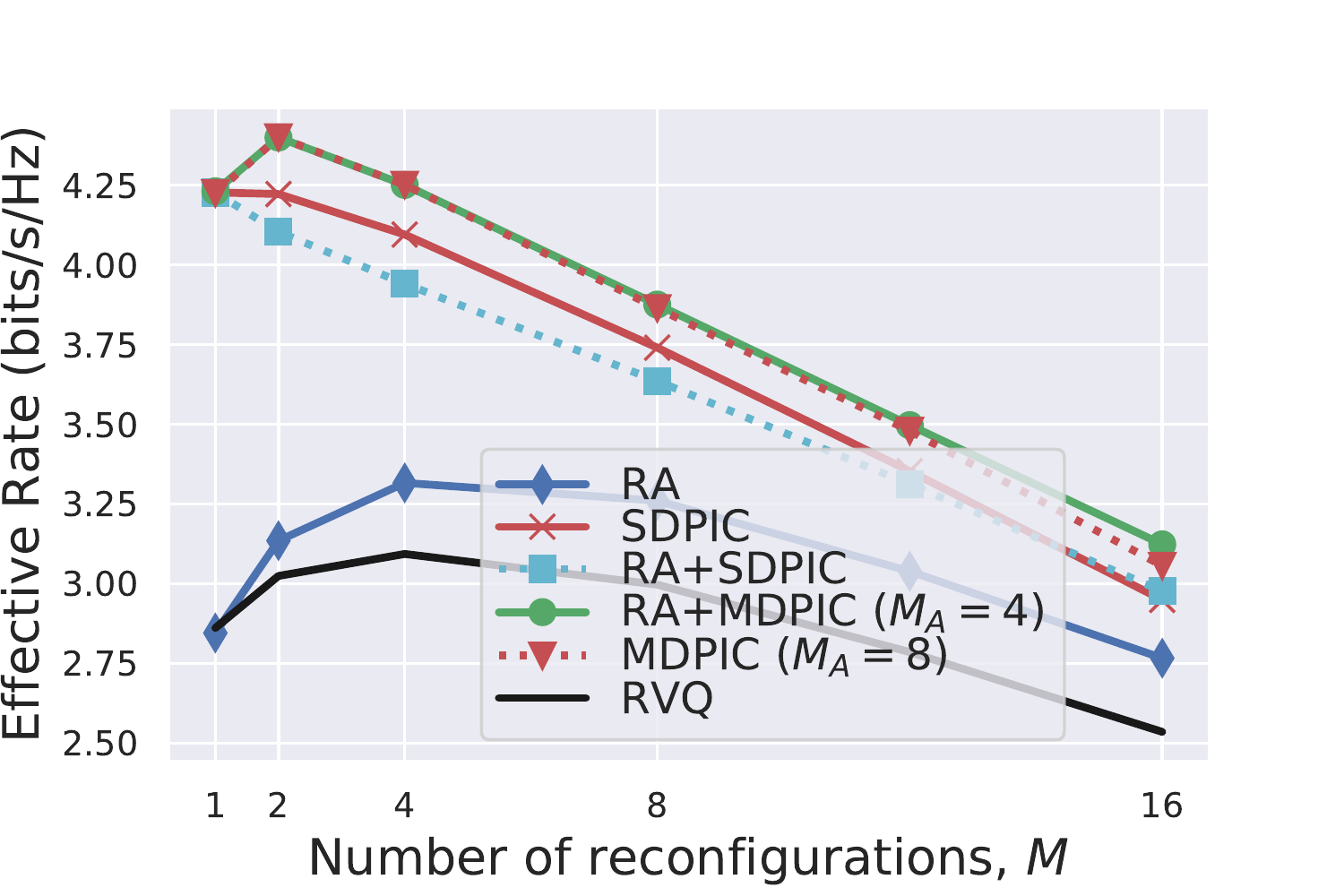}
  \vspace{-10mm}
  \caption{ Effective data rate along $M$
  }
  \label{fig:LoS:effrate_M}
\end{subfigure}
\begin{subfigure}{.333\linewidth}
  \centering
  \includegraphics[width=\linewidth]{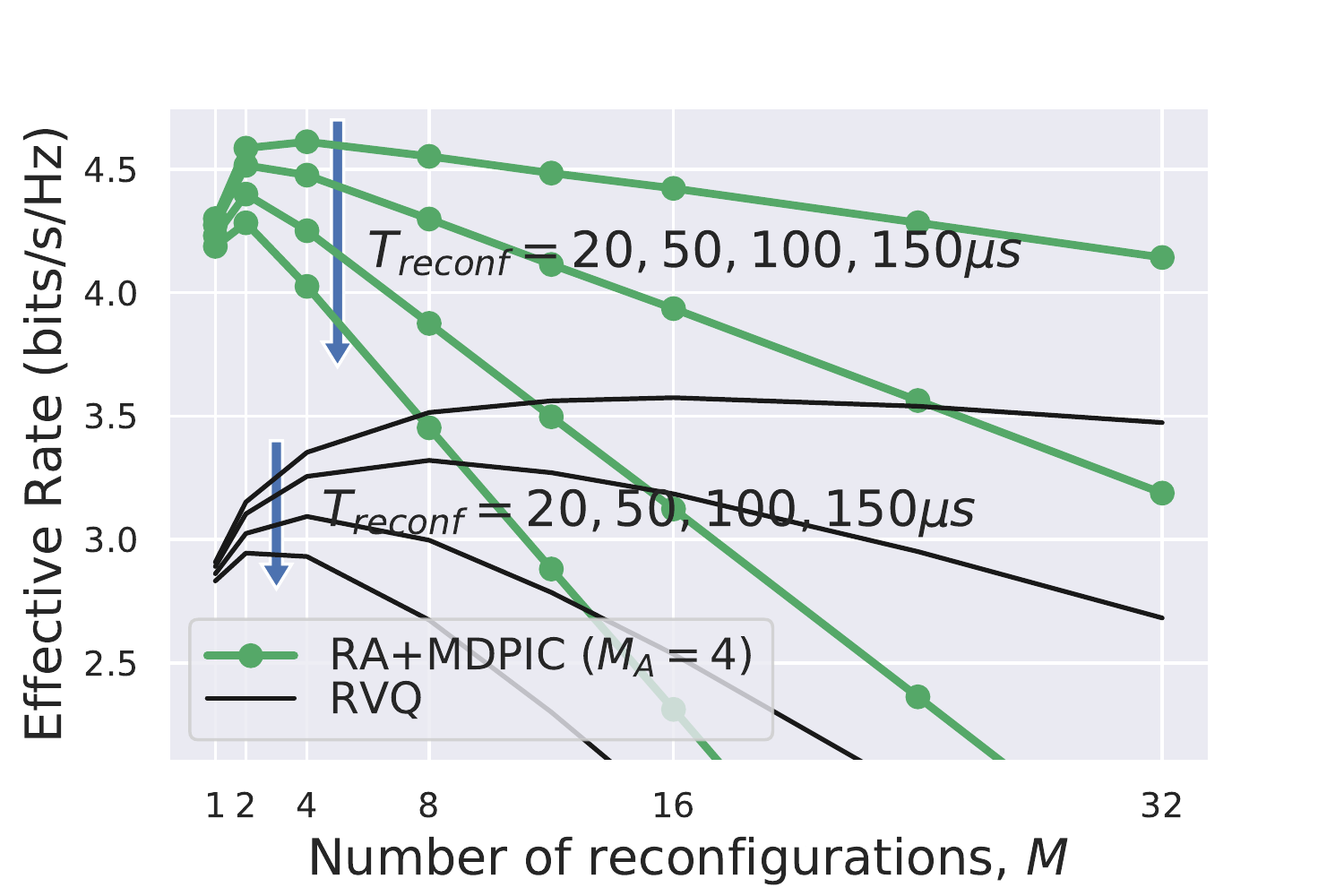}
  \vspace{-10mm}
  \caption{ Effective data rate with different $T_{\rm reconf}$}
  \label{fig:LoS:effrate_Tr}
\end{subfigure}
\begin{subfigure}{.32\linewidth}
  \centering
  \includegraphics[width=\linewidth]{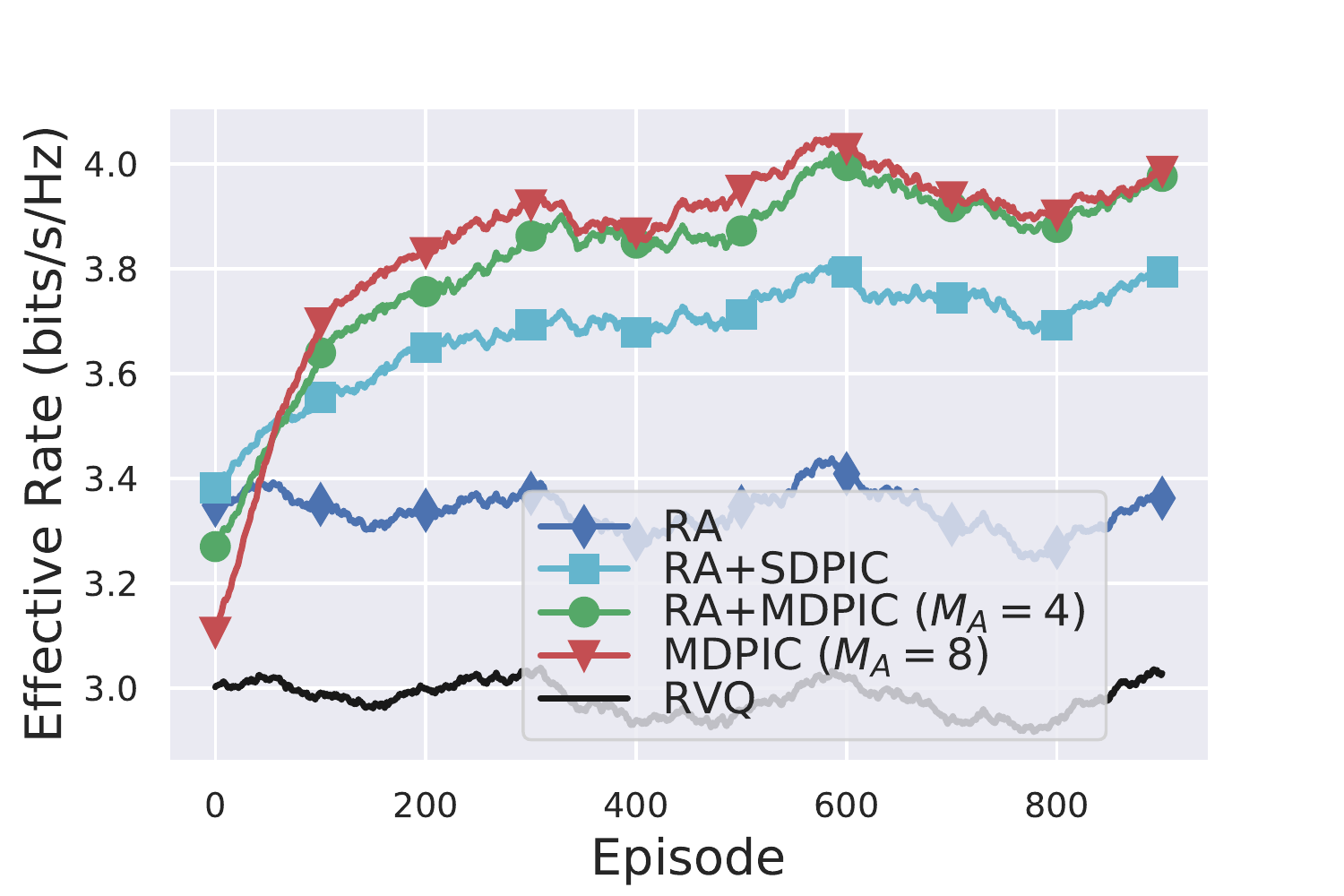}
  \vspace{-10mm}
  \caption{Effective data rate along episodes}
  \label{fig:LoS:effrate_train}
\end{subfigure}
\vspace{-3mm}
\caption{Performance evaluation of our methodology  in Scenario 2. The plots in (\subref{fig:LoS:rate_time})-(\subref{fig:LoS:effrate_Tr}) correspond to the utilization phase, while the plot in (\subref{fig:LoS:effrate_train}) to the training phase.
% In (a), the proposed schemes obtain high data rate within 4-5 timesteps. In (b),  the data rate increases as $M$ increases. In (c), the effective data rate peaks around $M^\star=4$.
% % because the increased time overhead outweighs the data rate improvement when $M>4$. 
% In (d), the peak point varies for the IRS reconfiguration time $T_{\rm reconf}$. 
% In (e), 
% % the hybrid approaches can supplement the data rate  during the beginning of the training, while t
% the RA+MDPIC and MDPIC yield the high effective data rate after the multiple agents are fully trained.
% %The data rate in (a) and the effective data rate in (b) and (c) along the number of IRS reconfiguration, $M$, in the utilization period of Scenario 2. In (c), the best $M^\star$ yielding the highest effective data rate varies by the IRS configuration time $T_{\rm reconf}$.
}
\label{fig:LoS}
\vspace{-8mm}
\end{figure*}
%%%%%%%%%%%%%%%%%%%%
% a-d:2-0, e:2-1

Scenario 2 represents an {\it outdoor UE} with an LoS link to the IRS, for which we follow the same order of the simulations and the same configuration for the proposed schemes as in the previous scenario.
% Fig. \ref{fig:LoS}(a) shows the average data rate along the timesteps over $2000$ episodes during the utilization period.
% Although there exists an LoS link from the UE to the IRS,
While the UE-IRS channel in Scenario 2 is subject to less variations than in Scenario 1, it still corresponds to a dynamic environment which is challenging to address, because the LoS channel between the UE and IRS is different per episode due to the different UE position, and varies over timesteps due to the UE mobility.
% In this scenario,
% the UE-IRS channel is subject to less variations  than in Scenario 1,
% However, the LoS channel is different per episode due to the different UE position, and varies over timesteps due to the UE mobility. This implies that, although an LoS link is presumed in Scenario 2, it still corresponds to a dynamic environment which is challenging to address.
% Nevertheless, due to the existence of dominant path,
% we expect that the BS (or its agents) with the DPIC would perform better in this scenario.
% This is revealed by Fig. \ref{fig:LoS} in which augmented strategies that use multiple agents outperform other methods.
% \rev{The specific configuration of the proposed schemes is the same as in Scenario 1.}

We consider $2000$ episodes each containing $30$ timesteps for the utilization phase.
Fig. \ref{fig:LoS}(\subref{fig:LoS:rate_time}) shows the average data rate along the timesteps over $2000$ episodes.
The performances of the proposed schemes are improved and converge only within $4$-$5$ timesteps. 
The RA+MDPIC and MDPIC yield better performances than other methods by a large margin due to the exploitation of the multiple trained agents.
% The MDPIC outperforms the other methods. The RA+MDPIC yields a good performance along with the MDPIC  due to the exploitation of the multiple trained agents.
%because the trained agents perform well for the codebook update. 
%In Scenario 2, the agents are expected to learn better because a dominant path of the UE-IRS channel is less dynamically varying than in Scenario 1.
Fig. \ref{fig:LoS}(\subref{fig:LoS:rate_M}) shows the data rate along the number of IRS reconfiguration $M$, where each data is averaged over $2000$ episodes and $30$ timesteps. 
The MDPIC outperforms other methods, and is slightly better than the RA+MDPIC due to using more agents.

% As demonstrated in Fig. \ref{fig:LoS}(a),
% MDPIC with multiple agents ($M_{\rm A}=8$) yields the best performance than other methods. 
% The RA+MDPIC ($M_{\rm A} \negmedspace  = \negmedspace  4$) yields a good performance along with the MDPIC  due to the exploitation of the multiple trained agents.
%Its performance is way better than DPIC with a single agent ($M_{\rm A}=1$) due to the advantage of the ensemble learning, as the same as in Fig. \ref{fig:NLoS:util_M}.
In Fig. \ref{fig:LoS}(\subref{fig:LoS:effrate_M}), the MDPIC and RA+MDPIC 
% ($M_{\rm A} \negmedspace = \negmedspace  4$) 
yield better effective data rate performances than those of the RA and RA+SDPIC, although an additional feedback overhead is required 
due to the DPIC updates.
However, the RA+MDPIC is slightly better than the MDPIC because less agents in RA+MDPIC require less feedback overhead for the codebook update.
At $M^\star \negmedspace  = \negmedspace 2$ or $M^\star \negmedspace  = \negmedspace4$ depending on the method, the effective data rate is the highest, which means that, as 
% $M(> \negmedspace  4)$
% increases, 
$M$ grows larger than $2$ or $4$,
the increased overhead outweighs the improvement of the data rate.
% , leading to the effective data rate decreased. 
% In this scenario, training many agents $M_{\rm A} \negmedspace  = \negmedspace  8$ or $M_{\rm A} \negmedspace  = 
Fig. \ref{fig:LoS}(\subref{fig:LoS:effrate_Tr}) shows the effective data rate along $M$ with different $T_{\rm reconf}$. 
With the same reason in Fig. \ref{fig:NLoS}(\subref{fig:NLoS:effrate_Tr}),
$M^\star$ would decrease as $T_{\rm reconf}$ increases, where  $M^\star = 4,2,2,2$ for $T_{\rm reconf} =20, 50, 100, 150 \mu s$, respectively.
We next focus on the training phase with $1000$ episodes each consisting of $500$ timesteps and $M=8$.
% during the training period. 
Fig. \ref{fig:LoS}(\subref{fig:LoS:effrate_train}) shows the effective data rate averaged over $500$ timesteps per episode. Each data point is a moving average of the previous $100$ episodes.
% Thanks to the high performance of the inference from the agents,
%
% In Scenario 2, training more agents, i.e., $M_{\rm A} \negmedspace = \negmedspace 8$, is preferred.
Similar to Scenario 1, 
% in Fig. \ref{fig:NLoS}(\subref{fig:NLoS:effrate_train}), 
the RA+MDPIC yields the best performance during training.

\textbf{Discussion on comprehensive strategy.} Utilizing our learning-based method, i.e.,  the DPIC algorithm, is preferred for both the NLoS and LoS scenarios.
In practice, the BS may not be aware of whether there exists an LoS link from the UE to the IRS. 
The BS can thus select the RA+MDPIC as a comprehensive strategy 
because it yields a satisfactory performance in either of the scenarios.
The RA+MDPIC also has advantages during the training phase since (i) it only requires just a few agents to be trained, leading to less burden for training, and (ii) 
it exhibits a high performance during the training phase.

%% file: conc.tex
\vspace{-3mm}
% \section{Conclusion and Future Work}
\section{Conclusion}
\label{sec:conc}

In this paper, we introduced a novel signal model that takes into account the practical IRS reflection behavior.
To address the design challenges associated with (i) the practical IRS reflection response, (ii) multi-path time-varying channels, and (iii) low-overhead feedback requirement,
we proposed an adaptive codebook-based limited feedback protocol for the IRS-assisted communication.
We proposed two adaptive codebook design approaches: random adjacency (RA) and deep neural network policy-based IRS control (DPIC). Then, we discussed the computational complexity of the RA and DPIC.
Further, we developed several augmented schemes based on the DPIC.
Throughout the simulations, we showed that the data rate performance is improved by the proposed schemes.
In addition, we demonstrated that the average data rate over one coherence time is degraded when the time overhead for the IRS reconfiguration and feedback increases.